\documentstyle[aps,epsfig]{revtex}

\newcommand{\be}{\begin{eqnarray}}
\newcommand{\ee}{\end{eqnarray}}
\newcommand{\bea}{\begin{eqnarray}}
\newcommand{\eea}{\end{eqnarray}}
\newcommand{\beq}{\begin{equation}}
\newcommand{\eeq}{\end{equation}}
\newcommand{\inli}{\int\limits}
\newcommand{\lam}{\lambda}
\newcommand{\vk}{{\bf k}}

\def\fun#1#2{\lower3.6pt\vbox{\baselineskip0pt\lineskip.9pt
\ialign{$\mathsurround=0pt#1\hfil##\hfil$\crcr#2\crcr\sim\crcr}}}

\begin{document}

\title{Quark-antiquark composite systems:\\
the Bethe--Salpeter equation in the spectral-integration
 technique}

\author{A.V. Anisovich, V.V. Anisovich, V.N. Markov,\\
M.A. Matveev ,and A.V. Sarantsev }

\maketitle

\begin{abstract}

The Bethe--Salpeter equations for the quark-antiquark composite
systems, $q\bar q$, are written in terms of  spectral integrals.
For the $q\bar q$-mesons characterized by the mass $M$, spin $J$,
and radial quantum number $n$, the equations are presented for the
following $(n,M^2)$-trajectories: $\pi_J$, $\eta_J$, $a_J$, $f_J$,
$\rho_J$, $\omega_J$, $h_J$, and $b_J$.

\end{abstract}

\section{Introduction}

The relativistic description of composite systems was always an
actual and challenging  task. The most frequently used technique,
which takes into account the relativism of the constituents, is
the Bethe--Salpeter equation \cite{bethe}. But in the standard
formulation of the Bethe--Salpeter equation, when the Feynman
integration technique with mass-off-shell amplitudes is used, one
faces problems of the description of multiparticle channels and
high-spin states. More appropriate technique for the high-spin
composite systems is the dispersion relation approach, in
particular, the most developed $N/D$-method \cite{chew}. However,
our experience and intuition are based on the consideration of the
potential type interactions, i.e., those which are associated with
the particle-exchange mechanism. In terms of the $N/D$-method, one
can easily relate the nearest left-hand side singularity of the
$N$-function to the $t$-channel (or $u$-channel) meson-exchange
diagram, but the reconstruction of the full set of left-hand
singularities, when the interaction is given by the particle
exchange or potential forces, is not a simple problem. Here we
present the Bethe--Salpeter equation in terms of the
spectral-integral technique which has advantages of both
approaches discussed  above: \\ (i) in the spectral integrals, the
mass-on-shell amplitudes are used,  \\ (ii) the interaction terms
are written in the potential or particle-exchange form. \\
Moreover, in the spectral-integral technique one can use the
energy-dependent forces as well.

In \cite{deut1,deut2,AS}, the dispersion-relation approach was
applied to the description of the deuteron, being two-nucleon
composite system, by treating nucleon-nucleon forces in terms of
separable interactions. By using the interaction in a separable
form one can work with mass-on-shell amplitudes and meson-exchange
interactions. The expansion of the one-meson exchange diagrams in
a series of separable interaction factors was developed in
\cite{kapitanov}. The principal points in the transformation of
the standard Bethe--Salpeter equation to the dispersion-relation
representation for the  case of separable vertices were clarified
in \cite{petry}. However, the representation of the meson-exchange
diagram as a finite set of the separable vertices works
successfully for mesons with nonzero mass only. For the long-range
interaction, like confinement forces, the separable-vertex
approach fails, thus forcing us to use not the standard
$N/D$-method but the spectral-integral technique.

The important ingredient of the dispersion relation and
spectral-integration methods is the moment-operator expansion. The
elements of the moment-operator expansion technique were presented
in \cite{deut1, deut2,nato}, and a systematic presentation of the
technique may be found in \cite{operators}.

The Bethe--Salpeter equation in the spectral integral
representation is written here for quark-antiquark systems. Our
attention is focused on the light-quark bound states, $q\bar q$,
where $q=u,d,s$: for the simplicity's sake we consider here the
systems built by quark and antiquark with equal masses: $u\bar d$,
$u\bar u$, $d\bar d$, $s\bar s$. The treatment of heavy-quark
composite systems, $c\bar c$ and $b\bar b$, can be performed
similarly.

The necessity to deal with a full set of  equations for the
light-quark composite systems is governed by the rich information
on the light-meson radiative decays that appeared recently
\cite{kienzele,schegelsky,novosibirsk}.The radiative decay data make it
possible to restore the wave functions of mesons involved into
reactions. Corresponding relativistic technique based on the
consideration of the form-factor amplitudes in terms of the double
spectral integrals was developed in \cite{raddecay1} for
pseudoscalar $q\bar q$-mesons, and it was generalized for scalar
and tensor $q\bar q$-mesons in \cite{raddecay2}. Finding out the
meson wave functions in the spectral-integral form (or in the
light-cone variables, see \cite{raddecay1, raddecay2} for details)
opens the way for direct reconstruction of the quark-antiquark
interactions.  The spectral-integral representation of the
Bethe--Salpeter equation gives us the possibility to find out
directly the interaction forces, provided the masses and wave
functions of composite systems are known: this problem is
discussed in Sect. 2 by using a simplified example of composite
particles with spinless constituents.

For the reconstruction of $q\bar q$ forces it is important for the
light-quark $q\bar q$-states to lay on linear trajectories in the
$(n,M^2)$ plane, where $n$ is the radial quantum number of the
meson with mass $M$ \cite{syst}. In more detail, the $q\bar
q$-states can be classified, within  spectroscopic notations, as
the $n^{2S+1}L_J$ levels, where $S$, $L$, and $J$ refer to the
spin, orbital, and total momenta, respectively. The analysis of
spectra in the mass region 1950--2400 MeV performed in \cite{RAL}
had fixed more than thirty new mesons  which belong to the meson
groups with positive and negative charge parities, $(C=+/-)$.
Namely,for the $(C=+)$-states one has:
\be \label{in1}
&&^1S_0q\bar q\, \to \qquad \pi\mbox{-mesons}\, ,\quad
                               \eta\mbox{-mesons}\, ,\quad
               \eta'\mbox{-mesons}\, ,\\
\nonumber
&&^1D_2q\bar q\, \to\qquad \pi_2\mbox{-mesons}\, ,\quad
                         \eta_2\mbox{-mesons}\, ,
\\
\nonumber
&&^3P_0q\bar q\, \to\qquad a_0\mbox{-mesons}\, ,\quad
                         f_0\mbox{-mesons}\, ,
\\
\nonumber
&&^3P_2q\bar q\, \to\qquad a_2\mbox{-mesons}\, ,\quad
                         f_2\mbox{-mesons}\, ,
\\
\nonumber
&&^3P_1q\bar q\, \to\qquad a_1\mbox{-mesons}\, ,
\\
\nonumber
&&^3F_2q\bar q\, \to\qquad a_2\mbox{-mesons}\, ,\quad
                         f_2\mbox{-mesons}\, ,
\\
\nonumber
&&^3F_3q\bar q\, \to\qquad a_3\mbox{-mesons}\, ,
\\
\nonumber
&&^3F_4q\bar q\, \to\qquad a_4\mbox{-mesons}\, ,\quad
                         f_4\mbox{-mesons}\, ,
\ee
and for the $(C=-)$-states:
\be
\label{in2}
 &&^3S_1q\bar q\, \to\qquad
\rho\mbox{-mesons}\, ,\quad \omega\mbox{-mesons}\, ,\quad
       \phi\mbox{-mesons}\, ,  \\
\nonumber
&&^3D_1q\bar q\, \to\qquad \rho\mbox{-mesons}\, ,
\\
\nonumber
&&^3D_3q\bar q\, \to\qquad \rho_3\mbox{-mesons}\, ,
\\
\nonumber
&&^1P_1q\bar q\, \to\qquad h_1\mbox{-mesons}\, ,\quad
                         b_1\mbox{-mesons}\, ,
\\
\nonumber &&^1F_3q\bar q\, \to\qquad b_3\mbox{-mesons}\, . \ee The
mesons measured in \cite{RAL} as well as those accumulated in the
compilation \cite{PDG-00}, being classified versus radial quantum
number $n$,  can be put, with sufficiently good accuracy, on
linear trajectories in the $(n,M^2)$-plane: \be
M^2=M^2_0+\mu^2(n-1)\, , \qquad n=1,2,3,4,...\, , \ee with the
universal slope $\mu^2\simeq 1.3$ GeV$^2$ \cite{syst}. The
linearity of trajectories, leading and daughter ones, was observed
for the $(J,M^2)$ plane too \cite{syst}.

The linearity of trajectories on the
$(n,M^2)$ and $(J,M^2)$ planes is in a good agreement with
large-$r$ behaviour of the confinement potential,
$V(r) \sim \alpha r$, e.g., see \cite{petry},
where the calculation of $q\bar q$ states from the groups
(\ref{in1}) and (\ref{in2}) has been carried out.

At the same time it is necessary to emphasize that for the
low-mass states one can expect a violation of the trajectory
linearity. For example, the $\pi$-meson is just an exception that
is not surprising because of a particular role of the pion. The
standard explanation is that the pion, being  a low-mass particle,
is  determined by the instanton-induced forces, see
\cite{petry2,inst} and references therein, although one cannot
exclude an alternative modelling of the short-range forces. The
problem of short-range forces is stressed by systematics of scalar
states: the $K$-matrix analysis of $\pi\pi$, $K\bar K$,
$\eta\eta$,and $\eta\eta'$ spectra \cite{K} tells us that the lightest
scalar--isoscalar state  belongs to the flavor octet, but in the
model calculations \cite{petry2,inst} the lightest state is close
to the flavor singlet. We hope that a precise reconstruction of
the $q\bar q$ forces can be facilitated by using the
Bethe--Salpeter equation for the $q\bar q$ states in the
spectral-integral form.

So, we focus our attention on the reconstruction of the $q\bar q$
interaction, on the basis of the following triad:
\\ (1) the Bethe--Salpeter equation in the spectral-integral form,
\\ (2) the linearity of  trajectories on the $(n,M^2)$ and $(J,M^2)$ planes,
\\ (3) the  use of  wave functions for low- and moderate-mass $q\bar q$
states found in the study of the meson radiative decays.

The important point is that radiative decays can give us the
information about  meson wave functions which are now  studied in
the mass region 1000--1800 MeV: just the mesons from this region
are determined by the short-$r$ and intermediate-$r$ forces, and
only the forces from this $r$-region are not known sufficiently
well thus being a subject of discussions and hypotheses.

The paper is organized as follows.  In Section 2 we recall  basic
statements of the Bethe--Salpeter equation written in terms of the
standard Feynman diagram technique, give the elements of the
dispersion relation $N/D$-method, and clarify the interplay of
these two methods by using separable vertices. The spectral
integral representation of the Bethe--Salpeter equation is also
written in this Section for the case of scalar constituents.

In Sect. 3 the $q\bar q$ system is
considered: the Bethe--Salpeter equations are written for the light-quark
mesons which belong to the following $(n,M^2)$-trajectories: $\pi_J$,
$\eta_J$, $a_J$, $f_J$, $\rho_J$, $\omega_J$, $h_J$, and $b_J$.

In Appendices A, B, and C, the necessary auxillary formulae are
presented which were used for deriving the equations.
In Appendix D we collect equations which are rather cumbersome,
these are the Bethe--Salpeter equations for $\omega$, $\phi$, $a_2$, and
$f_2$ trajectories.

\section{Scalar constituents: dispersion relation \\method and the
Bethe--Salpeter equation for composite particles }

In this Section, we compare the  Bethe--Salpeter equation for
composite particles written with the use of Feynman diagrams with
the equation  in terms of the dispersion relations with separable
vertices. This comparison gives us a guide for the transformation
of Bethe--Salpeter equation with separable vertices into the
spectral-integral  Bethe--Salpeter equation with
arbitrary meson-exchange-type interaction.

To simplify the consideration we deal here with scalar particles as
constituents.

\subsection{Bethe--Salpeter equation in the Feynman-diagram
technique }

Written in terms of the Feyman diagrams, the nonhomogeneous
 Bethe--Salpeter equation in the momentum representation reads:
\begin{eqnarray}
\label{3.26}
A(k_1,k_2;k''_1,k''_2)&=&V(k_1,k_2;k''_1,k''_2)+
\int\frac{d^4k'_1\,d^4k'_2}{i(2\pi)^4}\;V(k_1,k_2;k'_1,k'_2)
\nonumber
\\
&\times&\frac{\delta^4(k'_1+k'_2-P)}{(m^2-k'^2_1-i0)(m^2-k'^2_2-i0)}
A(k'_1,k'_2;k''_1k''_2) \, .
\end{eqnarray}
It is shown in Fig. 1 in the diagram form, and one can see there
the notations for particle momenta. In (\ref{3.26}) the
constituents obey the momentum-conservation constraint:

$$
\label{3.28}
k_1\; +\; k_2\;=\;k'_1\; +\; k'_2\;=\;k''_1+k''_2\; =\;P,
$$
and $V(k_1,k_2;k'_1,k'_2)$ is the irreducible kernel, i.e., the
block without two-particle intermediate states (dashed block in
Fig. 1).

The scattering amplitude $A(k_1,k_2;k''_1,k''_2)$ determined by the
Bethe--Salpeter equation (\ref{3.26}) is the mass-off-shell
amplitude. Even if we set
$k^2_1=k''^2_1=k^2_2=k''^2_2=m^2$ in the left-hand side of
Eq. (\ref{3.26}), the right-hand side contains the amplitude
$A(k'_1,k'_2;k''_1,k''_2)$ for $k'^2_1\neq m^2,\; k'^2_2\neq m^2.$

Let us draw the kernel $V$ as a meson-exchange diagram; then, by
iterating Eq. (\ref{3.26}), we represent $A(k_1,k_2;k''_1,k''_2)$
as an infinite set of ladder diagrams of Fig. 2a. For further
investigation it is important to fix intermediate states in the
scattering amplitude.  The ladder diagrams have two-particle
intermediate states which can appear as real states at the c.m.
energies squared $s=P^2 >4m^2$,that corresponds to the cutting of
ladder diagrams across the lines related to the constituents, see
Fig. 2b.

 Such a two-particle state manifests itself as a singularity of the
scattering amplitude at $s=4m^2$. However, the amplitude
$A(k_1,k_2;k''_1,k''_2)$ considered as a function of $s$ has not
only this singularity but also an infinite set of singularities
which correspond to the ladder-diagram cuts across meson lines
associated with the forces: the example of such a cutting is shown
in Fig. 2c. The diagrams, which appear after the cut procedure,
are the meson-production diagrams, e.g., see Fig. 2d.

So, in the complex $s$-plane the amplitude
$A(k_1,k_2;k''_1,k''_2)$ has the following singularity: \beq s\;
=\; 4m^2, \eeq which is related to the rescattering process. The
other singularities are related to the meson production processes
with  cuts originating at \beq \label{3.50} s\; =\; (2m+n\mu)^2 \,
, \qquad n=1,2,3,\ldots\, . \eeq The  four-point amplitude depends
on six variables as follows:
\begin{eqnarray}
 & & k^2_1,\; k^2_2,\; k''^2_1,\; k''^2_2,
\nonumber
\\
s&=&(k_1+k_2)^2\; =\; (k''_1 + k''_2)^2\, ,
\\
t & = & (k_1-k''_1)^2\; =\; (k_2-k''_2)^2.
\nonumber
\end{eqnarray}
The seventh variable, $u=(k_1-k''_2)^2 = (k''_1-k_2)^2$, is not
independent because of the relation
\beq
s+t+u \; =\; k^2_1 +k^2_2+k''^2_1 +k''^2_2.
\eeq
If the interaction creates a bound state, then the infinite set of
ladder diagrams should produce the pole singularity in the
amplitude. Near the pole, the scattering amplitude is determined by the
diagrams of Fig. 3a type that means that
 in the graphical form the equation for composite
system reads as Fig. 3b. In terms of the Feynman
integral, it is as follows:
\be
\label{3.51}
 A(k_1,k_2;P)&=&
\int\frac{d^4k'_1\,d^4k'_2}{i(2\pi)^4}\;V(k_1,k_2;k'_1,k'_2)\times
\nonumber
\\
&\times&\frac{\delta^4(k'_1+k'_2-P)}{(m^2-k'^2_1-i0)(m^2-k'^2_2-i0)}
A(k'_1,k'_2,P).
\ee

The homogeneous Bethe--Salpeter equation (\ref{3.51}), like
nonhomogeneous one, works upon the mass-off-shell amplitudes; the
multimeson production channels in (\ref{3.51}) exist, and they are
strongly related to the meson-exchange forces.

\subsection{Scattering amplitude in the dispersion relation
$N/D$-method}

Let us summarize analytical properties of the discussed scattering
amplitudes for two spinless particles (with the mass $m$) which
interact through the exchange of  another spinless particle (with
the mass $\mu$), Fig. 2a.  This amplitude, $A(s,t)$, has $s$- and
$t$-channel singularities. In the $t$-plane, there are
singularities at $t=\mu^2, 4\mu^2, 9\mu^2$, etc., which correspond
to the one- or multi-meson exchanges. In the $s$-plane the
amplitude has the singularity at $s=4m^2$ (elastic rescattering)
and singularities at $s=(2m+n\mu)^2$, where $n=1,2,\ldots,$that
corresponds to the production of $n$ mesons with the mass $\mu$.
For the bound state with mass $M$, there exists  a pole
singularity at $s=M^2$. If a mass of this bound state  $M>2m$,
this is a resonance, and corresponding pole is located on the
second sheet of the complex $s$-plane.

The dispersion-relation $N/D$-method deals with partial-wave
amplitudes. The $s$-channel partial amplitudes depend on $s$ only. They
have all the $s$-channel singularities of $A(s,t)$, namely, the
right-hand-side singularities at $s=M^2$, $s=4m^2,$ $s=(2m+\mu)^2$, and
so on, see Fig. 4.

Left-hand-side singularities of the partial amplitudes are related
to the $t$-channel singularities of $A(s,t)$. The $S$-wave partial
amplitude is equal to
\beq \label{3.1} A(s)=\int \limits^1_{-1}
\frac{dz}2 \,A(s,t(z)),
\eeq
where $t(z)=-2(s/4\,-m^2)(1-z)$ and
$z=\cos\theta$. The left-hand singularities correspond to
\beq
\label{3.2} t(z=-1)=(n\mu)^2,
\eeq
and they are located at $s=4m^2-\mu^2, s=4m^2-4\mu^2$, and so on.

The $N/D$-method provides us with the possibility to construct the
relativistic two-particle scattering amplitude in the region of
low and intermediate energies, where multiparticle production
processes are not important; this region is shown in Fig. 4 by
dashed line. If the threshold singularity at $s=(2m+\mu)^2$ is not
strong (one-meson production process is suppressed), the region of
partial amplitude under consideration can be expanded up to
the next threshold.

The unitarity condition  for the partial-wave  scattering
amplitude (we consider the $S$-wave amplitude as an example)
reads:
\beq
\label{3.3} \mbox{ Im }A(s)=\rho(s)\mid A(s)\mid^2.
\eeq
Here $\rho(s)$ is the two-particle phase space integrated at
fixed $s$:
\begin{eqnarray}
\label{3.4}
\rho(s)&=&\int d\Phi_2(P;k'_1,k'_2)=\frac1{16\pi}
\sqrt{\frac{s-4m^2}s}, \\
d\Phi_2(P;k'_1,k'_2)&=&\frac12
(2\pi)^4\delta^4(P-k'_1-k'_2)\frac{d^3k'_1}{(2\pi)^32k'_{10}}
\frac{d^3k'_2}{(2\pi)^32k'_{20}}  \ .
\nonumber
\end{eqnarray}
In the $N/D$-method the amplitude $A(s)$ is represented as
\beq
\label{3.5} A(s)=\frac{N(s)}{D(s)},
\eeq
where $N(s)$ has
left-hand singularities only, whereas $D(s)$ has the right-hand
ones only. So the $N$-function is real in the physical region,
$s>4m^2$.  The unitarity condition can be rewritten as follows:
\beq
\label{3.6} \mbox{ Im }D(s)=-\rho(s)N(s).
\eeq
The solution of this equation is
\beq
\label{3.7} D(s)=1-\int
\limits^\infty_{4m^2}\frac{ds'}\pi \frac{\rho(s')
N(s')}{s'-s}\equiv 1-B(s).
\eeq
In Eq. (\ref{3.7}) we suppose that
the CDD poles \cite{CDD} are absent and we normalize $N(s)$ by the
condition $D(s)\to1$ as $s\to\infty$.

In principle, Eqs. (\ref{3.5}), (\ref{3.7}) provide us with a
complete  description of partial amplitude in the low-$s$ region:
the amplitude is determined by the $N$-function being a set of the
left-hand singularities which are due to the one-meson exchange
($s=4m^2-\mu^2$), two-meson exchange ($s=4m^2-4\mu^2$), and so on.
The right-hand singularities in Eqs. (\ref{3.5}), (\ref{3.7}) are
uncoupled to the left-hand ones,  opposite to the Feynman-diagram
approach given by (\ref{3.26}). It is important for the
description of the realistic processes to have the left-hand and
right-hand singularities uncoupled: a well-known example provides
us with the $pn$-amplitude, with the deuteron quantum numbers,
where the production of pions is suppressed (right-hand
singularity at $s=(2m+\mu_\pi)^2$ is weak), while the forces
related to the pion exchange are significant (left-hand
singularity at $s=4m^2-\mu^2_\pi$ is strong).

\subsection{$N/D$-method and separable interaction}
The $N/D$-method gives us the mass-on-shell partial amplitude,
provided the $N$-function is known. However, the $N$-functions have
rather intricate properties: they depend on the total number of
the $t$- and $u$-channel exchanges and do not obey the
factorization constraints, i.e.,for different reactions the
$N$-functions may be independently different. As was stressed
above, the spectral integral representation for the
Bethe--Salpeter equation, keeping advantages of the dispersion
relation method, is free from this problem:  it uses $t$- and $u$-
channel exchanges, with  universal interaction blocks.

As the first step in rewriting the Bethe--Salpeter equation in the
spectral-integral form, let us consider separable interaction as
an example. For this purpose, we rewrite Eqs. (\ref{3.5}),
(\ref{3.7}) introducing the vertex function \beq \label{3.8}
g(s)=\sqrt{N(s)}. \eeq Here we assume that $N(s)$ is positive (the
cases with negative $N(s)$ or with changing-\-sign $N(s)$ need the
introduction of several vertices). So, the partial wave amplitude
$A(s)$ written in terms of the separable vertex $g(s)$ is given by
the following series: \beq \label{3.9}
A(s)=g(s)\left[1+B(s)+B^2(s)+B^3(s)+\cdots \right]g(s). \eeq Its graphic
interpretation is shown by Fig. 5.

This set of diagrams can be rewritten in the form of the
Bethe--Salpeter equation:
\be
A(s)=g^2(s)+g(s)\int \limits^\infty_{4m^2}\frac{ds'}{\pi}g(s')
\frac{\rho (s')}{s'-s}A(s',s),
\ee
where $A(s',s)$ is the
energy-off-shell amplitude which enters the intermediate state of
the diagrams of Fig. 5; one has for the energy-on-shell amplitude
$A(s,s)=A(s)$.The interaction block is written as
follows:
\be
V(s',s)=g(s')\, g(s)
\ee
Therefore, the Bethe--Salpeter equation is to be
applied for the amplitude $A(s,s')$; it reads:
\be \label{3.10}
A(s',s)=g(s')g(s)+g(s')\int
\limits^\infty_{4m^2}\frac{ds''}{\pi}g(s'') \frac{\rho
(s'')}{s''-s}A(s'',s).
\ee

If the bound state exists, the amplitude contains a pole
singularity at $s=M^2$. Considering Eq. (\ref{3.10}) near the pole
and neglecting the non-pole terms, we have the following equation
for the bound state vertex:
\be
\label{3.11} G(s',M^2)=g(s')\int
\limits^\infty_{4m^2}\frac{ds''}{\pi}g(s'') \frac{\rho
(s'')}{s''-M^2}G(s'',M^2),
\ee
where
\be
G(s',M^2)=\left[\frac{A(s',s)(M^2-s)}{G(s)}\right]_{s\to M^2}.
\ee

The Bethe--Salpeter equation (\ref{3.11}) gives
us a guide for the consideration of
 general case, when the interaction is of the meson-exchange type.
But before we need to consider in  more detail the representation
of the loop diagram.

\subsection{Loop diagram}

The loop diagram $B(s)$ plays the decisive role for the two-meson
amplitude, so let us  compare in details  the dispersion and
Feynman integral expressions for $B(s)$.

Namely, the Feynman expression for $B_F(s)$, with a special choice
of separable interaction $G(4k^2_{\perp}+4m^2)$, is proved to be
equal to the dispersion integral representation, where the
four-\-vector $k_\perp$ is defined as follows:
\beq \label{3.12}
2k_\perp=k_1-k_2-\frac{k^2_1-k^2_2}{P^2}(k_1+k_2).
\eeq

In this Section, we use the total-momentum vector $P=k_1+k_2$, so
it is convenient to write here $P^2$ but not $s$.

The Feynman expression for the loop diagram reads:
\beq
\label{3.13}
B_F(P^2)=\frac1{(2\pi)^4i}\int\frac{d^4k_2\,G^2\left(4(Pk_2)^2/P^2\,-4k^2_2
+4m^2\right)}{(m^2-k^2_2-i0)(m^2-(P-k_2)^2-i0)}.
\eeq
Since it is more  convenient to treat  composite
system with  light-\-cone variables, they are hereafter:
\beq \label{3.14}
k_-=\frac1{\sqrt{2}}(k_{20}-k_{2z});\quad
k_+=\frac1{\sqrt{2}}(k_{20}+k_{2z});\quad \vec k_{2\perp}=\vec k_T\, .
\eeq
We choose the reference frame, where $P_T=0$. Then
\beq
\label{3.15} Pk_2=P_+k_- +P_-k_+,
\eeq
and Eq. (\ref{3.13}) takes the form
\bea
\label{3.16}
B_F(P^2)=\frac1{(2\pi)^4i} \;\times
\eea
\bea
\times\int\frac{dk_+dk_-d^2k_T}{(2k_+k_- -m^2_T+i0)(P^2-2(P_+k_-
+P_-k_+)+2k_+k_--m^2_T+i0)}\, ,
\nonumber
\eea
where $m^2_T=m^2+k^2_T\,$. It should be mentioned,that if
$G\equiv1$, one can perform the integration over $k_-$ right now,
closing the integration contour around the pole
\beq \label{3.17}
k_- =\frac{m^2_T-i0}{2k_+}, \eeq
and we obtain the standard dispersion representation for the Feynman loop graph
$(x=k_+/P_+)$:
\be
\label{3.18} \frac1{(2\pi)^4i}\int d^2k_T \int \limits^1_0 dx
\frac{(-2\pi i)}{2x(P^2-m^2_T/x-P^2x+i0)}=
\int\frac{ds}{\pi(s-P^2-i0)}\times
\\
\nonumber
\times\qquad \int\frac{dxdk^2_T}{x(1-x)} \frac{\delta(s-m^2_T/[x(1-x)])}{16\pi}
=\int \limits^\infty_{4m^2} \frac{ds\cdot\rho(s)}{\pi(s-P^2-i0)}.
\ee

The dispersion integral (\ref{3.18}) is divergent at $s\to\infty$
due to $G=1$, and it is the function  $G$ which makes $B_F$
convergent in Eq. (\ref{3.13}). Convergence of the integral
(\ref{3.18}) can be restored by the subtraction procedure.

For $G\neq1$, some additional steps are needed to obtain the
dispersion representation, namely, we introduce new variables $\xi_+$
and $\xi_-$:
\begin{eqnarray}
\label{3.19}
P_+k_- +P_-k_+ &=& \sqrt{P^2} \xi_+
\nonumber
\\
P_+k_--P_-k_+ &=& \sqrt{P^2}\xi_-.
\end{eqnarray}
With these variables, Eq. (\ref{3.13}) takes the form
\begin{eqnarray}
\label{3.20}
B_F(P^2)&=&\frac1{(2\pi)^4i} \;\;\;\times
\nonumber
\\
&\times&\int\frac{G^2(4(\xi^2_-+m^2_T))d\xi_+ d\xi_-
d^2k_T}{(\xi^2_+ -\xi^2_--m^2_T+i0)(P^2-2\sqrt{P^2}
\xi_++\xi^2_+-\xi^2_- -m^2_T+i0)} =
\nonumber
\\
&=&\int dk^2_T \int \limits^\infty_0 2d\xi_-\pi G^2(4(\xi^2_-+m^2_T)) \times\\
&\times&\int
\limits^\infty_{-\infty}\frac{d\xi_+}{[\xi^2_+-(\xi^2_-+m^2_T) +i0]
[(\xi_+-\sqrt{P^2})^2-(\xi^2_-+m^2_T)+i0)]}.
\nonumber
\end{eqnarray}
The integration over $\xi_+$ is performed by closing the
integration contour in the upper half-\-plane, so the  two poles,
$\xi_+=-\sqrt{\xi^2_-+m^2_T}+i0$ and
$\xi_+=\sqrt{P^2}-\sqrt{\xi^2_-+m^2_T}+i0$, contribute.
The introduction of a new variable $s=4(\xi^2_-+m^2_T)$ yields
\beq
\label{3.22} B_F(P^2)=\int
\limits^\infty_{4m^2}\frac{ds\,G^2(s)}{\pi(s-P^2)}\frac1{16\pi}
\sqrt{1-\frac{4m^2}s},
\eeq
that is the dispersion representation of Eq. (\ref{3.7}).

So the hypothesis of the separable interaction gives us an
opportunity  to
solve the Bethe--Salpeter equation  easily. Within this hypothesis we can use
different techniques: either Feynman integration, or spectral-integral
representation, or light-cone variables.

\subsection{Spectral-integral representation and  interaction
forces}

The introduction of a separable interaction is not the only way to
make the Bethe--Salpeter equation easily solvable. The main point in
handling the Bethe--Salpeter equation is to control the right-hand side
singularities, especially those related to  multimeson production,
at $s=(2m+\mu)^2$, $s=(2m+2\mu)^2$,... , and it is the spectral
integration technique which enables us to control the multimeson
production processes.

The spectral integral representation is based on the following
corner stones:\\
(i) constituent particles in the intermediate states are mass-on-shell
 ($k'^2_1=m^2$ and $k'^2_2=m^2$ in Fig. 1);
(ii) there is no energy conservation in the interaction processes
($s \ne s' \ne s''$ in Fig.~5).

Based on these statements we consider potential interaction, or the particle-exchange
interaction, by using the spectral-integral diagrams. Consider as
an example the interaction associated with the $t$-channel exchange by
a meson with the mass $\mu$:
\be \label{3.30}
V(k_1,k'_1)=\frac{g^2}{\mu^2-t}\, , \qquad t=(k_1-k'_1)^2 \, .
\ee

In the c.m. system, which is the most convenient for the consideration,
one has for the four-momenta of the constituent particles:
\be
\label{3.31}
k_1=(k_0,\vec k)=\left(\frac{\sqrt{s}}{2},\vec n
\sqrt{\frac{s}{4}-m^2} \right),
\\
\nonumber
k'_1=(k'_0,\vec k')=\left(\frac{\sqrt{s'}}{2},\vec n'
\sqrt{\frac{s'}{4}-m^2} \right),
\ee
where $\vec n^2=\vec n'^2=1$. The interaction block $V(k_1,k'_1)$ can be
expanded in a series with respect to $z=( \vec n \vec n')$.
In this way, we may obtain the interaction for
different partial waves. For example, the interaction in the wave with
the angular momentum $L=0$ is equal to:
\be
\label{3.32}
V_0(s,s')=\int\limits^1_{-1} \frac{dz}{2}V(k_1,k'_1)\, .
\ee

Actually, Eqs. (\ref{3.30}),(\ref{3.31}),(\ref{3.32}) allow us to
generalize  the procedure with the separable interaction
considered above. Indeed, expanding (\ref{3.32}) in a series with
respect to orthogonal functions, one has
\be \label{3.33}
V_0(s,s')=\sum_n g_n(s)g_n(s'),
\ee
that is
 a separable interaction in a generalized form,
supposing the choice of the functions allows one to use a
restricted number of terms in (\ref{3.33}). Separable interaction
taken in such a form was used in \cite{deut1,deut2} for the
description of  nucleon--nucleon interactions by considering the
deuteron within dispersion-relation technique.

\subsection{Spectral-integral representation of the Bethe--
\\Salpeter equation for  composite system }

First, we consider the case of $L=0$ for  scalar constituents with
equal masses, though not identical. The bound system
is treated as a composite system of these constituents.
Furthermore the case $L\ne 0$ is considered in detail.

\subsubsection{Bethe--Salpeter equation  for  vertex function with $L=0$}

The equation for the vertex $composite\; system \to constituents$,
shown graphically in Fig. 3b, reads:
\be
G(s)=\int \limits^{\infty}_{4m^2}\frac{ds'}{\pi}
\int
d\Phi_2(P';k'_1,k'_2)V(k_1,k_2;k'_1,k'_2) \frac{G(s')}{s'-M^2-i0},
\label{1.1}
\ee
where the phase space is determined by Eq. (\ref{3.4}).
Scalar constituents are supposed to be not identical, so
we do not write  additional identity factor $1/2$ in the phase space.

The equation (\ref{1.1}) written in the spectral-representation
form deals with the energy off-shell states $s'=(k'_1+k'_2)^2 \ne
M^2$, $s=(k_1+k_2)^2 \ne M^2$ and $s\ne s'$; the constituents are
mass-on-shell, $k'^2_1=m^2$ and $k'^2_2=m^2$. We can use
alternative expression for the phase space:
\be
d\Phi_2(P';k'_1,k'_2) = \rho (s') \frac{d z}{2} \equiv d\Phi
(k'), \qquad z=\frac {(kk')}{\sqrt{k^2}\sqrt{k'^2}},
\label{1.3}
\ee
where $k=(k_1-k_2)/2$ and $k'=(k'_1-k'_2)/2$. Then
\be
G(s)=\int \limits_{4m^2}^{\infty}\frac{ds'}{\pi} \int d\Phi
(k') \; V \left (s,s',(kk')\right ) \frac{G(s')}{s'-M^2-i0}.
\label{1.4}
\ee
In the c.m. system $(kk')=-(\vk \vk')$ and
$\sqrt{k^2} =\sqrt{-\vk^2}= i|\vk|$ and $\sqrt{k'^2}
=\sqrt{-\vk'^2}= i|\vk'|$ so $z=(\vk \vk')/(|\vk||\vk'|)$. The
phase space and spectral integrations can be written as follows:
\be
\int \limits^{\infty}_{4m^2}\frac{ds'}{\pi} \int
d\Phi_2(P';k'_1,k'_2)= \int \frac{d \vk'}{(2\pi)^3 k'_{0}} \ ,
\label{1.5}
\ee
where $k'_0=\sqrt{m^2+\vk'^2}$. In the  c.m.
system Eq. (\ref{1.1}) reads
\be
G(s)=\int \frac{d \vk'}{(2\pi)^3
k'_{0}} V \left (s,s',-(\vk\vk')\right ) \frac{G(s')}{s'-M^2-i0}\
. \label{1.6}
\ee

\subsubsection{Bethe--Salpeter equation  for the $(L=0)$-wave function}

Now consider the wave function of composite system,
\be
\psi(s)=\frac{G(s)}{s-M^2}. \label{eq5}
\ee
 To this aim, the
identity transformation upon the equation (\ref {1.6}) should be
done as follows:
\be
\label{eq5.1} (s-M^2)\frac{G(s)}{s-M^2}= \int
\limits_{4m^2}^{\infty} \frac{ds'}{\pi} \int d\Phi (k') V\left(s,
s',(kk')\right) \frac{G(s')}{s'-M^2}\ .
\ee
Using the wave
functions, the equation (\ref{eq5.1}) can be written as follows:
\be (s-M^2)\psi (s)= \int \limits_{4m^2}^{\infty} \frac{ds'}{\pi}
\int d\Phi_2 (k') V\left(s, s',(kk')\right)\psi (s')\ . \label{eq6}
\ee
Finally, using $\vk'^2$ and $\vk^2$ instead of $s'$ and $s$,
$$\psi (s) \to \psi (\vk^2), $$ we have:
\be
(4\vk^2+4m^2-M^2)\psi
(\vk^2)= \int \frac{d \vk'}{(2\pi)^3 k'_{0}} V\left(s,
s',-(\vk\vk')\right)\psi (\vk'^2). \label{eq7}
\ee
This is a basic equation for the set of states with $L=0$. The set is formed
by the levels with different radial excitations $n=1,2,3,...$, and
relevant wave functions are as follows: $$ \psi_1(\vk^2),\
\psi_2(\vk^2),\ \psi_3(\vk^2),... $$ The wave functions are
normalized and orthogonal to each other. The
normalization/orthogonality condition reads:
\be
\int
\frac{d \vk}{(2\pi)^3 k_{0}}
\psi_n(\vk^2)\psi_{n'}(\vk^2)=\delta_{nn'}. \label{eq9}
\ee
Here $\delta_{nn'}$ is the Kronecker symbol. The equation (\ref{eq9})
is due to the consideration of the charge form factors of
composite systems with the gauge-invariance requirement imposed,
see for detail \cite{raddecay1}. This normalization/\\orthogonality
condition looks as in quantum mechanics.

Therefore, the Bethe--Salpeter equation for the $S$-wave mesons reads:
\be
4(\vk^2+m^2)\psi_n(\vk^2)-
\inli^\infty_0 \frac{d\vk'^2}{\pi}V_0(\vk^2,\vk'^2)
\phi(\vk'^2)\psi_n(\vk'^2)=
M_n^2\psi_n(\vk^2)\ ,
\label{eq2.1}
\ee
where
\be
\phi(\vk'^2)=\frac1{4\pi}\frac{|\vk'|}{k'_0}\ .
\label{eq2.2}
\ee
The $\psi_n(\vk^2)$ presents a full set of wave functions which are
orthogonal and normalized:
\be
\inli^\infty_0\frac{d\vk^2}{\pi}
\psi_a(\vk^2)\phi(\vk^2)\psi_b(\vk^2)=\delta_{ab}\ .
\label{eq2.3}
\ee
The function  $V_0(\vk^2,\vk'^2)$ is the projection of potential
$V(s,s',(kk'))$ on the $S$-wave:
\be
V_0(\vk^2,\vk'^2)=\int \frac{d\Omega_{\vk}}{4\pi} \int
\frac{d\Omega_{\vk'}}{4\pi} V\left (s,s',-(\vk\vk')\right ).
\label{eq2.4}
\ee
Let us expand $V_0(\vk^2,\vk'^2)$ with respect to full set of wave
functions:
\be
V_0(\vk^2,\vk'^2)=\sum_{a,b}\psi_a(\vk^2)
v_{ab}^{(0)}\psi_b(\vk'^2),
\label{eq2.5}
\ee
where numerical coefficients $v_{ab}^{(0)}$ are defined by the inverse
transformation as follows:
\be
v_{ab}^{(0)}=
\inli^\infty_0\frac{d\vk^2}{\pi}\frac{d\vk'^2}{\pi}
\psi_a(\vk^2)\phi(\vk^2)
V_0(\vk^2,\vk'^2)\phi(\vk'^2)\psi_b(\vk'^2).
\ee
Taking account of a series (\ref{eq2.5}), the equation (\ref{eq2.1}) is
rewritten as follows:
\be
4(\vk^2+m^2)\psi_n(\vk^2)-
\sum_{a}\psi_a(\vk^2)v_{an}^{(0)}=M_n^2\psi_n(\vk^2).
\label{eq2.7}
\ee
Such a transformation should be carried out upon the kinetic-energy
term, it is also expanded in a series with respect to a full set of
wave functions:
\be
4(\vk^2+m^2)\psi_n(\vk^2)=\sum_a K_{na}\psi_a(\vk^2)\ ,
\label{eq2.8}
\ee
where
\be
K_{na}=\inli^\infty_0 \frac{d\vk^2}{\pi} \psi_a(\vk^2)
\phi(\vk^2)\;
4(\vk^2+m^2)\psi_n(\vk^2)\ .
\label{eq2.9}
\ee
Finally, the Bethe--Salpeter equation takes the form:
\be
\sum_a K_{na}\psi_a(\vk^2)-\sum_a
v_{na}^{(0)}\psi_a(\vk^2)=M^2_n \psi_n(\vk^2)\ .
\label{eq2.10}
\ee
We take into account that
$v_{na}^{(0)}=v_{an}^{(0)}$.

The equation (\ref{eq2.10}) is a standard homogeneous equation:
\be
\sum_a s_{na}\psi_a(\vk^2)=M^2_n\psi_n(\vk^2)\ ,
\label{eq2.11}
\ee
with $s_{na}=K_{na}-v_{na}^{(0)}$.
The values $M^2$ are defined as zeros of the determinant
\be
det|\hat s-M^2I|=0\ ,
\label{eq2.12}
\ee
where $I$ is the unity matrix.

\subsubsection{The Bethe--Salpeter
equation for the states with arbitary angular momentum $L$}

For the wave function with arbitrary  angular momentum $\psi_{(n)\mu_1,\ldots,\mu_L}^{(L)}(s)$, we use the following ansatz:
\be
\psi_{(n)\mu_1,\ldots,\mu_L}^{(L)}(s)
=X^{(L)}_{\mu_1,\ldots,\mu_L}(k)\psi_n^{(L)}(s)\ .
\label{eq3.1}
\ee
The momentum operator $X^{(L)}_{\mu_1,\ldots,\mu_L}(k)$ was
introduced in \cite{operators}, we remind its features in Appendix A.

The Bethe--Salpeter equation
for the $(L,n)$-state, presented in the form similar to
(\ref{eq2.1}), reads:
$$
4(\vk^2+m^2)X^{(L)}_{\mu_1,\ldots,\mu_L}(k)\psi^{(L)}_n(\vk^2)-
X^{(L)}_{\mu_1,\ldots,\mu_L}(k)\times
$$
\be
\times \inli^\infty_0 \frac{d\vk'^2}{\pi}V_L(s,s')
 X^2_L(k'^2)\phi(\vk'^2)\psi_n^{(L)}(\vk'^2)=M_n^2 X^{(L)}_{\mu_1,\ldots,\mu_L}(k)\psi_n^{(L)}(\vk^2),
\label{eq3.2}
\ee
where
\be
X^2_L(k'^2)=\int \frac{d\Omega_{\vk'}}{4\pi}
\left(X^{(L)}_{\nu_1,\ldots,\nu_L}(k')\right ) ^2
=\alpha (L) (k'^2)^L=\alpha (L) (-\vk'^2)^L,
\label{eq3.3}
\ee
\be
\alpha (L)=\frac{(2L-1)!!}{L!}\,\,\,\, , \alpha(0)=1.
\label{eq3.3.1}
\ee
The potential is expanded in a series with respect to the
product of operators
$X^{(L)}_{\mu_1,\ldots,\mu_L}(k) X^{(L)}_{\mu_1,\ldots,\mu_L}(k')$,
that is,
$$
V\left (s,s',(kk')\right )=\sum_{L,\mu_1 \ldots \mu_L}
X^{(L)}_{\mu_1,\ldots,\mu_L}(k) V_L(s,s')
X^{(L)}_{\mu_1,\ldots,\mu_L}(k'),
$$
\be
X^2_L(k^2)V_L(s,s')X^2_L(k'^2)=\label{eq3.4}
\ee
\be
=\int\frac{d\Omega_{\vk}}{4\pi} \frac{d\Omega_{\vk'}}{4\pi}
X^{(L)}_{\nu_1,\ldots,\nu_L}(k) V\left (s,s',(kk')\right )
X^{(L)}_{\nu_1,\ldots,\nu_L}(k').\nonumber
\ee
Therefore,formula (\ref{eq3.2}) reads as follows:
\be
4(\vk^2+m^2)\psi_n^{(L)}(\vk^2)- \inli^\infty_0
\frac{d\vk'^2}{\pi}V_L(s,s')\alpha(L) (-\vk'^2)^L\phi(\vk'^2)
\psi_n^{(L)}(\vk'^2)= \label{eq3.5}
\ee
\be
=M_n^2 \psi_n^{(L)}(\vk^2)& . \nonumber
\ee
As compared to (\ref{eq2.1}) this equation contains additional
factor $X^2_L(k'^2)$; still, the same factor is in the
normalization condition, so it would be reasonable to insert it
into the phase space. Finally, we have:
\be
4(\vk^2+m^2)\psi_n^{(L)}(\vk^2)- \inli^\infty_0
\frac{d\vk'^2}{\pi} \widetilde V_L(s,s')
\phi_L(\vk'^2)\psi_n^{(L)}(\vk'^2) =  M_n^2 \psi_n^{(L)}(\vk^2)\ ,
\label{eq3.6}
\ee
where
\be \phi_L(\vk'^2)=\alpha (L)(\vk'^2)^L
\phi (\vk'^2), \qquad \widetilde V_L(s,s')=(-1)^L V_L(s,s')\, .
\label{eq3.7}
\ee
The normalization condition for  a set of wave
functions with orbital momentum $L$ reads:
\be \inli^\infty_0
\frac{d\vk^2}{\pi} \psi_a^{(L)}(\vk^2)
\phi_L(\vk^2)\psi_b^{(L)}(\vk^2) =\delta_{ab}. \label{eq3.8}
\ee
One can see that it is similar to the case of $L=0$, the only
difference consists in  the  redefinition of the phase space $\phi
\to \phi_L$.
The Bethe--Salpeter equation reads:
\be
\sum_a s^{(L)}_{na}\psi^{(L)}_a(\vk^2)=M^2_{n,L}\psi^{(L)}_n(\vk^2),
\label{eq3.9}
\ee
with
\bea
s^{(L)}_{na}&=&
K^{(L)}_{na}-v_{na}^{(L)},
\nonumber \\
v_{ab}^{(L)}&=&\inli^\infty_0 \frac{d\vk^2}{\pi}\frac{d\vk'^2}{\pi}
\psi^{(L)}_a(\vk^2)\phi_L(\vk^2)
\widetilde V_L(s,s')\phi_L(\vk'^2)\psi^{(L)}_b(\vk'^2),
\nonumber \\
K^{(L)}_{na}&=&\inli^\infty_0 \frac{d\vk^2}{\pi}
\psi^{(L)}_a(\vk^2) \phi_L(\vk) 4(\vk^2+m^2)\psi^{(L)}_n(\vk^2).
\label{eq3.10}
\eea
Using radial-excitation levels one can
reconstruct the potential in the $L$-wave, then to reconstruct,
with the help of (\ref{eq3.4}), the $t$-dependent potential.

\section{Quark-antiquark composite systems}

For the $q \bar q$ system, the Bethe--Salpeter equation for the
wave function with the total momentum $J$, angular momentum
$L=|\vec J -\vec S|$ and quark-antiquark spin $S$ can be
conventionally written as follows:
\be
\left(s-M^2\right)
\widehat\Psi^{(S,L,J)}_{(n)\,\mu_{1}\cdots\mu_{J}} (k)=
\int\frac{d^3 k'}{(2\pi)^3 k'_0}\widehat V\left(
s,s',(kk')\right) \widehat
\Psi^{(S,L,J)}_{(n)\,\mu_{1}\cdots\mu_{J}} (k') \ , \label{bs1}
\ee
where
\be
k=\frac12\left(k_1-k_2\right), \,\,\,\, s=(k_1+k_2)^2 ,
\qquad k'=\frac12\left(k'_1-k'_2\right), \,\,\,\, s'=(k'_1+k'_2)^2  .
\label{bs2}
\ee
The wave-function operator with the fixed quantum numbers is presented as
\be
\widehat \Psi^{(S,L,J)}_{(n)\,\mu_{1}\cdots\mu_{J}}(k)=
\widehat Q^{(S,L,J)}_{\mu_{1}\cdots\mu_{J}}(k)\,\psi^{(S,L,J)}_n (k^2)\ ,
\label{bs3}
\ee
where $\widehat Q$ is the moment operator for the
$q\bar q$ system.

The potential operator can be decomposed as follows:
\be
\widehat V\left(s, s', (kk')\right)=
\sum_{I} V^{(0)}_I\left(s, s', (kk')\right)
\widehat O_I \otimes \widehat O_I\ ,
\label{bs4}
\ee
where $I=S,V,T,A,P$ is a full set of Dirac matrices  in the $t$-channel:
\be
\widehat O_I
={\rm I}, \;\; \gamma_{\mu},\;\; i\sigma_{\mu\nu},\;\;
i\gamma_{\mu}\gamma_5, \;\; \gamma_5 .
\label{bs5}
\ee

The potential operator
$\widehat V\left(s, s', (kk')\right)$
can be decomposed in the $s$ channel by using the Fierz
transformation:
\be
\widehat V\left(s, s', (kk')\right)=
\sum_I\sum_c  \widehat V^{(0)}_I\left(s, s', (kk')\right)
C_{Ic}\, (\widehat O_c\otimes \widehat O_c),
\label{bs6}
\ee
where $C_{Ic}$ are coefficients of the Fierz matrix:
\be
C_{Ic}=
\left(
   \begin{array}{ccccc}
\frac14 & \frac14  & \frac18  & \frac14  & \frac14   \\ \\
1       & -\frac12 & 0        & \frac12  & -1        \\ \\
3       & 0        & -\frac12 & 0        &  3        \\ \\
1       & \frac12  & 0        & -\frac12 & -1        \\ \\
\frac14 & -\frac14 & \frac18  & -\frac14 & \frac14   \\ \\
   \end{array}
\right).
\ee

Denoting
\be
V_c\left(s, s', (kk')\right)=
\sum_I  \widehat V^{(0)}_I\left(s, s', (kk')\right)
C_{Ic}\ ,
\label{bs7}
\ee
we have
\be
\widehat V\left(s, s', (kk')\right)=
\sum_c (\widehat O_c\otimes\widehat  O_c)\,
V_c\left(s, s', (kk')\right)=
\label{bs8}
\ee
$$
=({\rm I}\otimes {\rm I})\,V_S\left(s, s', (kk')\right) +
(\gamma_{\mu}\otimes\gamma_{\mu})\,
V_V\left(s, s', (kk')\right) +
(i\sigma_{\mu\nu}\otimes i\sigma_{\mu\nu})\,\times
\nonumber
$$
$$
\times V_T\left(s, s', (kk')\right)+(i\gamma_{\mu}\gamma_5\otimes i\gamma_{\mu}\gamma_5)\,
V_A\left(s, s', (kk')\right)
+(\gamma_5\otimes\gamma_5)\,
V_P\left(s, s', (kk')\right) .
\nonumber
$$
Let us multiply Eq. (\ref{bs1}) by the operator
$ \widehat Q^{(S,L,J)}_{\mu_1\ldots\mu_J}(k)$
and convolute over the spin-momentum indeces. After re-definition
$\widehat V(s,s',(kk'))\to (\hat k_1'+m') \widehat
V(s,s',(kk'))(-\hat k_2'+m')$ one has:
\be &&\left(s-M^2\right) Sp\left[ \widehat
\Psi^{(S,L,J)}_{(n)\,\mu_1\ldots\mu_J}(k) (\widehat{k}_1+m) \widehat
Q^{(S,L,J)}_{\mu_1\ldots\mu_J}(k) (-\widehat{k}_2+m)\right]= \\
\nonumber
&&=\sum_{c}
Sp\left[\widehat{O}_c\, (\widehat{k}_1+m)
\widehat Q^{(S,L,J)}_{\mu_1\ldots\mu_J}(k)(-\widehat{k}_2+m)\right]
\int\frac{d^3 k'}{(2\pi)^3 k'_0}
V_c\left(s, s', (kk')\right)\times
\\
\nonumber
&&\times
Sp\left[(\widehat{k}'_1+m')
\widehat{O}_c\, (-\widehat{k}'_2+m')
\widehat \Psi^{(S,L,J)}_{(n)\,\mu_1\ldots\mu_J}(k') \right].
\label{bs9}
\ee
Here we define $m'$ as ${k'}_1^2={k'}_2^2={m'}^2$.
We have four states with the $q\bar q$ spins $S=0$ and $S=1$: \\
1) $S=0$; $L=J$, \\
2) $S=1$; $L=J+1,\,J,\,J-1$. \\
These states are constructed from the operators \cite{operators} as
follows:
\be
&& \widehat Q^{(0,J,J)}_{\mu_1\ldots\mu_J}(k)=i\gamma_5
X^{(J)}_{\mu_1\ldots\mu_J}(k),
\\
&& \widehat Q^{(1,J+1,J)}_{\mu_1\ldots\mu_J}(k)=\gamma^{\perp}_\alpha
X^{(J+1)}_{\mu_1\ldots\mu_{J}\alpha}(k),
\\
&& \widehat Q^{(1,J,J)}_{\mu_1\ldots\mu_J}(k)=
\varepsilon_{\alpha\nu_1\nu_2\nu_3}\gamma^{\perp}_{\alpha}P_{\nu_1}
Z^{(J)}_{\nu_2\mu_1\ldots\mu_J,\nu_3} (k),
\\
&& \widehat Q^{(1,J-1,J)}_{\mu_1\ldots\mu_J}(k)=
\gamma^{\perp}_\alpha
Z^{(J-1)}_{\mu_1\ldots\mu_J,\alpha}(k).
\label{bs10}
\ee
For these operators, the wave functions read:

for $J=L+1$,
\be
\widehat \Psi^{(1,J-1,J)}_{(n)\,\mu_1\ldots\mu_J}(k)=
\widehat Q^{(1,J-1,J)}_{\mu_1\ldots\mu_J}(k)\,\psi^{(1,J-1,J)}_n(k^2) \, ,
\label{bs13}
\ee

for $J=L-1$,
\be \widehat \Psi^{(1,J+1,J)}_{(n)\,\mu_1\ldots\mu_J}(k)=
\widehat Q^{(1,J+1,J)}_{\mu_1\ldots\mu_J}(k)\,\psi^{(1,J+1,J)}_n(k^2) \, ,
\label{bs14}
\ee

for $S=1,\, L=J\pm 1,J$,
\be
\widehat \Psi^{(1,(J\pm 1),J)}_{(n)\,\mu_1\ldots\mu_J}(k)=
A_j
\widehat \Psi^{(1,J-1,J)}_{(n)\,\mu_1\ldots\mu_J}(k)
+B_j
\widehat \Psi^{(1,J+1,J)}_{(n)\,\mu_1\ldots\mu_J}(k).
\label{bs15}
\ee
where $A_j$ and $B_j$ are the mixing coefficients with $j=1,2$.

These wave functions are normalized as follows:
\be
\int\frac{d^3k}{(2\pi)^3 k_0}
(-1)Sp\left[
\widehat \Psi^{(S',L'_{j'},J')}_{(n')\,\mu_1\ldots\mu_J}(k)
(\widehat{k}_1+m)
\widehat \Psi^{(S,L_j,J)}_{(n)\,\mu_1\ldots\mu_J}(k)
(-\widehat{k}_2+m)\right]
\ee
$$
=(-1)^{J} \delta_{S',S}
\delta_{L'_{j'},L_{j}}
\delta_{J',J}\delta_{n',n}\, .
\label{bs12}
$$

\subsection{ Equation for ($S=0$, $J=L$)-state}

The equation for the state with $S=0$, $J=L$ reads:
\be
\left(s-M^2\right) \,X^{(J)}_{\mu_1\ldots\mu_J}(k)\,
Sp\left[i\gamma_5 (\widehat{k}_1+m)
i\gamma_5(-\widehat{k}_2+m)\right] \times
\label{3.1.1}
\ee
$$
\times X^{(J)}_{\mu_1\ldots\mu_J}(k)\,
\psi^{(0,J,J)}_n(k^2)=X^{(J)}_{\mu_1\ldots\mu_J}(k)\,
\sum_c Sp\left[\widehat{F}_c\, (\widehat{k}_1+m)
i\gamma_5(-\widehat{k}_2+m)\right] \times
\nonumber
$$
$$
\times\int\frac{d^3 k'}{(2\pi)^3 k'_0}
V_c \left(s, s', (kk')\right)
Sp\left[i\gamma_5(\widehat{k}'_1+m')
\widehat{F}_c\, (-\widehat{k}'_2+m')\right] \times
\nonumber
$$
$$
\times X^{(J)}_{\mu_1\ldots\mu_J}(k')\,
\psi^{(0,J,J)}_n(k'^2).
\nonumber
$$
Now consider the left-hand side of the  Eq. (\ref{3.1.1}).
Using the traces presented in
Appendix B and   convolution of
operators from Appendix C, we have:
\be
X^{(J)}_{\mu_1\ldots\mu_J}(k)
Sp\left[i\gamma_5 (\widehat{k}_1+m)i\gamma_5(-\widehat{k}_2+m)\right]
X^{(J)}_{\mu_1\ldots\mu_J}(k)=
\nonumber
\label{3.1.2}
\ee
$$
=X^{(J)}_{\mu_1\ldots\mu_J}(k)\, (-2s) \,
X^{(J)}_{\mu_1\ldots\mu_J}(k)=-2s\,\alpha (J)\, k^{2J} \ .
\nonumber
$$
The right-hand side of the equation is calculated in two steps:
first, we summarize with respect to $c$:
\be
\label{3.1.3}
A\left(s, s', (kk')\right)=
\sum_{c=T,A,P} A_c\left(s, s', (kk')\right)V_c\left(s, s', (kk')\right)=
\ee
$$
=
\sum_{c=T,A,P}  Sp\left[\widehat{F}_c\, (\widehat{k}_1+m)
i\gamma_5(-\widehat{k}_2+m)\right]
Sp\left[i\gamma_5(\widehat{k}'_1+m')
\widehat{F}_c\, (-\widehat{k}'_2+m')\right]\times\nonumber
$$
$$
\times V_c\left(s, s', (kk')\right).
\nonumber
$$
In Appendix B the trace calculations are presented, and the values\\
$A_c\left(s, s', (kk')\right)$
are given. In this way, the sum is written as follows:
\be
\label{3.1.4}
A\left(s, s', (kk')\right)=
\sum_{c=T,A,P}  A_c\left(s, s', (kk')\right)
V_c\left(s, s', (kk')\right)=-4\sqrt{ss'}\times
\ee
$$
\times \left[\sqrt{ss'}\, V_P\left(s, s', (kk')\right)+
4mm'\, V_A\left(s, s', (kk')\right)
+8(kk')\, V_T\left(s, s', (kk')\right)\right].
\nonumber
$$
At the second step, the convolution of operators is performed by using
equations of
Appendix C and recurrent formulae for the Legendre polynomials:
$$
zP_J (z)=\frac{J+1}{2J+1}P_{J+1}(z)+\frac{J}{2J+1}P_{J-1}(z)\, ,
\nonumber
$$
that allows us to write the Bethe--Salpeter equation in terms of the
Legendre polynomials (recall that $z=(kk')/(\sqrt{k^2}\sqrt{k'^2})$
and
$\sqrt{k^2}=i \sqrt{s/4-m^2}$,
$\sqrt{k'^2}=i \sqrt{s'/4-m'^2}$).
For the exceptional case $J=0$ we put $P_{-1}(z)=0$
As a result we get:
\be
\label{3.1.5}
X^{(J)}_{\mu_1\ldots\mu_J}(k)\, A\left(s, s', (kk')\right)
X^{(J)}_{\mu_1\ldots\mu_J}(k')=
\alpha(J)\left(\sqrt{k^2}\sqrt{k'^2}\right)^J (-4\sqrt{ss'})\times
\ee
$$
\times \left[
8\frac{J+1}{2J+1}\sqrt{k^2}\sqrt{k'^2}P_{J+1}(z)\,
V_T\left(s, s', (kk')\right)+ \right .
\nonumber
$$
$$
\left .
+\left(\sqrt{ss'}\,
V_P\left(s, s', (kk')\right)+4mm'\,
V_A\left(s, s', (kk')\right)\right)P_J(z)+
\right .
\nonumber
$$
$$
\left .
+8\frac{J}{2J+1}\sqrt{k^2}\sqrt{k'^2}P_{J-1}(z)\,
V_T\left(s, s', (kk')\right)
\right].
\nonumber
$$
Substituting the obtained expressions into Eq.
(\ref{3.1.1}), we obtain:
\be
\label{3.1.6}
\left(s-M^2\right)\, (-2s)\,
\alpha(J)\, k^{2J}
\psi^{(0,J,J)}_n(k^2)
=\int\frac{d^3 k'}{(2\pi)^3 k'_0}\times
\ee
$$
\nonumber
\times (-4\sqrt{ss'})\,
\alpha(J)\left(\sqrt{k^2}\sqrt{k'^2}\right)^J
\left [
8\frac{J+1}{2J+1}\sqrt{k^2}\sqrt{k'^2}P_{J+1}(z)\,
V_T\left(s, s', (kk')\right)+
\right .
$$
$$
\nonumber
\left .
+\left(\sqrt{ss'}\,
V_P\left(s, s', (kk')\right)+4mm'\,
V_A\left(s, s', (kk')\right)\right)P_J(z)+ \right .
$$
$$
\nonumber
\left .
+8\frac{J}{2J+1}\sqrt{k^2}\sqrt{k'^2}P_{J-1}(z)\,
V_T\left(s, s', (kk')\right)
\right ]
\psi^{(0,J,J)}_n(k'^2).
$$
Expanding the interaction block in the Legendre polynomial
series,
\be
\label{3.1.7}
V_c\left(s, s', (kk')\right)=
\sum_{J} V^{(J)}_c\left(s, s' \right) P_J(z)=
\ee
$$
=\sum_{J}\widetilde
 V^{(J)}_c\left(s, s' \right)
\alpha(J)\left(-\sqrt{k^2}\sqrt{k'^2}\right)^J P_J(z),
$$
and integrating over angle variables in the right-hand side
by taking into account
the standard normalization condition
$\int ^{1}_{-1}dz/2 \;  P^2_J(z)=1/(2J+1)$,
we have finally:
\be
\label{3.1.8}
\left(s-M^2\right) \psi^{(0,J,J)}_n(s)
=\int\limits_{4m'^2}^{\infty}\frac{ds'}{\pi}\rho (s')
(-k'^2)^J 2 \sqrt{\frac{s'}{s}}\times
\ee
$$
\times
\left[
 -8\frac{J+1}{2J+1}\,\xi (J+1) k^2k'^2 \,
   \widetilde V^{(J+1)}_T\left(s, s' \right)
 +\sqrt{ss'}\, \xi (J)\,
   \widetilde V^{(J)}_P\left(s, s' \right)+
\right .
$$
$$
\left .
 +4mm'\,\xi (J)\,
   \widetilde  V^{(J)}_A\left(s, s' \right)
 -8\frac{J}{2J+1}\,\xi (J-1)
   \widetilde V^{(J-1)}_T \left(s, s'\right)
\right]
\psi^{(0,J,J)}_n(s').
$$
where:
$$
\int ^{1}_{-1}\frac{dz}{2} P_J (z) V_c\left(s, s', (kk')\right)=
\int ^{1}_{-1}\frac{dz}{2} P_J (z)
\sum_{J'}
\widetilde V^{(J')}_c\left(s, s' \right)P_{J'}(z)\alpha(J')\times
$$
$$
\times\left(-\sqrt{k^2}\sqrt{k'^2}\right)^{J'}=
\frac{\alpha(J)}{2J+1}\,\widetilde V^{(J)}_c\left(s, s' \right)
\left(-\sqrt{k^2}\sqrt{k'^2}\right)^J=
$$
$$
=\xi (J)\,\widetilde V^{(J)}_c\left(s, s' \right)
\left(-\sqrt{k^2}\sqrt{k'^2}\right)^J \, ,
$$
\be
\xi (J)=\frac{\alpha(J)}{2J+1}=
\frac{(2J-1)!!}{(2J+1)\cdot J!}\ .
\ee

\subsubsection{ Equation for the pion ($M^2,n$)-trajectory}

The pion states which belong to the pion ($M^2,n$)-trajectory
obey the following equation:
\be
\label{3.1.9}
\left(s-M^2\right) \psi^{(0,0,0)}_{\pi,n}(s)
=\int\limits_{4m^2}^{\infty}\frac{ds'}{\pi}\rho (s')
\;
2 \sqrt{\frac{s'}{s}} \times
\ee
$$
\nonumber
\times
\left [
  -\frac{8}{3}k^2 k'^2\,
     \widetilde V^{(1)}_T\left(s, s' \right)
  +\sqrt{ss'}\,\widetilde V^{(0)}_P\left(s, s' \right)
  +4m^2\, \widetilde  V^{(0)}_A\left(s, s' \right)
\right ]
\psi^{(0,0,0)}_{\pi,n}(s').
\nonumber
$$
Recall that $n$ is the radial quantum number, and the
following states with different $n$ are located on the discussed
($M^2,n$)-trajectory: $\pi (140)$ with $n=1$, $\pi (1300)$ with $n=2$,
$\pi (1800)$ with $n=3$, $\pi (2070)$ with $n=4$, $\pi (2360)$ with
$n=5$, and so on.

The wave functions of the states laying on the ($M^2,n$)-trajectory
satisfy  the orthogonality/normalization constraint:
\be
\int\limits_{4m^2}^{\infty}\frac{ds}{\pi}\rho (s) \; 2s\;
\psi^{(0,0,0)}_{\pi,n'}(s)\psi^{(0,0,0)}_{\pi,n}(s)=
\delta_{n',n} \; .
\ee
The factor $2s$ is due to summing over the spin variables
of quarks.

Expanding the interaction block over full set of the radial wave
functions, we can transform (\ref{3.1.9}) into a system of the linear
equations of the type of (\ref{eq2.11}).

\subsubsection{ Equation for the $\eta$ ($M^2,n$)-trajectory}

The $\eta$-states have two components, $n\bar n =(u\bar u+d\bar
d)/\sqrt 2$ and $s\bar s$. We write
$\eta_n = \cos \Theta_n \; n\bar n +\sin \Theta_n\; s\bar s$,
and
$\eta'_n =- \sin \Theta_n \; n\bar n +\cos \Theta_n \; s\bar s$.
For the lightest mesons $\eta (550)$ and $\eta'(958)$, one
has
$\cos \Theta_1=\simeq 0.8$ and $\sin \Theta_1 \simeq -0.6$.

Correspondingly, we have two equations for the wave
functions which describe the $n\bar n$ and $s\bar s$ components:
\be
\left(s-M^2\right) \psi^{(0,0,0)}_{\eta (n\bar n),n}(s) \cos\Theta_n
=\int\limits_{4m^2}^{\infty}\frac{ds'}{\pi}\rho (s')
\;
2 \sqrt{\frac{s'}{s}}\times
\ee
$$
\times\left [
  -\frac{8}{3} k^2 k'^2
     \widetilde V^{(1)}_{(n\bar n \to n\bar n), T}\left(s, s' \right)
  +\sqrt{ss'}
     \widetilde V^{(0)}_{(n\bar n \to n\bar n), P}\left(s, s' \right)
  +4m^2
     \widetilde  V^{(0)}_{(n\bar n \to n\bar n), A}\left(s, s'\right)
\right ]\times
$$
$$
\times \psi^{(0,0,0)}_{\eta (n\bar n) ,n}(s')\cos\Theta_n+\int\limits_{4m_s^2}^{\infty}\frac{ds'}{\pi}\rho_s (s')
\;
2 \sqrt{\frac{s'}{s}}
\left [
  -\frac{8}{3}k^2 k'^2_s\,
     \widetilde V^{(1)}_{(s\bar s\to n\bar n ), T}\left(s, s' \right)+
\right .
$$
$$
\left .
   +\sqrt{ss'}\,
     \widetilde V^{(0)}_{(s\bar s\to n\bar n ), P}\left(s, s' \right)
   +4mm_s\,
     \widetilde  V^{(0)}_{(s\bar s\to n\bar n ), A}\left(s, s' \right)
\right ]
\psi^{(0,0,0)}_{\eta (s\bar s) ,n}(s')\sin\Theta_n \, ,
\nonumber
$$
where $\rho_s(s')$ refers to the $s\bar s$ phase space and
$\sqrt{k'^2_s}=i\sqrt{s'/4-m^2_s}$. The second equation, for the
$s\bar s$-component, reads:
\be
\left(s-M^2\right) \psi^{(0,0,0)}_{\eta (s\bar s),n}(s)
\sin\Theta_n =\int\limits_{4m^2}^{\infty}\frac{ds'}{\pi}\rho (s') \;
2 \sqrt{\frac{s'}{s}}\times
\ee
$$
\nonumber
\left [
  -\frac{8}{3}k_s^2k'^2
     \widetilde V^{(1)}_{(n\bar n \to s\bar s), T}\left(s, s' \right)
  +\sqrt{ss'}
     \widetilde V^{(0)}_{(n\bar n \to s\bar s), P}\left(s, s' \right)
  +4mm_s
     \widetilde  V^{(0)}_{(n\bar n \to s\bar s), A}\left(s, s' \right)
\right ]\times
\nonumber
$$
$$
\times \psi^{(0,0,0)}_{\eta (n\bar n) ,n}(s')\cos\Theta_n+
\int\limits_{4m_s^2}^{\infty}\frac{ds'}{\pi}\rho_s (s')
\;
2 \sqrt{\frac{s'}{s}}
\left [
  -\frac{8}{3}k^2_s k'^2_s\,
     \widetilde V^{(1)}_{(s\bar s \to s\bar s), T}\left(s, s' \right)+
\right .
$$
$$
\nonumber
\left .
  +\sqrt{ss'}\,
     \widetilde V^{(0)}_{(s\bar s \to s\bar s), P}\left(s, s' \right)
  +4m_s^2 \,
     \widetilde  V^{(0)}_{(s\bar s \to s\bar s), A}\left(s, s'
\right)
\right ]
\psi^{(0,0,0)}_{\eta (s\bar s) ,n}(s')\sin\Theta_n .
\nonumber
$$
The wave functions $\psi^{(0,0,0)}_{\eta (n\bar n) ,n}(s)$
and $\psi^{(0,0,0)}_{\eta (s\bar s) ,n}(s)$ satisfy the
normalization condition within an obvious change of
the integration region for the $s\bar s$ component: $4m^2\to 4m^2_s$.

The following states are located on the $\eta$ and $\eta'$
($M^2,n$)-trajectories:\\
1) $\eta$-trajectory: $\eta (550)$ with $n=1$, $\eta (1295)$ with $n=2$,
$\eta (1700)$ with $n=3$, $\eta (2010)$ with $n=4$,
$\eta (2320)$ with $n=5$, and so on.\\
2) $\eta'$-trajectory: $\eta' (958)$ with $n=1$, $\eta (1440)$ with $n=2$,
$\eta (1820)$ with $n=3$, and so on.\\

\subsubsection{ Equation for the $b_1$ ($M^2,n$)-trajectory}

The equation for the states with $S=0$, $L=1$, $J=1$ reads:
\be
\left(s-M^2\right) \psi^{(0,1,1)}_{b_1,n}(s)
=-\int\limits_{4m^2}^{\infty}\frac{ds'}{\pi}\rho (s')
\; k'^2\, \frac23 \sqrt{\frac{s'}{s}}\times
\ee
$$
\times
\left [
  -\frac{24}{5}k^2 k'^2\,
     \widetilde V^{(2)}_T\left(s, s' \right)
  +\sqrt{ss'}\,
     \widetilde V^{(1)}_P\left(s, s' \right)
  +4m^2\, \widetilde  V^{(1)}_A\left(s, s' \right)
  -8\,\widetilde  V^{(0)}_T\left(s, s' \right)
\right ]\times
$$
$$
\times \psi^{(0,1,1)}_{b_1,n}(s').
$$
The following states are located on the $b_1$
($M^2,n$)-trajectories: $b_1 (1235)$ with $n=1$, $b_1 (1640)$ at
$n=2$, $b_1 (1970)$ with $n=3$, $b_1 (2210)$ with $n=4$, and so on.

The wave functions of the $b_1$-states laying on the
($M^2,n$)-trajectory satisfy the orthogonality/normalization
constraint:
\be
\int\limits_{4m^2}^{\infty}\frac{ds}{\pi}\rho (s) \;
2s k^2\; \psi^{(0,1,1)}_{b_1,n'}(s)
\psi^{(0,1,1)}_{b_1,n}(s)=\delta_{n',n}.
\ee
\subsubsection{ Equation for the $h_1$ ($M^2,n$)-trajectory}

The $h_1$-states have two components,
$n\bar n =(u\bar u+d\bar d)/\sqrt 2$ and $s\bar s$; we write
$h_{1,n} = \cos \Theta_n \; n\bar n +\sin \Theta_n\; s\bar s$.
Correspondingly, we have two equations for the wave-function
 $n\bar n$-component:
\be
\left(s-M^2\right) \psi^{(0,1,1)}_{h_1 (n\bar n),n}(s) \cos\Theta_n
=-\int\limits_{4m^2}^{\infty}\frac{ds'}{\pi}\rho (s')
\; k'^2\,
\frac23 \sqrt{\frac{s'}{s}}\times
\ee
$$
\nonumber
\times \left [
  -\frac{24}{5} k^2 k'^2\,
     \widetilde V^{(2)}_{(n\bar n \to n\bar n), T}\left(s, s' \right)
  +\sqrt{ss'}\,
     \widetilde V^{(1)}_{(n\bar n \to n\bar n), P}\left(s, s' \right)+
\right .
$$
$$
\left .
 +4m^2\,
     \widetilde  V^{(1)}_{(n\bar n \to n\bar n), A}\left(s, s'\right)-8\,
     \widetilde V^{(0)}_{(n\bar n \to n\bar n), T}\left(s, s' \right)
\right ] \psi^{(0,1,1)}_{h_1 (n\bar n) ,n}(s')\cos\Theta_n-
$$
$$
-\int\limits_{4m_s^2}^{\infty}\frac{ds'}{\pi}\rho_s (s')
\;k'^2_s
\frac23 \sqrt{\frac{s'}{s}}
 \left [
  -\frac{24}{5}k^2 k'^2_s
     \widetilde V^{(2)}_{(s\bar s\to n\bar n ), T}\left(s, s' \right)
   +\sqrt{ss'}
     \widetilde V^{(1)}_{(s\bar s\to n\bar n ), P}\left(s, s' \right)+
   \right .
$$
$$
\left .
+4mm_s\,
     \widetilde  V^{(1)}_{(s\bar s\to n\bar n ), A}\left(s, s' \right)
  -8\,
     \widetilde V^{(0)}_{(s\bar s \to n\bar n), T}\left(s, s' \right)
\right ]
\psi^{(0,1,1)}_{h_1 (s\bar s) ,n}(s')\sin\Theta_n,
$$
where $\rho_s(s')$ refers to the $s\bar s$ phase space
and
$\sqrt{k'^2_s}=i\sqrt{s'/4-m^2_s}$.
For the $s\bar s$-component we have:
\be
\left(s-M^2\right) \psi^{(0,1,1)}_{h_1 (s\bar s),n}(s)
\sin\Theta_n =-\int\limits_{4m^2}^{\infty}\frac{ds'}{\pi}\rho (s')
\; k'^2
\frac23 \sqrt{\frac{s'}{s}}\times
\ee
$$
\times\left [
  -\frac{24}{5}k_s^2k'^2\,
     \widetilde V^{(2)}_{(n\bar n \to s\bar s), T}\left(s, s' \right)
  +\sqrt{ss'}\,
     \widetilde V^{(1)}_{(n\bar n \to s\bar s), P}\left(s, s' \right)+
\right.
$$
$$
\left.
  +4mm_s\,
     \widetilde  V^{(1)}_{(n\bar n \to s\bar s), A}\left(s, s' \right)
  -8\,
     \widetilde V^{(0)}_{(n\bar n \to s\bar s), T}\left(s, s' \right)
\right ]
\psi^{(0,1,1)}_{h_1 (n\bar n) ,n}(s')\cos\Theta_n-
$$
$$
-\int\limits_{4m_s^2}^{\infty}\frac{ds'}{\pi}\rho_s (s')
\; k'^2_s
\frac23 \sqrt{\frac{s'}{s}}
\left [
  -\frac{24}{5}k^2_s k'^2_s\,
     \widetilde V^{(2)}_{(s\bar s \to s\bar s), T}\left(s, s' \right)
      +\sqrt{ss'}\,
     \widetilde V^{(1)}_{(s\bar s \to s\bar s), P}\left(s, s' \right)+
\right .
$$
$$
\left .
  +4m_s^2 \,
     \widetilde  V^{(1)}_{(s\bar s \to s\bar s), A}\left(s, s'\right)
  -8\,
     \widetilde V^{(0)}_{(s\bar s \to s\bar s), T}\left(s, s' \right)
\right ]
\psi^{(0,1,1)}_{h_1 (s\bar s) ,n}(s')\sin\Theta_n \, .
$$
The wave functions $\psi^{(0,1,1)}_{h_1 (n\bar n) ,n}(s)$
and $\psi^{(0,1,1)}_{h_1 (s\bar s) ,n}(s)$ satisfy the
normalization condition within the obvious change
the integration region for the $s\bar s$ component: $4m^2\to 4m^2_s$.

The following states are located on the $h_1$
($M^2,n$)-trajectories:\\
1) $h_1 (1170)$ with $n=1$, $h_1 (1600)$ with $n=2$,
$h_1 (2000)$ with $n=3$, $h_1 (2270)$ with $n=4$, and so on; \\
2) $h_1 (1390)$ with $n=1$, $h_1 (1780)$ with $n=2$,
$h_1 (2120)$ with $n=3$, and so on.

\subsubsection{ Equation for the $\pi_2$ ($M^2,n$)-trajectories}

The equation for the $\pi_2$-states ($S=0$, $L=2$, $J=2$) reads:
\be
\left(s-M^2\right) \psi^{(0,2,2)}_{\pi_2,n}(s)
=\int\limits_{4m^2}^{\infty}\frac{ds'}{\pi}\rho (s')k'^4
\;
\frac35 \sqrt{\frac{s'}{s}}\times
\ee
$$
\times
\left [
  -\frac{40}{7}k^2 k'^2\,
     \widetilde V^{(3)}_T\left(s, s' \right)
  +\sqrt{ss'}\widetilde V^{(2)}_P\left(s, s' \right)
  +4m^2\widetilde  V^{(2)}_A\left(s, s' \right)
  -\frac{32}{9}
     \widetilde V^{(1)}_T\left(s, s' \right)
\right ]\times
$$
$$
\times \psi^{(0,2,2)}_{\pi_2,n}(s').
$$
The following states are located on the $\pi_2$ ($M^2,n$)-trajectory
\cite{syst}:
$\pi_2(1670)$ with $n=1$, $\pi_2(2005)$ with $n=2$, $\pi_2(2245)$ with
$n=3$,and so on.

The wave functions of the $\pi_2$-states
satisfy the orthogonality/normalization constraint:
\be
\label{3.1.10}
\int\limits_{4m^2}^{\infty}\frac{ds}{\pi}\rho (s)\, k^4 \;
2s\,\alpha (2) \psi^{(0,2,2)}_{\pi,n'}(s)\psi^{(0,2,2)}_{\pi,n}(s)
= \delta_{n',n}.
\ee
where $\alpha (2)$ is determined by  Eq. (\ref{eq3.3.1}).

\subsubsection{ Equation for the $\eta_2$ ($M^2,n$)-trajectory}

The $\eta_2$-states have two components,
$n\bar n =(u\bar u+d\bar d)/\sqrt 2$ and $s\bar s$. We write
$\eta_{2,n} = \cos \Theta_n \; n\bar n +\sin \Theta_n\; s\bar s$, and,
correspondingly, we have two equations for the wave
functions:
\be
\left(s-M^2\right) \psi^{(0,2,2)}_{\eta_2 (n\bar n),n}(s) \cos\Theta_n
=\int\limits_{4m^2}^{\infty}\frac{ds'}{\pi}\rho (s')
\; k'^4\,
\frac35 \sqrt{\frac{s'}{s}}\times
\ee
$$
\nonumber
\times \left [
  -\frac{40}{7} k^2 k'^2\,
     \widetilde V^{(3)}_{(n\bar n \to n\bar n), T}\left(s, s' \right)
  +\sqrt{ss'}\,
     \widetilde V^{(2)}_{(n\bar n \to n\bar n), P}\left(s, s' \right)+
\right .
\nonumber
$$
$$
\left .
+4m^2\,
     \widetilde  V^{(2)}_{(n\bar n \to n\bar n), A}\left(s, s'\right)-\frac{32}{9}\,
     \widetilde V^{(1)}_{(n\bar n \to n\bar n), T}\left(s, s' \right)
\right ] \psi^{(0,2,2)}_{\eta_2 (n\bar n) ,n}(s')\cos\Theta_n+
$$
$$
+\int\limits_{4m_s^2}^{\infty}\frac{ds'}{\pi}\rho_s (s')
\; k'^4_s\,
\frac35 \sqrt{\frac{s'}{s}}
\left [
  -\frac{40}{7}k^2 k'^2_s\,
     \widetilde V^{(3)}_{(s\bar s\to n\bar n ), T}\left(s, s' \right)
  +\sqrt{ss'}\,
     \widetilde V^{(2)}_{(s\bar s\to n\bar n), P}\left(s, s' \right)+
\right .
$$
$$
\left .
  +4mm_s\,
     \widetilde  V^{(2)}_{(s\bar s\to n\bar n), A}\left(s, s' \right)
  -\frac{32}{9} \,
     \widetilde V^{(1)}_{(s\bar s \to n\bar n), T}\left(s, s' \right)
\right ]
\psi^{(0,2,2)}_{\eta_2 (s\bar s) ,n}(s')\sin\Theta_n \, ,
$$

and
\be
\left(s-M^2\right) \psi^{(0,2,2)}_{\eta_2 (s\bar s),n}(s)
\sin\Theta_n
=\int\limits_{4m^2}^{\infty}\frac{ds'}{\pi}\rho (s')
\; k'^4\,
\frac35 \sqrt{\frac{s'}{s}}\times
\ee
$$
 \times \left [
  -\frac{40}{7}k_s^2k'^2\,
     \widetilde V^{(3)}_{(n\bar n \to s\bar s), T}\left(s, s' \right)
  +\sqrt{ss'}\,
     \widetilde V^{(2)}_{(n\bar n \to s\bar s), P}\left(s, s' \right)+
\right .
$$
$$
\left .
+4mm_s\,
     \widetilde  V^{(2)}_{(n\bar n \to s\bar s), A}\left(s, s' \right)
  -\frac{32}{9}\,
     \widetilde V^{(1)}_{(n\bar n \to s\bar s), T}\left(s, s' \right)
\right ]
\psi^{(0,2,2)}_{\eta_2 (n\bar n) ,n}(s')\cos\Theta_n+
$$
$$
+\int\limits_{4m_s^2}^{\infty}\frac{ds'}{\pi}\rho_s (s')
\; k'^4_s\,
\frac35 \sqrt{\frac{s'}{s}}
\left [
  -\frac{40}{7}k^2_s k'^2_s\,
     \widetilde V^{(3)}_{(s\bar s \to s\bar s), T}\left(s, s' \right)
  +\sqrt{ss'}\,
     \widetilde V^{(2)}_{(s\bar s \to s\bar s), P}\left(s, s' \right)+
\right .
$$
$$
\left .
  +4m_s^2 \,
     \widetilde  V^{(2)}_{(s\bar s \to s\bar s), A}\left(s, s' \right)
  -\frac{32}{9} \,
     \widetilde V^{(1)}_{(s\bar s \to s\bar s), T}\left(s, s' \right)
\right ]
\psi^{(0,2,2)}_{\eta_2 (s\bar s) ,n}(s')\sin\Theta_n \, .
$$
The wave functions $\psi^{(0,2,2)}_{\eta_2 (n\bar n) ,n}(s)$
and $\psi^{(0,2,2)}_{\eta_2 (s\bar s) ,n}(s)$ satisfy the
normalization condition (\ref{3.1.10}), with obvious change of
the integration region for the $s\bar s$ component: $4m^2\to 4m^2_s$.

The following states are located on the $\eta_2$ ($M^2,n$)-trajectory
\cite{syst}: $\eta_2 (1645)$ with $n=1$, $\eta_2 (2030)$ with $n=2$,
$\eta_2 (2250)$ with $n=3$, and so on.

\subsubsection{ Equation for the $b_3$ ($M^2,n$)-trajectory}

The equation for the $b_3$-mesons ($S=0$, $L=3$, $J=3$) is as  follows:
\be
\left(s-M^2\right) \psi^{(0,3,3)}_{b_3,n}(s)
=-\int\limits_{4m^2}^{\infty}\frac{ds'}{\pi}\rho (s')
\; k'^6\, \frac57 \sqrt{\frac{s'}{s}} \times
\ee
$$
\times
\left [
  -\frac{56}{9}k^2 k'^2\,
     \widetilde V^{(2)}_T\left(s, s' \right)
  +\sqrt{ss'}
     \widetilde V^{(1)}_P\left(s, s' \right)
  +4m^2 \widetilde  V^{(1)}_A\left(s, s' \right)
  -\frac{72}{25}\widetilde  V^{(0)}_T\left(s, s' \right)
\right]\times
$$
$$
\times \psi^{(0,3,3)}_{b_3,n}(s').
$$

According to \cite{syst}, the following states are located on the $b_3$
($M^2,n$)-trajectory in the mass region below $2400$ MeV:
 $b_3 (2020)$ with $n=1$, $b_3 (2245)$ with $n=2$, and so on.

The wave functions of the states laying on the ($M^2,n$)-trajectory
satisfy the orthogonality/normalization constraint:
\be
\int\limits_{4m^2}^{\infty}\frac{ds}{\pi}\rho (s) \;
k^6\, 2s\; \alpha (3)\,
\psi^{(0,3,3)}_{b_3,n'}(s)
\psi^{(0,3,3)}_{b_3,n}(s)
=\delta_{n',n}.
\ee
The factor $\alpha (3)$ is given by the Eq. (\ref{eq3.3.1}).

\subsection{ Equation for the ($S=1$, $J=L$)-state}

The equation for the ($S=1$, $J=L$) state reads:
\be
\left(s-M^2\right)
\varepsilon_{\beta \nu_1\nu_2\nu_3} P_{\nu_1}
Z^{(J)}_{\nu_2 \mu_1\cdots\mu_J,\nu_3}(k)
Sp\left[\gamma^{\perp}_\alpha (\widehat{k}_1+m)
\gamma^{\perp}_\beta(-\widehat{k}_2+m)\right]\times
\label{3.2.1}
\ee
$$
\times
\varepsilon_{\alpha \xi_1\xi_2\xi_3}
P_{\xi_1} Z^{(J)}_{\xi_2 \mu_1\cdots\mu_J,\xi_3}(k)
\,\psi^{(1,J,J)}_n(s)=
$$
$$
=\varepsilon_{\beta' \nu_1\nu_2\nu_3} P_{\nu_1}
        Z^{(J)}_{\nu_2 \mu_1\cdots\mu_J,\nu_3}(k)
\sum_c Sp\left[\widehat{F}_c\, (\widehat{k}_1+m)
\gamma^{\perp}_{\beta'}(-\widehat{k}_2+m)\right]\times
$$
$$
\times\int\frac{d^3 k'}{(2\pi)^3 k'_0}
V_c\left (s,s',(kk')\right )
Sp\left[\gamma^{\perp}_{\alpha'}(\widehat{k}'_1+m')
\widehat{F}_c\, (-\widehat{k}'_2+m')\right]\times
$$
$$
\times
\varepsilon_{\alpha'\xi_1\xi_2\xi_3}
 P_{\xi_1} Z^{(J)}_{\xi_2\mu_1\cdots\mu_J,\xi_3}(k')\,
\psi^{(1,J,J)}_n(s').
$$
The left-hand side of the equation is calculated by using the trace
and operator convolutions given in Appendices B and C:
$$
\varepsilon_{\beta \nu_1\nu_2\nu_3} P_{\nu_1}
        Z^{(J)}_{\nu_2 \mu_1\cdots\mu_J,\nu_3}(k)
Sp\left[\gamma^{\perp}_\alpha (\widehat{k}_1+m)
\gamma^{\perp}_\beta(-\widehat{k}_2+m)\right] \times
$$
\be
\label{3.2.2}
\times \varepsilon_{\alpha \xi_1\xi_2\xi_3} P_{\xi_1}Z^{(J)}_{\xi_2 \mu_1\cdots\mu_J,\xi_3}(k)
=-2s^2\, \frac{J(2J+3)^2}{(J+1)^3} \alpha (J)\, k^{2J}.
\ee
As before, the right-hand side is calculated in two steps.

1) We calculate traces:
\be
\label{3.2.3}
B_{\beta'\alpha'}\left (s,s',(kk')\right )=
\sum_{c=T,A,V,S} (B_c)_{\beta'\alpha'}\left (s,s',(kk')\right )
V_c\left (s,s',(kk')\right )=
\ee
$$
=\sum_{c=T,A,V,S} Sp\left[\widehat{F}_c\, (\widehat{k}_1+m)
\gamma^{\perp}_{\beta'}(-\widehat{k}_2+m)\right]
Sp\left[\gamma^{\perp}_{\alpha'}(\widehat{k}'_1+m')
\widehat{F}_c\, (-\widehat{k}'_2+m')\right]\times
$$
$$
\times
V_c\left (s,s',(kk')\right ).
$$
Following  the items presented in Appendix B, we write:
\be
\label{3.2.4}
B_{\beta'\alpha'} \left(s,s',(kk')\right )=g^{\perp}_{\beta'\alpha'}\,
 4\sqrt{ss'}\left[\sqrt{ss'}\,
V_V\left (s,s',(kk')\right)+
\right .
\ee
$$
\left .
+8mm'\, V_T\left (s,s',(kk')\right )
+4\sqrt{k^2}\sqrt{k'^2}\, z\, V_A\left (s,s',(kk')\right )\right]+
$$
$$
+64 mm'k^{\perp}_{\beta'}k'^{\perp}_{\alpha'}\,
V_S\left (s,s',(kk')\right )
-16k'^{\perp}_{\beta'}k^{\perp}_{\alpha'}\, \sqrt{ss'}\,
V_A\left (s,s',(kk')\right )+
$$
$$
+16\left[s'k^{\perp}_{\beta'}k^{\perp}_{\alpha'}
+sk'^{\perp}_{\beta'}k'^{\perp}_{\alpha'}
+4\, z\,\sqrt{k^2}\sqrt{k'^2} k^{\perp}_{\beta'}k'^{\perp}_{\alpha'}
\right]V_V\left (s,s',(kk')\right ).
$$
2) The convolutions of the trace
factor $B_{\beta'\alpha'} \left (s,s',(kk')\right )$ with
angular momentum wave functions are presented in Appendix C; we have:
\be
\label{3.2.5}
\varepsilon_{\beta' \nu_1\nu_2\nu_3} P_{\nu_1}
Z^{(J)}_{\nu_2 \mu_1\cdots\mu_J,\nu_3}(k)
\, B_{\beta'\alpha'}\left (s,s',(kk')\right )\,\times
\ee
$$
\times \varepsilon_{\alpha' \xi_1\xi_2\xi_3} P_{\xi_1}
Z^{(J)}_{\xi_2 \mu_1\cdots\mu_J,\xi_3}(k')=
\alpha(J)\left(\sqrt{k^2}\sqrt{k'^2}\right)^J(-4ss')
\frac{J(2J+3)^2}{(J+1)^3}\times
$$
$$
\times\left[
4\frac{J}{2J+1}\sqrt{k^2}\sqrt{k'^2}P_{J+1}(z)
 V_A\left (s,s',(kk')\right )
+\left(\sqrt{ss'} V_V\left (s,s',(kk')\right )+ \right . \right .
$$ $$ \left . \left . +8mm' V_T\left (s,s',(kk')\right
)\right)P_{J}(z) +4\frac{J+1}{2J+1}\sqrt{k^2}\sqrt{k'^2}
P_{J-1}(z) V_A\left (s,s',(kk')\right ) \right].
$$
Inserting
these expressions into  Eq. (\ref{3.2.1}), we obtain:
\be
\label{3.2.6}
\left(s-M^2\right)\,(-2s^2)\,
 k^{2J}\;
\psi^{(1,J,J)}_n(s)=
\int\frac{d^3 k'}{(2\pi)^3 k'_0}
(-4ss')
\left(\sqrt{k^2}\sqrt{k'^2}\right)^J\times
\ee
$$
\times
\left[
 4\frac{J}{2J+1}\sqrt{k^2}\sqrt{k'^2}P_{J+1}(z)
   V_A\left (s,s',(kk')\right )
 +\sqrt{ss'} V_V\left (s,s',(kk')\right ) P_{J}(z)+
\right .
$$
$$
\left .
+8mm' V_T\left (s,s',(kk')\right )P_{J}(z)
+4\frac{J+1}{2J+1}\sqrt{k^2}\sqrt{k'^2} P_{J-1}(z)
V_A\left (s,s',(kk')\right )
\right]\times
$$
$$
\times
\psi^{(1,J,J)}_n(s').
$$
Expanding the interaction block according to (\ref{3.1.7}) and
integrating both sides over $\int^1_{-1} dz/2$, we get:
\be
\label{3.2.7}
\left(s-M^2\right)\; \psi^{(1,J,J)}_n(s)
=\int\limits_{4m'^2}^{\infty}\frac{ds'}{\pi} \rho (s') (-k'^2)^J\,
2\, \frac{s'}{s}\times
\ee
$$
\times
\left[
 -4\frac{J}{2J+1}\,\xi (J+1)\, k^2 k'^2\,
   \widetilde V^{(J+1)}_A (s,s')
 +\sqrt{ss'}\,\xi (J)\,
   \widetilde V^{(J)}_V(s,s')+
\right .
$$
$$
\left .
 +8mm'\,\xi (J)\,
   \widetilde V^{(J)}_T(s,s')
 -4\frac{J+1}{2J+1}\,\xi (J-1)\,
   \widetilde V^{(J-1)}_A (s,s')
\right]
\psi^{(1,J,J)}_n(s').
$$
Normalization condition for the ($S=1$,$J=L$) wave functions reads:
\be
\label{3.2.8}
\int\limits_{4m^2}^{\infty}\frac{ds}{\pi}\rho (s) k^{2J} \;
2\, s^2 \frac{J(2J+3)^2}{(J+1)^3} \alpha (J)
\psi^{(1,J,J)}_{n'}(s)\psi^{(1,J,J)}_{n}(s)=
\delta_{n',n}.
\ee

\subsubsection{ Equation for the $a_1$ ($M^2,n$)-trajectory}

The $a_1$ states ($S=1$, $L=1$, $J=1$) obey the Bethe--Salpeter
equation:
\be
\left(s-M^2\right) \psi^{(1,1,1)}_{a_1,n}(s)
=-\int\limits_{4m^2}^{\infty}\frac{ds'}{\pi}\rho (s')
\; k'^2\, \frac23 \frac{s'}{s}\times
\ee
$$
\times
\left [
  -\frac{6}{5}k^2 k'^2\,
     \widetilde V^{(2)}_A\left(s, s' \right)
  +\sqrt{ss'}\,
     \widetilde V^{(1)}_V\left(s, s' \right)
  +8m^2\, \widetilde  V^{(1)}_T\left(s, s' \right)
  -8\,\widetilde  V^{(0)}_A\left(s, s' \right)
\right ]
\times
$$
$$
\times
\psi^{(1,1,1)}_{a_1,n}(s').
$$

The following states are located on the $a_1$
($M^2,n$)-trajectory \cite{syst}:
$a_1 (1230)$ with $n=1$, $a_1 (1640)$ with $n=2$,
$a_1 (1960)$ with $n=3$, $a_1 (2270)$ with $n=4$, and so on.

The wave functions of the $a_1$-states satisfy  the
orthogonality/normalization condition:
\be
\int\limits_{4m^2}^{\infty}\frac{ds}{\pi}\rho (s)
\; k^2\, 2s^2\; \frac{25}{8} \alpha (1)
\psi^{(1,1,1)}_{a_1,n'}(s)
\psi^{(1,1,1)}_{a_1,n}(s)
=\delta_{n',n}.
\ee

\subsubsection{ Equation for the $a_3$ ($M^2,n$)-trajectory}

For the $a_3$-mesons ($S=1$, $L=3$, $J=3$) the Bethe--Salpeter equation reads:
\be
\left(s-M^2\right) \psi^{(1,3,3)}_{a_3,n}(s)
=-\int\limits_{4m^2}^{\infty}\frac{ds'}{\pi}\rho (s')
\; k'^6\, \frac57 \frac{s'}{s}\times
\ee
$$
\times
\left [
  -\frac{7}{3}k^2 k'^2\,
     \widetilde V^{(4)}_A\left(s, s' \right)
  +\sqrt{ss'}\,
     \widetilde V^{(3)}_V\left(s, s' \right)
  +8m^2\, \widetilde  V^{(3)}_T\left(s, s' \right)
  -\frac{48}{25}\,\widetilde  V^{(2)}_A\left(s, s' \right)
\right ]\times
$$
$$
\times\psi^{(1,3,3)}_{a_3,n}(s').
$$

Two $a_3$-states were seen: $a_3 (2030)$ with $n=1$, $a_3 (2275)$ with $n=2$
\cite{syst}.

The orthogonality/normalization constraint reads:
\be
\int\limits_{4m^2}^{\infty}\frac{ds}{\pi}\rho (s) \;
k^6\, s\; \frac{243}{32}\alpha (3)\,
\psi^{(1,3,3)}_{a_3,n'}(s)
\psi^{(1,3,3)}_{a_3,n}(s)
=\delta_{n',n} .
\ee

\subsection{ Equations for the $(S=1,J=L\pm 1)$-states}

We have two equations for two states with $S=1$ and
$J=L\pm 1$ for $J>0$. Corresponding wave functions are denoted as
$ A_j \widehat \Psi^{(1,J-1,J)}_{(n)\,\mu_1\ldots\mu_J}(k)
+B_j \widehat \Psi^{(1,J+1,J)}_{(n)\,\mu_1\ldots\mu_J}(k) $
with $j=1,2$. These wave functions are orthogonal
to one another. Normalization and
orthogonality conditions give three constraints for four
mixing parameters $A_j$ and $B_j$.

Each wave function obeys two equations:
\be
\label{3.3.1}
\left(s-M^2\right)
X^{(J+1)}_{\mu_1\ldots\mu_{J}\beta}(k)
Sp\left[\gamma^{\perp}_\alpha (\widehat{k}_1+m)
\gamma^{\perp}_\beta(-\widehat{k}_2+m)\right]\times
\ee
$$
\times
\left(
A_j  Z^{(J-1)}_{\mu_1\ldots\mu_J,\alpha}(k)\,
        \psi^{(1,J-1,J)}_n(k^2) +
B_j  X^{(J+1)}_{\mu_1\ldots\mu_{J}\alpha}(k)\,
        \psi^{(1,J+1,J)}_n(k^2) \right)=
$$
$$
=X^{(J+1)}_{\mu_1\ldots\mu_J\beta'}(k)
\sum_c Sp\left[\widehat{F}_c (\widehat{k}_1+m)
\gamma^{\perp}_{\beta'} (-\widehat{k}_2+m)\right]\times
$$
$$
\times
\int\frac{d^3 k'}{(2\pi)^3 k'_0}
V_c \left( s,s', (kk')\right)
Sp\left[\gamma^{\perp}_{\alpha'}(\widehat{k}'_1+m')
\widehat{F}_c(-\widehat{k}'_2+m')\right]\times
$$
$$
\times
\left(
A_j  Z^{(J-1)}_{\mu_1\ldots\mu_J,\alpha'}(k')\,
        \psi^{(1,J-1,J)}_n(k'^2)  +
B_j  X^{(J+1)}_{\mu_1\ldots\mu_J\alpha'}(k')\,
        \psi^{(1,J+1,J)}_n(k'^2)\right)\, ,
$$
and
\be
\label{3.3.2}
\left(s-M^2\right)
Z^{(J-1)}_{\mu_1\ldots\mu_J,\beta}(k)\,
Sp\left[\gamma^{\perp}_\alpha (\widehat{k}_1+m)
\gamma^{\perp}_\beta(-\widehat{k}_2+m)\right]\times
\ee
$$
\nonumber
\times
\left(
A_j  Z^{(J-1)}_{\mu_1\ldots\mu_J,\alpha}(k)\,
        \psi^{(1,J-1,J)}_n(k^2) +
B_j X^{(J+1)}_{\mu_1\ldots\mu_{J}\alpha}(k)\,
        \psi^{(1,J+1,J)}_n(k^2) \right)=
$$
$$
\nonumber
=Z^{(J-1)}_{\mu_1\ldots\mu_J,\beta'}(k)\,
\sum_c Sp\left[\widehat{F}_c (\widehat{k}_1+m)
\gamma^{\perp}_{\beta'} (-\widehat{k}_2+m)\right]\times
$$
$$
\nonumber
\times
\int\frac{d^3 k'}{(2\pi)^3 k'_0}
V_c \left( s,s', (kk')\right)
Sp\left[\gamma^{\perp}_{\alpha'}(\widehat{k}'_1+m')
\widehat{F}_c(-\widehat{k}'_2+m')\right]\times
$$
$$
\nonumber
\times
\left(
A_j  Z^{(J-1)}_{\mu_1\ldots\mu_J,\alpha'}(k')\,
        \psi^{(1,J-1,J)}_n(k'^2)  +
B_j  X^{(J+1)}_{\mu_1\ldots\mu_J \alpha'}(k')\,
        \psi^{(1,J+1,J)}_n(k'^2)\right).
$$
First, let us consider (\ref{3.3.1});
in the left-hand side of (\ref{3.3.1})
one has two convolutions:
$$
X^{(J+1)}_{\mu_1\ldots\mu_{J}\beta}(k)
Sp\left[\gamma^{\perp}_\alpha (\widehat{k}_1+m)
\gamma^{\perp}_\beta(-\widehat{k}_2+m)\right]
X^{(J+1)}_{\mu_1\ldots\mu_{J}\alpha}(k)=
$$
$$
=2\alpha (J)k^{2(J+1)}\left[\frac{2J+1}{J+1}s+4k^2\right],
$$
\be
\label{3.3.3}
X^{(J+1)}_{\mu_1\ldots\mu_{J}\beta}(k)
Sp\left[\gamma^{\perp}_\alpha (\widehat{k}_1+m)
\gamma^{\perp}_\beta(-\widehat{k}_2+m)\right]
Z^{(J-1)}_{\mu_1\ldots\mu_{J},\,\alpha}(k)=
\ee
$$
=8\alpha (J)k^{2(J+1)}.
$$
The left-hand side of (\ref{3.3.2}) contains also
two convolutions:
\be
\label{3.3.4}
Z^{(J-1)}_{\mu_1\ldots\mu_{J},\beta}(k)
Sp\left[\gamma^{\perp}_\alpha (\widehat{k}_1+m)
\gamma^{\perp}_\beta(-\widehat{k}_2+m)\right]
X^{(J+1)}_{\mu_1\ldots\mu_{J}\alpha}(k)=
\ee
$$
=8\alpha (J) k^{2(J+1)},
$$
$$
Z^{(J-1)}_{\mu_1\ldots\mu_{J},\beta}(k)
Sp\left[\gamma^{\perp}_\alpha (\widehat{k}_1+m)
\gamma^{\perp}_\beta(-\widehat{k}_2+m)\right]
Z^{(J-1)}_{\mu_1\ldots\mu_{J},\alpha}(k)=
$$
$$
=2\alpha (J)k^{2(J-1)}\left[\frac{2J+1}{J}s+4k^2\right].
$$
The right-hand side of Eqs. (\ref{3.3.1})
and (\ref{3.3.2}) is determined by  convolutions
of the trace factor
$ B_{\beta'\alpha'}\left (s,s',(kk')\right )$,
see Eqs. (\ref{3.2.3}) and (\ref{3.2.4}), with
angular momentum wave functions;  corresponding
formulae are presented in Appendix C. Following  them,
one has for the right-hand side of (\ref{3.3.1}):
\be
X^{(J+1)}_{\mu_1\ldots\mu_{J}\beta'}(k)
\, B_{\beta'\alpha'}\left (s,s',(kk')\right )\,
X^{(J+1)}_{\mu_1\ldots\mu_{J}\alpha'}(k')=
4\alpha (J)\left(\sqrt{k^2}\sqrt{k'^2}\right)^{J+1}\times
\ee
$$
\nonumber
\times\left(\left[
\frac{2J+1}{J+1}\sqrt{ss'}\left(\sqrt{ss'}\, V_V (s,s',(kk'))
+8mm'\, V_T (s,s',(kk'))\right)+
\right .\right .
$$
$$
\nonumber
\left .
+4s'k^2\, V_V (s,s',(kk'))+4sk'^2\,  V_V (s,s' )
+16\frac{J+1}{2J+1}k^2k'^2\, V_V (s,s',(kk'))\right]\times
$$
$$
\nonumber
\times P_{J+1}(z)+\left[16mm'\, V_S (s,s',(kk') )
+4\frac{J}{J+1}\sqrt{ss'}\, V_A (s,s',(kk'))\right]
\times
$$
$$
\left .
\times \sqrt{k^2}\sqrt{k'^2}P_{J}(z)
+16\frac{J}{2J+1}k^2k'^2\, V_V (s,s',(kk'))  P_{J-1}(z) \right)
$$
and
\be
X^{(J+1)}_{\mu_1\ldots\mu_{J}\beta'}(k)
\, B_{\beta'\alpha'}\left (s,s',(kk')\right )\,
Z^{(J-1)}_{\mu_1\ldots\mu_{J},\,\alpha'}(k')=
\ee
$$
=16\alpha (J) k^2\left(\sqrt{k^2}\sqrt{k'^2}\right)^{J-1}
\left(
\left[s+4\frac{J+1}{2J+1}k^2\right]k'^2P_{J+1}(z)\, V_V (s,s',(kk'))+
\right .
$$
$$
+\left[-\sqrt{ss'}\, V_A(s,s',(kk'))
+4mm'\, V_S(s,s',(kk'))
\right]
\sqrt{k^2}\sqrt{k'^2}P_{J}(z)+
$$
$$
\nonumber
\left .
+\left[s'+4\frac{J}{2J+1}k'^2\right]k^2P_{J-1}(z)\, V_V(s,s',(kk'))
\right).
$$
For the right-hand side of (\ref{3.3.2}) one has:
\be
Z^{(J-1)}_{\mu_1\ldots\mu_{J},\beta'}(k)
\, B_{\beta'\alpha'}\left (s,s',(kk')\right )\,
X^{(J+1)}_{\mu_1\ldots\mu_{J}\alpha'}(k')=
\ee
$$
=16\alpha (J)k'^2\left(\sqrt{k^2}\sqrt{k'^2}\right)^{J-1}
\left(
\left[s'+4\frac{J+1}{2J+1}k'^2\right]k^2P_{J+1}(z)\, V_V(s,s',(kk'))+
\right .
$$
$$
+\left[-\sqrt{ss'}\, V_A(s,s',(kk'))
+4mm'\, V_S(s,s',(kk'))\right]
\sqrt{k^2}\sqrt{k'^2}P_{J}(z)+
$$
$$
\left .
+\left[s+4\frac{J}{2J+1}k^2\right]k'^2P_{J-1}(z)\, V_V(s,s',(kk'))
\right)
$$
and
\be
Z^{(J-1)}_{\mu_1\ldots\mu_{J},\beta'}(k)
\, B_{\beta'\alpha'}\left (s,s',(kk')\right )\,
Z^{(J-1)}_{\mu_1\ldots\mu_{J},\alpha'}(k')=
4\alpha (J)\left(\sqrt{k^2}\sqrt{k'^2}\right)^{J-1}\times
\ee
$$
\times\left(
16\frac{J+1}{2J+1}k^2k'^2P_{J+1}(z)\, V_V(s,s',(kk'))+
\right .
$$
$$
+\left[16mm'\, V_S(s,s',(kk'))
+4\sqrt{ss'}\frac{J+1}{J}\, V_A(s,s',(kk'))\right]
\sqrt{k^2}\sqrt{k'^2}P_{J}(z)+
$$
$$
+\left[\frac{2J+1}{J}\sqrt{ss'}\left(\sqrt{ss'}\, V_V(s,s',(kk'))
+8mm'\, V_T(s,s',(kk'))\right)+
\right .
$$
$$
\left .\left .
+4s'\,k^2\, V_V(s,s',(kk'))+4s\,k'^2\, V_V(s,s',(kk'))+
\right . \right .
$$
$$
\left . \left .
+16\frac{J}{2J+1}k^2k'^2\, V_V(s,s',(kk'))\right] P_{J-1}(z)
\right).
$$
In the right-hand sides of Eqs. (\ref{3.3.1}) and (\ref{3.3.2}),
we expand the interaction blocks in the Legendre polynomial
series (\ref{3.1.7})
and integrate over angle variables
$\int ^{1}_{-1} dz/2$.
As a result, Eq. (\ref{3.3.1}) reads:
\be
(s-M^2) \left[ 4\psi^{(1,J-1,J)}_n(k^2)A_j  +
      \left(\frac{2J+1}{J+1}s+4k^2\right)
        \psi^{(1,J+1,J)}_n(k^2)B_j\right]=
\ee
$$
=\int\limits_{4m'^2}^{\infty}\frac{ds'}{\pi}\rho(s')
\, 8\, (-k'^2)^{J-1}
\psi^{(1,J-1,J)}_n(k'^2)A_j\times
$$
$$
\times
\left[
 \xi (J+1)\,
   \left( s+4\frac{J+1}{2J+1}k^2 \right)k'^4 \,
      \widetilde V^{(J+1)}_V(s,s')
 +\xi (J) \sqrt{ss'}k'^2\,
      \widetilde V^{(J)}_A(s,s')-
\right .
$$
$$
\left .
 -4mm'\, k'^2\,\xi (J)\,
   \widetilde V^{(J)}_S(s,s')
 +\xi (J-1)\,\left( s'+4\frac{J}{2J+1}k'^2\right)\,
   \widetilde V^{(J-1)}_V(s,s')
\right]+
$$
$$
+\int\limits_{4m'^2}^{\infty}\frac{ds'}{\pi}\rho(s')
\, 2\, (-k'^2)^{J+1}
 \psi^{(1,J+1,J)}_n(k'^2)B_j\times
$$
$$
\times
\left[
 8mm'\frac{2J+1}{J+1}\,\xi (J+1)\,\sqrt{ss'}\,
  \widetilde V^{(J+1)}_T(s,s')+
\right .
$$
$$
\left .
 +\xi (J+1)\,
  \left(
    \frac{2J+1}{J+1}ss'+4s'k^2+4sk'^2+16\frac{J+1}{2J+1}k^2k'^2
  \right)
      \widetilde V^{(J+1)}_V(s,s')-
\right .
$$
$$
\left .
 -16mm'\xi (J)\,
   \widetilde V^{(J)}_S(s,s')
 -4\frac{J}{J+1}\,\xi (J)\, \sqrt{ss'}
   \widetilde V^{(J)}_A(s,s')+
\right .
$$
$$
\left .
+16\frac{J}{2J+1}\,\xi (J-1)\,
   \widetilde V^{(J-1)}_V(s,s')
    \right].
$$

The second equation, (\ref{3.3.2}), reads:
\be
(s-M^2) \left[ \left(\frac{2J+1}{J}s+4k^2\right)
\psi^{(1,J-1,J)}_n(k^2)A_j +
4k^4\psi^{(1,J+1,J)}_n(k^2)B_j\right]=
\ee
$$
=\int\limits_{4m'^2}^{\infty}\frac{ds'}{\pi}\rho(s')
\, 2\, (-k'^2)^{J-1}
 \psi^{(1,J-1,J)}_n(k'^2)A_j\times
$$
$$
\times
\left[
 8mm'\frac{2J+1}{J}\,\xi (J-1)\,\sqrt{ss'}\,
   \widetilde V^{(J-1)}_T(s,s')+
\right .
$$
$$
\left .
 +\xi (J-1)
  \left(
    \frac{2J+1}{J}ss'+4s'k^2+4sk'^2+16\frac{J}{2J+1}k^2k'^2
  \right)
      \widetilde  V^{(J-1)}_V(s,s')-
\right .
$$
$$
 -16mm'\,\xi (J)\,k^2k'^2
   \widetilde V^{(J)}_S(s,s')
 -4\frac{J+1}{J}\,\xi (J)\,\sqrt{ss'}k^2k'^2
   \widetilde V^{(J)}_A(s,s')+
$$
$$
\left .
 +16\frac{J+1}{2J+1}\,\xi (J+1)\, k^4k'^4
   \widetilde V^{(J+1)}_V(s,s') \right ]+
$$
$$
+\int\limits_{4m'^2}^{\infty}\frac{ds'}{\pi}\rho(s')
8 (-k'^2)^{J+1} \psi^{(1,J+1,J)}_n(k'^2)B_j\times
$$
$$
\times \left[
 \xi (J+1)\,\left( s'+4\frac{J+1}{2J+1}k'^2 \right) k^4 \,
   \widetilde V^{(J+1)}_V(s,s')
 +\xi (J)\, \sqrt{ss'}k^2 \,
   \widetilde V^{(J)}_A(s,s')-
\right .
$$
$$
\left .
 -4mm'\,\xi (J) k^2\,
   \widetilde V^{(J)}_S(s,s')
 +\xi (J-1)\,\left( s+4\frac{J}{2J+1}k^2\right)\,
   \widetilde V^{(J-1)}_V(s,s')
\right].
$$
Normalization and orthoganality
conditions determined by Eq. (\ref{bs15}) are as follows:
\be
\int\limits_{4m^2}^{\infty}\frac{ds}{\pi}\rho(s)
\left [
A^2_j \left( \psi^{(1,J-1,J)}_n(k^2)\right)^2
2\alpha (J)(-k^2)^{(J-1)}\left(\frac{2J+1}{J}s+4k^2\right)+
\right .
\ee
$$
\left .
+2A_j B_j  \psi^{(1,J-1,J)}_n(k^2)\psi^{(1,J+1,J)}_n(k^2)
8\alpha (J) (-k^2)^{(J+1)}+
\right .
$$
$$
\left .
+
B_j^2 \left( \psi^{(1,J+1,J)}_n(k^2) \right)^2
2\alpha (J)(-k^2)^{(J+1)}\left( \frac{2J+1}{J+1}s+4k^2 \right)
\right] =1, \qquad j=1,2,
$$
and
\be
\int\limits_{4m^2}^{\infty}\frac{ds}{\pi}\rho(s)
\left [
A_1 A_2\left( \psi^{(1,J-1,J)}_n(k^2)\right)^2
2\alpha (J)(-k^2)^{(J-1)}\left(\frac{2J+1}{J}s+4k^2\right)+
\right .
\ee
$$
\left .
+(A_1 B_2+A_2 B_1)   \psi^{(1,J-1,J)}_n(k^2)\psi^{(1,J+1,J)}_n(k^2)
8\alpha (J) (-k^2)^{(J+1)}+
\right .
$$
$$
\left .
+
B_1B_2 \left( \psi^{(1,J+1,J)}_n(k^2) \right)^2
2\alpha (J)(-k^2)^{(J+1)}\left( \frac{2J+1}{J+1}s+4k^2 \right)
\right] =0.
$$

Let us emphasize again: all the above equations are written for the
case $J>0$.

\subsubsection{ Equation for the $a_0$ ($M^2,n$)-trajectory}

For the ($S=1$, $L=1$, $J=0$) state, we have only one level $L=J+1$;
the wave function of this state obeys the equation:
\be
(s-M^2) \left(s+4k^2\right) \psi^{(1,1,0)}_{a_0,n}(s)\,
=-\int\limits_{4m'^2}^{\infty}\frac{ds'}{\pi}\rho(s')
2 \, k'^2\, \psi^{(1,1,0)}_{a_0,n}(s')\times
\ee
$$
\times
\left[
  \frac83 m^2\,\sqrt{ss'}\,  \widetilde V^{(1)}_T(s,s')
 +\frac13
    \left( ss'+4s'k^2+4sk'^2
      +16\, k^2k'^2\right)\widetilde V^{(1)}_V(s,s')-
\right .
$$
$$
\left .
 -16m^2\, \widetilde V^{(0)}_S(s,s')
\right].
$$

According to \cite{syst}, the following states are located on the $a_0$
($M^2,n$)-trajectory:\\
$a_0 (980)$ with $n=1$, $a_0 (1520)$ with $n=2$, $a_0 (1830)$ with $n=3$,
$a_0 (2120)$ with $n=3$, and so on.

The normalization reads:
\be
\int\limits_{4m^2}^{\infty}\frac{ds}{\pi}\rho(s)
\left(\psi^{(1,1,0)}_{a_0,n}(s)\right)^2
2 (-k^2)\left(s+4k^2 \right)=1.
\ee

\subsubsection{ Equation for the $f_0$ ($M^2,n$)-trajectory}

The $f_0$-states have two flavor components, $n\bar n$ and $s\bar s$,
correspondingly we have two equations for two wave functions:
\be
\left(s-M^2\right)
\left(s+4k^2\right)
\psi^{(1,1,0)}_{a_0 (n\bar n),n}(s)\cos\Theta_n =\,
\ee
$$
=-\int\limits_{4m^2}^{\infty}\frac{ds'}{\pi}\rho(s')
2\, k'^2\,\psi^{(1,1,0)}_{a_0 (n\bar n),n}(s')\cos\Theta_n \,
\left[
  \frac83 m^2\,\sqrt{ss'}\,
   \widetilde V^{(1)}_{(n\bar n \to n\bar n), T}(s,s')+
\right .
$$
$$
\left .
 +\frac13
    \left( ss'+4s'k^2+4sk'^2
      +16\, k^2k'^2\right)
   \widetilde V^{(1)}_{(n\bar n \to n\bar n), V}(s,s')-
\right .
$$
$$
\left .
-16m^2\, \widetilde V^{(0)}_{(n\bar n \to n\bar n), S}(s,s')
\right]
-\int\limits_{4m_s^2}^{\infty}\frac{ds'}{\pi}\rho_s(s')
2\, k'^2_s\,\psi^{(1,1,0)}_{a_0 (s\bar s),n}(s')\sin\Theta_n  \times \,
$$
$$
\times
\left[
  \frac83 mm_s\,\sqrt{ss'}\,
   \widetilde V^{(1)}_{(s\bar s \to n\bar n), T}(s,s')
 +\frac13
    \left( ss'+4s'k^2+4sk'^2_s
      +16\, k^2k'^2_s\right)
  \times
\right .
$$
$$
\left .
\times
\widetilde V^{(1)}_{(s\bar s \to n\bar n), V}(s,s')
-16mm_s\, \widetilde V^{(0)}_{(s\bar s \to n\bar n),  S}(s,s')
\right] \ ,
$$
and
\be
\left(s-M^2\right)
\left(s+4k^2\right)
\psi^{(1,1,0)}_{a_0 (s\bar s),n}(s)\sin\Theta_n =\,
\ee
$$
=-\int\limits_{4m^2}^{\infty}\frac{ds'}{\pi}\rho(s')
2\, k'^2\,\psi^{(1,1,0)}_{a_0 (n\bar n),n}(s')\cos\Theta_n \,
\left[
  \frac83 mm_s\,\sqrt{ss'}\,
   \widetilde V^{(1)}_{(n\bar n \to s\bar s), T}(s,s')+
\right .
$$
$$
\left .
 +\frac13
    \left( ss'+4s'k_s^2+4sk'^2
      +16\, k_s^2k'^2\right)
   \widetilde V^{(1)}_{(n\bar n \to s\bar s), V}(s,s')-
\right .
$$
$$
\left .
-16mm_s\, \widetilde V^{(0)}_{(n\bar n \to s\bar s), S}(s,s')
\right]
-\int\limits_{4m_s^2}^{\infty}\frac{ds'}{\pi}\rho_s(s')
2\, k'^2_s\,\psi^{(1,1,0)}_{a_0 (s\bar s),n}(s')\sin\Theta_n \times \,
$$
$$
\times\left[
  \frac83 m^2_s\,\sqrt{ss'}\,
   \widetilde V^{(1)}_{(s\bar s \to s\bar s), T}(s,s')
 +\frac13
    \left( ss'+4s'k_s^2+4sk'^2_s
      +16\, k_s^2k'^2_s\right)\times
\right .
$$
$$
\left .
\times
   \widetilde V^{(1)}_{(s\bar s \to s\bar s), _V}(s,s')
 -16m^2_s\, \widetilde V^{(0)}_{(s\bar s \to s\bar s),  S}(s,s')
\right] .
$$

The following states are located on two $f_0$
($M^2,n$)-trajectories \cite{syst}:\\
1) $f_0 (980)$ with $n=1$, $f_0 (1500)$ with $n=2$,
$f_0 (2005)$ with $n=3$, $f_0 (2240)$ with $n=4$, and so on,\\
2) $f_0 (1300)$ with $n=1$, $f_0 (1750)$ with $n=2$,
$f_0 (2105)$ with $n=3$, $f_0 (2330)$ with $n=4$, and so on.

\subsubsection{ Equation for the $\rho$ ($M^2,n$)-trajectory}

Two equations read:
\be
(s-M^2) \left[ 4\psi^{(1,0,1)}_{\rho ,n}(s)A_j  +
      \left(\frac32s+4k^2\right)
        \psi^{(1,2,1)}_{\rho ,n}(s)B_j\right]=
\ee
$$
=\int\limits_{4m^2}^{\infty}\frac{ds'}{\pi}\rho(s')
\, 8 \,
 \psi^{(1,0,1)}_{\rho ,n}(s')A_j\times
$$
$$
\times
\left[
  \frac{3}{10}\left( s+\frac83 k^2 \right)k'^4 \,
     \widetilde V^{(2)}_V(s,s')
  +\frac13 \sqrt{ss'}k'^2\,
     \widetilde V^{(1)}_A(s,s')-
\right .
$$
$$
\left .
  - \frac43 m^2\, k'^2\,
     \widetilde V^{(1)}_S(s,s')
  +\left(s'+\frac43 k'^2\right)\,
     \widetilde V^{(0)}_V(s,s')
\right]+
$$
$$
+\int\limits_{4m^2}^{\infty}\frac{ds'}{\pi}\rho(s')\,
2\, k'^4
 \psi^{(1,2,1)}_{\rho ,n}(s')B_j\times
$$
$$
\times
\left[
   m^2\frac{18}{5}\,\sqrt{ss'}\,
     \widetilde V^{(2)}_T(s,s')
   +\frac{3}{10}\,\left(\frac32 ss'+4s'k^2+4sk'^2
    +\frac{32}{3} k^2k'^2\right)
     \widetilde V^{(2)}_V(s,s')-
\right .
$$
$$
\left .
  -m^2\frac{16}{3} \,
    \widetilde V^{(1)}_S(s,s')
  -\frac23 \, \sqrt{ss'}
    \widetilde V^{(1)}_A(s,s')
  +\frac{16}{3} \,
    \widetilde V^{(0)}_V(s,s')
    \right],
$$

and
\be
(s-M^2) \left[\left(3s+4k^2\right)\psi^{(1,0,1)}_{\rho ,n}(s)A_j
              + 4k^4\psi^{(1,2,1)}_{\rho ,n}(s)B_j\right]=
\ee
$$
=\int\limits_{4m^2}^{\infty}\frac{ds'}{\pi}\rho(s')\,
2\, \psi^{(1,0,1)}_{\rho ,n}(s')A_j\times
$$
$$
\times
\left[
   24\, m^2 \,\sqrt{ss'}\,
     \widetilde V^{(0)}_T(s,s')
  +\left(3ss'+4s'k^2+4sk'^2+\frac{16}{3} k^2k'^2\right)
      \widetilde  V^{(0)}_V(s,s')-
\right .
$$
$$
\left .
  -m^2\,\frac{16}{3} \,k^2k'^2
    \widetilde V^{(1)}_S(s,s')
  -\frac83 \,\sqrt{ss'}k^2k'^2
    \widetilde V^{(1)}_A(s,s')
  +\frac{16}{5}\, k^4k'^4
    \widetilde V^{(2)}_V(s,s') \right ]+
$$
$$
+\int\limits_{4m^2}^{\infty}\frac{ds'}{\pi}\rho(s')\,
8\, k'^4 \psi^{(1,2,1)}_{\rho ,n}(s')B_j\times
$$
$$
\times \left[
  \frac{3}{10}\,\left(s'+\frac83 k'^2 \right)k^4 \,
    \widetilde V^{(2)}_V(s,s')
  +\frac13 \,\sqrt{ss'}k^2 \,
    \widetilde V^{(1)}_A(s,s')-
\right .
$$
$$
\left .
  -m^2\,\frac43 k^2\,
    \widetilde V^{(1)}_S(s,s')
  +\left(s+\frac43 k^2\right)\,
    \widetilde V^{(0)}_V(s,s')
\right].
$$

The following states are located on the $\rho$
($M^2,n$)-trajectories \cite{syst}:\\
1) $\rho (770)$ with $n=1$, $\rho (1450)$ with $n=2$,
$\rho (1830)$ with $n=3$, $\rho (2110)$ with $n=4$, and so on,\\
2) $\rho (1700)$ with $n=1$, $\rho (1990)$ with $n=2$,
$\rho (2285)$ with $n=3$, and so on.

Normalization and orthogonality conditions are as follows:
\be
\int\limits_{4m^2}^{\infty}\frac{ds}{\pi}\rho(s)
\left [
A^2_j \left( \psi^{(1,0,1)}_{\rho ,n}(s)\right)^2
2\left(3s+4k^2\right)+
\right .
\ee
$$
\left .
+2A_j B_j  \psi^{(1,0,1)}_{\rho ,n}(s)\psi^{(1,2,1)}_{\rho ,n}(s)
\, 8 k^4+
\right .
$$
$$
\left .
+B_j^2 \left(\psi^{(1,2,1)}_{\rho ,n}(s) \right)^2
\, 2\, k^4\left(\frac32 s+4k^2 \right)
\right] =1, \qquad j=1,2,
$$
and
\be
\int\limits_{4m^2}^{\infty}\frac{ds}{\pi}\rho(s)
\left[
A_1 A_2\left(\psi^{(1,0,1)}_{\rho ,n}(s)\right)^2
\, 2\, \left(3s+4k^2\right)+
\right .
\ee
$$
\left .
+(A_1 B_2+A_2 B_1)\psi^{(1,0,1)}_{\rho ,n}(s)\psi^{(1,2,1)}_{\rho,n}(s)
\, 8\, k^4+
\right .
$$
$$
\left .
+ B_1B_2 \left(\psi^{(1,2,1)}_{\rho ,n}(s)\right)^2 \,
2\, k^4\left(\frac32 s+4k^2 \right) \right] =0.
$$
\section*{ACKNOWLEDGEMENTS}
We are grateful to D.V.Bugg, L.G.Dakhno, D.I.Melikhov, V.A.Nikonov,
 and H.R.Petry for helpful and stimulating discussions.The work was
supported by RFBR grants 0102-17861 and 0102-17152.One of us (VNM)
thanks for support INTAS, project 2000-587, and  Prof. H.R.Petry for
the hospitality in ITKP(Bonn).

\def\thesection{Appendix \Alph{section}}
\def\theequation{\Alph{section}.\arabic{equation}}

\setcounter{equation}{0}
\appendix\section{Appendix: Angular--momentum operators  }

Here we present the angular momentum operator
$ X^{(L)}_{\mu_1\mu_2\ldots\mu_{L-1}\mu_L}(k)$ and briefly recall
its properties; a full presentation of the angular-momentum operators
can be found in \cite{operators}.

The operator $ X^{(L)}_{\mu_1\mu_2\ldots\mu_{L-1}\mu_L}(k)$
is constructed by using relative momentum
of mesons in the space orthogonal to
the total momentum $P$:
\be
k^\perp_\mu = k_\nu
g^\perp_{\nu\mu},  \qquad
g^\perp_{\nu\mu}=g_{\nu\mu}-\frac{P_\nu
P_\mu}{s}, \qquad
g_{\mu\nu}=(1,-1,-1,-1).
\label{A.0}
\ee
In the
center-of-mass system, where $P=(P_0,\vec P)=(\sqrt s,0)$, the vector
$k^\perp$ is space-like: $k^\perp=(0,\vec k)$.
We determine the operator
$ X^{(L)}_{\mu_1\mu_2\ldots\mu_{L-1}\mu_L}(k)$
as symmetrical and traceless. It is easy to construct it for the lowest
values of $L=0,1,2,3$:
\be
X^{(0)}=1, \qquad X^{(1)}_\mu=k^\perp_\mu, \qquad
X^{(2)}_{\mu_1 \mu_2}=\frac32\left(k^\perp_{\mu_1}
k^\perp_{\mu_2}-\frac13\, (k^\perp)^2 g^\perp_{\mu_1\mu_2}\right),
\label{A.2}
\ee
$$
X^{(3)}_{\mu_1\mu_2\mu_3}=\frac52\left[k^\perp_{\mu_1} k^\perp_{\mu_2 }
k^\perp_{\mu_3}-\frac{(k^\perp)^2}5\left(g^\perp_{\mu_1\mu_2}k^\perp
_{\mu_3}+g^\perp_{\mu_1\mu_3}k^\perp_{\mu_2}+
g^\perp_{\mu_2\mu_3}k^\perp_{\mu_1}
\right)\right].
$$
Correspondingly, the generalization of
$X^{(L)}_{\mu_1\ldots\mu_L}$ for
$L>1$ reads:
\be
\label{A.3}
X^{(L)}_{\mu_1\ldots\mu_L}&=&k^\perp_\alpha
Z^{(L-1)}_{\mu_1\ldots\mu_L, \alpha}, \\
Z^{(L-1)}_{\mu_1\ldots\mu_L, \alpha}&=&
\frac{2L-1}{L^2}\left (
\sum^L_{i=1}X^{{(L-1)}}_{\mu_1\ldots\mu_{i-1}\mu_{i+1}\ldots\mu_L}
g^\perp_{\mu_i\alpha} -\right .
\nonumber \\
&&\left . -\frac{2}{2L-1} \sum^L_{i,j=1 \atop i<j}
g^\perp_{\mu_i\mu_j}
X^{{(L-1)}}_{\mu_1\ldots\mu_{i-1}\mu_{i+1}\ldots\mu_{j-1}\mu_{j+1}
\ldots\mu_L\alpha} \right ).
\nonumber
\ee
It is seen that the operator
$ X^{(L)}_{\mu_1\mu_2\ldots\mu_{L-1}\mu_L}(k)$
constructed in accordance with (\ref{A.3})
is symmetrical,
\be
X^{(L)}_{\mu_1\ldots\mu_i\ldots\mu_j\ldots\mu_L}\; =\;
X^{(L)}_{\mu_1\ldots\mu_j\ldots\mu_i\ldots\mu_L},
\label{A.12}
\ee
and it works in the space orthogonal to $P$:
\be
P_{\mu_i}X^{(L)}_{\mu_1\ldots\mu_i\ldots\mu_L}\ =\ 0.
\label{A.13}
\ee
The angular-momentum operator
$X^{(L)}_{\mu_1\ldots\mu_L}$ is traceless over any two
indices:
\be
g_{\mu_i\mu_j}X^{(L)} _{\mu_1\ldots\mu_i\ldots\mu_j\ldots\mu_L}\
=\ g^\perp _{\mu_i\mu_j}X^{(L)} _{\mu_1\ldots\mu_i\ldots\mu_j\ldots\mu_L}
\ =\ 0.
\label{A.1}
\ee
The tracelessness property given by
(\ref{A.1}) is obvious for the lowest-order operators entering
(\ref{A.2}), for example, $g^\perp_{\mu_1\mu_2}X^{(2)}_{\mu_1\mu_2}=0$
(recall that $g^\perp_{\mu_1\mu_2} g^\perp_{\mu_1\mu_2}=3$).

The convolution equality reads:
\be
X^{(L)}_{\mu_1\ldots\mu_L}k^\perp_{\mu_L}=(k^\perp)^2
X^{(L-1)}_{\mu_1\ldots\mu_{L-1}}\ .
\label{A.4}
\ee
Using (\ref{A.4}) we rewrite the recurrent equation (\ref{A.3}) in the
form:
\be
X^{(L)}_{\mu_1\ldots\mu_L}=
\label{A.5}
\ee
$$
=\frac{2L-1}{L^2}
\sum^L_{i=1}k^\perp_{\mu_i}
X^{{(L-1)}}_{\mu_1\ldots\mu_{i-1}\mu_{i+1}\ldots\mu_L}
-\frac{2(k^\perp)^2}{L^2}\sum^L_{i,j=1 \atop i<j}
g_{\mu_i\mu_j}^\perp
X^{\left(L-2\right)}_{\mu_1\ldots\mu_{i-1}\mu_{i+1}\ldots\mu_{j-1}\mu_{j+1}
\ldots\mu_L}\ .
$$
On the basis of this recurrent equation and
taking into account the tracelessness of $X^{(L)}_{\mu_1\ldots\mu_L}$,
one can write the normalization condition for the moment-$L$ operator
as follows:
\be
X^{(L)}_{\mu_1\ldots\mu_L}(k)X^{(L)}_{\mu_1\ldots\mu_L}(k)
=\alpha(L)(k^\perp)^{2L} \,\,\,\ ,
\alpha(L)=\prod^L_{l=1}\frac{2l-1}{l}=\frac{(2L-1)!!}{L!}.
\label{A.10}
\ee
The iteration of (\ref{A.5}) gives us the
following expression for the operator $X^{(L)}_{\mu_1\ldots\mu_L}$:
\be
&&X^{(L)}_{\mu_1\ldots\mu_L}(k)
=\frac{(2L-1)!!}{L!}
\bigg [
k^\perp_{\mu_1}k^\perp_{\mu_2}k^\perp_{\mu_3}k^\perp_{\mu_4}
\ldots k^\perp_{\mu_L} -
\label{x-direct}
\\
&&-\frac{(k^\perp)^{2}}{2L-1}\left(
g^\perp_{\mu_1\mu_2}k^\perp_{\mu_3}k^\perp_{\mu_4}\ldots k^\perp_{\mu_L}
+g^\perp_{\mu_1\mu_3}k^\perp_{\mu_2}k^\perp_{\mu_4}\ldots
k^\perp_{\mu_L} + \ldots \right)+
\nonumber
\\
&&
+\frac{(k^\perp)^{4}}{(2L-1)
(2L-3)}\left(
g^\perp_{\mu_1\mu_2}g^\perp_{\mu_3\mu_4}k^\perp_{\mu_5}
k^\perp_{\mu_6}\ldots k^\perp_{\mu_L}+
\right .
\nonumber
\\
&&
\left .
+g^\perp_{\mu_1\mu_2}g^\perp_{\mu_3\mu_5}k^\perp_{\mu_4}
k^\perp_{\mu_6}\ldots k^\perp_{\mu_L}+
\ldots\right)+\ldots\bigg ]\ .
\nonumber
\ee

One can introduce a projection operator
$O^{\mu_1\ldots\mu_L}_{\nu_1\ldots \nu_L}$ for the partial wave with
the angular momentum $L$. The operator is defined by
the following relations:
\be
X^{(L)}_{\mu_1\ldots\mu_L}(k)O^{\mu_1\ldots\mu_L}_{\nu_1\ldots \nu_L}\
=\ X^{(L)}_{\nu_1\ldots \nu_L}(k), \qquad
O^{\mu_1\ldots\mu_L}_{\alpha_1\ldots\alpha_L} \
O^{\alpha_1\ldots\alpha_L}_{\nu_1\ldots \nu_L}\
=\ O^{\mu_1\ldots\mu_L}_{\nu_1\ldots \nu_L}.
\label{proj_op}
\ee
For the sets of indices $\mu_1\ldots\mu_L$ and
$\nu_1\ldots\nu_L$, the operator $ O$ has all the properties of
the operator $\hat X^{(L)}$: it is symmetrical and traceless,
\be
O^{\mu_1\mu_2\ldots\mu_L}_{\nu_1\nu_2\ldots \nu_L}=
O^{\mu_2\mu_1\ldots\mu_L}_{\nu_1\nu_2\ldots \nu_L}=
O^{\mu_1\mu_2\ldots\mu_L}_{\nu_2\nu_1\ldots \nu_L} \  , \qquad
O^{\mu_1\mu_1\ldots\mu_L}_{\nu_1\nu_2\ldots \nu_L}=
O^{\mu_1\mu_2\ldots\mu_L}_{\nu_1\nu_1\ldots \nu_L}=0\ .
\label{A.17}
\ee
The projection operator $ O$ can be consructed as a product of
the operators \\ $X^{(L)}_{\mu_1\ldots\mu_L}(k)
X^{(L)}_{\nu_1\ldots\nu_L}(k)$ integrated over angular variables of the
momentum $k^\perp$, so we have a
convolution of the $(2L+1)$-dimensional vectors, which provide us with
 irreducible representation of Lorentz group in the
$k^\perp/|k^\perp|$-space.
So,
\be
\xi(L)O^{\mu_1\ldots\mu_L}_{\nu_1\ldots \nu_L}=
\frac{1}{(k^\perp)^{2L}}\int\frac{d\Omega }{4\pi}\
X^{(L)}_{\mu_1\ldots\mu_L}(k)
X^{(L)}_{\nu_1\ldots\nu_L}(k),
\label{A.9}
\ee
where $\xi(L)$ is a normalization factor  fixed below.
Using the definition of the projection operator
$O^{\mu_1\ldots\mu_L}_{\nu_1\ldots \nu_L}$ we have:
\be
k_{\mu_1}\ldots
k_{\mu_L}O^{\mu_1\ldots\mu_L}_{\nu_1\ldots \nu_L}\ =
\frac{1}{\alpha(L)}
X^{(L)}_{\nu_1\ldots\nu_L}(k).
\label{A.11}
\ee
This equation represents the basic property of
the operator:
it projects any index-$L$ operator into partial-wave
operator with the angular momentum $L$.

Multiplying the equation
(\ref{A.9}) by the product $X^{(L)}_{\mu_1\ldots\mu_L}(q)\cdot
X^{(L)}_{\nu_1\ldots\nu_L}(q)$,
we get:
\be
\xi(L)
X^{(L)}_{\nu_1\ldots\nu_L}(q)X^{(L)}_{\nu_1\ldots\nu_L}(q)=
(q^\perp)^{2L} \alpha^2(L)
\int\limits_{-1}^{1}\frac{dz}{2}\; P^2_L(z),
\label{A.20}
\ee
that gives the normalization constant in (\ref{A.9}):
\be
\xi(L)\ =\frac{\alpha(L)}{2L+1}=
\frac{(2L-1)!!}{(2L+1)\cdot L!}.
\label{A.21}
\ee
Summation in the projection operator over upper and lower indices
performed in (\ref{A.9})
gives us the following reduction formula:
\be
O^{\mu_1\ldots\mu_{L-1}\mu_L}_{\nu_1\ldots \nu_{L-1}\mu_L}=
\frac{2L+1}{2L-1}\;O^{\mu_1\ldots\mu_{L-1}}_{\nu_1\ldots\nu_{L-1}}.
\label{A.22}
\ee
Likewise, the summation over all indices gives us:
\be
O^{\mu_1\ldots\mu_L}_{\mu_1\ldots\mu_L}=2L+1,
\label{A.23}
\ee
that can be proven using formula (\ref{A.9}).
On the basis of the equation (\ref{A.11}), one gets
\be
X^{(L)}_{\mu_1\ldots\mu_{L-1}\mu_L}
O^{\mu_1\ldots\mu_{L-1}}_{\nu_1\ldots\nu_{L-1}}=
X^{(L)}_{\nu_1\ldots\nu_{L-1}\mu_L}.
\label{A.24}
\ee
Generally, one can write:
\be
X^{(L)}_{\mu_1\ldots\mu_{i}\mu_{i+1}\ldots\mu_L}
O^{\mu_1\ldots\mu_{i}}_{\nu_1\ldots\nu_{i}}=
X^{(L)}_{\nu_1\ldots\nu_{i}\mu_{i+1}\mu_L}.
\label{A.25}
\ee

\setcounter{equation}{0}
\section{Appendix: Traces of the loop diagrams }
Here we present the traces which are used for the calculations of
loop diagrams. Recall that in the spectral-integral representation,
there is no energy conservation, $s\neq s'$, where $P^2=s$, $P'^2=s'$,
but all constituents are mass-on-shell:  $$ k^2_1=k^2_2=m^2, \qquad
k'^2_1=k'^2_2=m'^2.  $$
The following notations are used for the quark momenta:
\be
&& k_{\nu}=\frac12 (k_1-k_2)_{\nu} ,\,\,\,
k'_{\nu}=\frac12 (k'_1-k'_2)_{\nu},
\\
\nonumber
&& k^{\perp}_{\mu}=k_{\nu}g^{\perp}_{\nu\mu}=k_{\mu},\qquad
k'^{\perp}_{\mu}=k'_{\nu}g^{\perp}_{\nu\mu}=k'_{\mu}.
\ee

We follow the definition of  matrices:
$$
\gamma_5=-i\gamma^0 \gamma^1 \gamma^2 \gamma^3,\,\,\,\,
\sigma_{\mu\nu}=\frac12 \left[\gamma_{\mu}\gamma_{\nu}\right].
$$

\subsection{Traces for the $S=0$ states}

For the $S=0$ states we have the following non-zero traces:
\be
&&T'_P=Sp\left[i\gamma_5(k'_1+m')\gamma_5(-k'_2+m')\right]=2is',
\\
\nonumber
&&T'_A=Sp\left[i\gamma_5
(k'_1+m')i\gamma_{\mu}\gamma_5(-k'_2+m')\right]=-4m'\,P'_{\mu},
\\
\nonumber
&&T'_T=Sp\left[i\gamma_5
(k'_1+m')i\sigma_{\mu\nu}(-k'_2+m')\right]=
-4i\epsilon_{\mu\nu\alpha\beta}P'_{\alpha}k'_{\beta},
\label{B.1}
\ee
and
\be
&&T_P=Sp\left[i\gamma_5(-k_2+m)\gamma_5(k_1+m)\right]=2is,
\\
\nonumber
&&T_A=Sp\left[i\gamma_5(-k_2+m)i\gamma_{\mu}\gamma_5(k_1+m)\right]=
4mP_{\mu},
\\
\nonumber
&&T_T=Sp\left[i\gamma_5(-k_2+m)i\sigma_{\mu\nu}(k_1+m)\right]=
4i\epsilon_{\mu\nu\alpha\beta}P_{\alpha}k_{\beta}.
\label{B.2}
\ee

The convolutions of the traces $A_P=(T_P\, T'_P)$, $A_A=(T_A\, T'_A)$,\\
$A_T=(T_T\, T'_T)$ are equal to:
\be
&&A_P=-4ss',
\\
\nonumber
&&A_A=-16mm'(PP'),
\\
\nonumber
&&A_T=-32(PP')(kk').
\label{B.3}
\ee

\subsection{Traces for the $S=1$ states}

The case of the ($S=1$) states, the traces are equal to:
\be
&&T'_S=Sp\left[\gamma_{\alpha'}^{\perp} (k'_1+m')\cdot {\rm I} \cdot
(-k'_2+m')\right]=8m'k'_{\alpha'},
\\
\nonumber
&&T'_V=Sp\left[\gamma_{\alpha'}^{\perp}
(k'_1+m')\gamma_{\mu}(-k'_2+m')\right]
=2\left(g_{\alpha'\mu}^{\perp}s'+4k'_{\alpha'}k'_{\mu}\right),
\\
\nonumber
&&T'_A=Sp\left[\gamma_{\alpha'}^{\perp}
(k'_1+m')i\gamma^{\perp}_{\mu}\gamma_5(-k'_2+m')\right]=
4\epsilon_{\alpha'\mu\alpha\beta}k'_{\alpha}P'_{\beta},
\\
\nonumber
&&T'_T=Sp\left[\gamma_{\alpha'}^{\perp}
(k'_1+m')i\sigma_{\mu\nu}(-k'_2+m')\right]
=4m'i\left[g_{\alpha'\nu}^{\perp}P'_{\mu}
-g_{\alpha'\mu}^{\perp}P'_{\nu}\right],
\label{B.5}
\ee
and
\be
&&T_S=Sp\left[\gamma_{\beta}^{\perp}(-k_2+m)\cdot {\rm I} \cdot
(k_1+m)\right] = 8mk_{\beta},
\\
\nonumber
&&T_V=Sp\left[\gamma^{\perp}_{\beta}(-k_2+m)\gamma_{\mu}(k_1+m)\right]=
2\left[g_{\mu\beta}^{\perp}s+4k_{\beta}k_{\mu}\right],
\\
\nonumber
&&T_A=Sp\left[\gamma_{\beta}^{\perp}(-k_2+m)i\gamma_{\mu}\gamma_5
(k_1+m)\right]=
-4\epsilon_{\beta\mu\alpha'\beta'}k_{\alpha'}P_{\beta'},
\\
\nonumber
&&T_T=Sp\left[\gamma_{\beta}^{\perp}(-k_2+m)i\sigma_{\mu\nu}
(k_1+m)\right]
=4mi\left[g_{\beta\mu}^{\perp}P_{\nu}
-g_{\beta\nu}^{\perp}P_{\mu}\right].
\label{B.6}
\ee

Corresponding convolutions $B_c=(T_c\, T'_c)$ read:
\be
&&(B_S)_{\beta'\alpha'}=64mm'k_{\beta'}k'_{\alpha'},
\\
\nonumber
&&(B_V)_{\beta'\alpha'}=4\left[ss'g^{\perp}_{\beta'\alpha'}+
4s'k_{\beta'}k_{\alpha'}+
4sk'_{\beta'}k'_{\alpha'}+
16(kk')k_{\beta'}k'_{\alpha'}\right],
\\
\nonumber
&&(B_A)_{\beta'\alpha'}=-16(PP')
\left[k'_{\beta'}k_{\alpha'}-
(kk')g^{\perp}_{\beta'\alpha'}\right],
\\
\nonumber
&&(B_T)_{\beta'\alpha'}=32(PP')mm'g^{\perp}_{\beta'\alpha'}.
\label{B.7}
\ee

\section{Appendix: Convolutions of the trace factors}
\setcounter{equation}{0}
Here we present the convolutions of the angular-momentum factors. We
work with $ k^{\perp}_{\mu}=k_{\nu}g^{\perp}_{\nu\mu}=k_\mu$ and
$k'^{\perp}_{\mu}=k'_{\nu}g^{\perp}_{\nu\mu}=k'_\mu$, and introduce:
\be
z=\frac{(kk')}{\sqrt{k^2}\sqrt{k'^2}}.
\ee

The convolutions for the ($S=1$) states read:
\be
X^{(J)}_{\mu_1 \mu_2 \cdots\mu_J}(k)
X^{(J)}_{\mu_1 \mu_2 \cdots\mu_J}(k')=
\alpha (J)\left(\sqrt{k^2}\sqrt{k'^2}\right)^{J} P_{J} (z).
\label{C.1}
\ee

Analogous convolutions for ($S=1$)-states are written as follows:
\be
&&X^{(J+1)}_{\mu_1 \mu_2 \cdots\mu_J\beta}(k)
X^{(J+1)}_{\mu_1 \mu_2 \cdots\mu_J\alpha}(k')=
\frac{\alpha (J)}{J+1}\left(\sqrt{k^2}\sqrt{k'^2}\right)^{J}\times
\\
\nonumber
&&\times\left[
 \frac{\sqrt{k'^2}}{\sqrt{k^2}}A_{P_{J,J+1}}(z)\, k_\beta k_\alpha
+\frac{\sqrt{k^2}}{\sqrt{k'^2}}B_{P_{J,J+1}}(z)\, k'_\beta k'_\alpha+
\right .
\\
\nonumber
&&\left .
+C_{P_{J,J+1}}(z)\, k_\beta k'_\alpha +D_{P_{J,J+1}}(z)\, k'_\beta
k_\alpha +\left(\sqrt{k^2}\sqrt{k'^2}\right)E_{P_{J,J+1}}(z)\,
g^{\perp}_{\beta\alpha} \right],
\label{C.2}
\ee

\be
&&X^{(J+1)}_{\mu_1 \mu_2 \cdots\mu_J\beta}(k)
Z^{(J-1)}_{\mu_1 \mu_2 \cdots\mu_J,\alpha}(k')
= -\frac{\alpha (J)}{J}\frac{1}{k'^2}
\left(\sqrt{k^2}\sqrt{k'^2}\right)^{J}\times
\\
\nonumber
&&\times\left[
 \frac{\sqrt{k'^2}}{\sqrt{k^2}}A_{P_{J,J+1}}(z)\, k_\beta k_\alpha
+\frac{\sqrt{k^2}}{\sqrt{k'^2}}
  \left(B_{P_{J,J+1}}(z)-(2J+1)A_J (z)\right) k'_\beta k'_\alpha+
\right .
\\
\nonumber
&&
+(C_{P_{J,J+1}}(z)-(2J+1)B_J (z))\, k_\beta k'_\alpha+
\\
\nonumber
&&\left .
+D_{P_{J,J+1}}(z)\, k'_\beta k_\alpha
+\left(\sqrt{k^2}\sqrt{k'^2}\right)E_{P_{J,J+1}}(z)\,
g^{\perp}_{\beta\alpha} \right],
\label{C.3}
\ee

\be
&&Z^{(J-1)}_{\mu_1 \mu_2 \cdots\mu_J,\beta}(k)
Z^{(J-1)}_{\mu_1 \mu_2 \cdots\mu_J,\alpha}(k')=
\frac{J+1}{J^2}\alpha (J)\left(\sqrt{k^2}\sqrt{k'^2}\right)^{J-2}\times
\\
\nonumber
&&\times\left[
 \frac{\sqrt{k'^2}}{\sqrt{k^2}}
  \left(A_{P_{J,J+1}}(z)-(2J+1)A_J (z)\right)\, k_\beta k_\alpha+
\right .
\\
\nonumber
&&+\frac{\sqrt{k^2}}{\sqrt{k'^2}}
  \left(B_{P_{J,J+1}}(z)-(2J+1)A_J (z)\right)\, k'_\beta k'_\alpha+
\\
\nonumber
&&+\left(C_{P_{J,J+1}}(z)+\frac{(2J+1)^2}{J+1}P_J (z)-2(2J+1)B_J (z)
   \right)\, k_\beta k'_\alpha+
\\
\nonumber &&\left . +D_{P_{J,J+1}}(z)\, k'_\beta k_\alpha
+(\sqrt{k^2}\sqrt{k'^2})E_{P_{J,J+1}}(z)\, g^{\perp}_{\beta\alpha}
\right],
\label{C.4}
\ee

\be
&&\epsilon_{\beta \nu_1\nu_2\nu_3} P_{\nu_1}
    Z^{(J)}_{\nu_2 \mu_1\cdots\mu_J,\nu_3}(k)\,\,
  \epsilon_{\alpha \lam_1\lam_2\lam_3} P'_{\lam_1}
    Z^{(J)}_{\lam_2 \mu_1\cdots\mu_J,\lam_3}(k')=
\\
\nonumber
&&=\frac{(2J+3)^2}{(J+1)^3}\alpha(J)
\left(\sqrt{k^2}\sqrt{k'^2}\right)^{J-1}(PP')\times
\\
\nonumber
&&\times
\left[-\sqrt{k^2}\sqrt{k'^2}\left((z^2-1)
D_{P_{J,J+1}}(z)+zE_{P_{J,J+1}}(z)\right)
g^{\perp}_{\beta\alpha}-
\right .
\\
\nonumber
&&-D_{P_{J ,J+1}}(z)
\left(\frac{\sqrt{k'^2}}{\sqrt{k^2}}k_{\beta}k_{\alpha}
+\frac{\sqrt{k^2}}{\sqrt{k'^2}}k'_{\beta}k'_{\alpha}
-z\,k_{\beta}k'_{\alpha}\right)+
\\
\nonumber
&&\left .
+\left(zD_{P_{J ,J+1}}(z)
+E_{P_{J,J+1}}(z)\right)k'_{\beta}k_{\alpha}\right],
\label{C.5}
\ee
and
\be
&&X^{(J)}_{\mu_1\cdots\mu_J}(k)\,
\epsilon_{\alpha \nu_1\nu_2\nu_3} P'_{\nu_1}
Z^{(J)}_{\nu_2\mu_1\cdots\mu_J,\nu_3}(k')=
\\
\nonumber
&&=\frac{2J+3}{J+1}\alpha(J)
\left(\sqrt{k^2}\sqrt{k'^2}\right)^{J-1}A_{P_{J,J+1}}(z)\,\,
\epsilon_{\alpha P'kk'}.
\label{C.6}
\ee

Here,
\be
&&A_{P_{J,J+1}}(z)=B_{P_{J,J+1}}(z)=
\\
\nonumber
&&=-\frac{2zP_J (z)+\left[Jz^2-(J+2)\right]P_{J+1}(z)}{(1-z^2)^2},
\\
\nonumber
&&C_{P_{J,J+1}}(z) =\frac{\left[(1-J)z^2+(J+1)\right]P_J (z)}{(1-z^2)^2}+
\\
\nonumber
&& \ \ \ \ \ \ \ \ \ \ \ \ \ \ \ \ \ +\frac{\left[(2J+1)z^2-(2J+3)\right]z P_{J+1} (z)}{(1-z^2)^2},
\\
\nonumber
&&D_{P_{J,J+1}}(z) =\frac{\left[(J+2)z^2-J\right]P_J
(z)-2zP_{J+1} (z)}{(1-z^2)^2},
\\
\nonumber
&&E_{P_{J,J+1}}(z)
=\frac{zP_J (z)-P_{J+1}(z)}{(1-z^2)}.
\\
\nonumber
&&
A_J =\ \frac{P_{J+1}(z)-z\,P_J(z)}{1-z^2}\ , \qquad B_J=\
\frac{P_J(z)-zP_{J+1}(z)}{1-z^2},
\\
\nonumber
&&\epsilon_{\alpha P'kk'}=\epsilon_{\alpha \beta \mu \nu} P'^{\beta}k^{\mu}k'^{\nu}
\label{C.7}
\ee

We also need more complicated convolutions,namely,  for the
factors \\
$
K_\beta X^{(J+1)}_{\mu_1 \mu_2 \cdots\mu_J\beta}(k)
X^{(J+1)}_{\mu_1 \mu_2 \cdots\mu_J\alpha}(k') K_\alpha
$
where $K=k,k'$:
\be
&&k_\beta\, X^{(J+1)}_{\mu_1 \mu_2 \cdots\mu_J\beta}(k)
X^{(J+1)}_{\mu_1 \mu_2 \cdots\mu_J\alpha}(k')\, k_\alpha=
\\
\nonumber
&&=k^2 \alpha (J)\left(\sqrt{k^2}\sqrt{k'^2}\right)^{J+1}P_{J+1}(z) ,
\\
\nonumber
&&k_\beta\, X^{(J+1)}_{\mu_1 \mu_2 \cdots\mu_J\beta}(k)
X^{(J+1)}_{\mu_1 \mu_2 \cdots\mu_J\alpha}(k')\, k'_\alpha=
\alpha (J)\left(\sqrt{k^2}\sqrt{k'^2}\right)^{J+2}P_{J}(z) ,
\\
\nonumber
&&k'_\beta\, X^{(J+1)}_{\mu_1 \mu_2 \cdots\mu_J\beta}(k)
X^{(J+1)}_{\mu_1 \mu_2 \cdots\mu_J\alpha}(k')\, k'_\alpha=
k'^2 \alpha (J)\left(\sqrt{k^2}\sqrt{k'^2}\right)^{J+1}P_{J+1}(z),
\\
\nonumber
&&k'_\beta\, X^{(J+1)}_{\mu_1 \mu_2 \cdots\mu_J\beta}(k)
X^{(J+1)}_{\mu_1 \mu_2 \cdots\mu_J\alpha}(k')\, k_\alpha=
\\
\nonumber
&&=\alpha (J)\left(\sqrt{k^2}\sqrt{k'^2}\right)^{J+2}
\left[\frac{2J+1}{J+1}zP_{J+1}(z)-\frac{J}{J+1}P_{J}(z)\right],
\\
\nonumber
&&g^{\perp}_{\beta\alpha}\,
X^{(J+1)}_{\mu_1 \mu_2 \cdots\mu_J\beta}(k)
X^{(J+1)}_{\mu_1 \mu_2 \cdots\mu_J\alpha}(k')=
\frac{2J+1}{J+1}\alpha (J)
\left(\sqrt{k^2}\sqrt{k'^2}\right)^{J+1}P_{J+1}(z);
\label{C.10}
\ee

and for the factors
$\left(K_\beta X^{(J+1)}_{\mu_1 \mu_2 \cdots\mu_J\beta}(k)
Z^{(J-1)}_{\mu_1 \mu_2 \cdots\mu_J,\alpha}(k') K_\alpha\right)$:
\be
&&k_\beta\, X^{(J+1)}_{\mu_1 \mu_2 \cdots\mu_J\beta}(k)
Z^{(J-1)}_{\mu_1 \mu_2 \cdots\mu_J,\alpha}(k')\, k_\alpha=
\\
\nonumber
&&
=k^4 \alpha (J)\left(\sqrt{k^2}\sqrt{k'^2}\right)^{J-1}P_{J-1}(z),
\\
\nonumber
&&k_\beta\, X^{(J+1)}_{\mu_1 \mu_2 \cdots\mu_J\beta}(k)
Z^{(J-1)}_{\mu_1 \mu_2 \cdots\mu_J,\alpha}(k')\, k'_\alpha=
k^2\alpha (J)\left(\sqrt{k^2}\sqrt{k'^2}\right)^{J}P_{J}(z) ,
\\
\nonumber
&&k'_\beta\, X^{(J+1)}_{\mu_1 \mu_2 \cdots\mu_J\beta}(k)
Z^{(J-1)}_{\mu_1 \mu_2 \cdots\mu_J,\alpha}(k')\, k'_\alpha=
\alpha (J)\left(\sqrt{k^2}\sqrt{k'^2}\right)^{J+1}P_{J+1}(z),
\\
\nonumber
&&k'_\beta\, X^{(J+1)}_{\mu_1 \mu_2 \cdots\mu_J\beta}(k)
Z^{(J-1)}_{\mu_1 \mu_2 \cdots\mu_J,\alpha}(k')\, k_\alpha=
k^2\alpha (J)\left(\sqrt{k^2}\sqrt{k'^2}\right)^{J}P_{J}(z),
\\
\nonumber
&&g^{\perp}_{\beta\alpha}\, X^{(J+1)}_{\mu_1 \mu_2 \cdots\mu_J\beta}(k)
Z^{(J-1)}_{\mu_1 \mu_2 \cdots\mu_J,\alpha}(k')=0;
\label{C.11}
\ee

and for the factors $\left(K_\beta Z^{(J-1)}_{\mu_1 \mu_2 \cdots\mu_J,\beta}(k)
Z^{(J-1)}_{\mu_1 \mu_2 \cdots\mu_J,\alpha}(k') K_\alpha\right)$:

\be
&&k_\beta\, Z^{(J-1)}_{\mu_1 \mu_2 \cdots\mu_J,\beta}(k)
Z^{(J-1)}_{\mu_1 \mu_2 \cdots\mu_J,\alpha}(k')\, k_\alpha=
\\
\nonumber
&&
=k^2 \alpha (J)\left(\sqrt{k^2}\sqrt{k'^2}\right)^{J-1}P_{J-1}(z),
\\
\nonumber
&&k_\beta\, Z^{(J-1)}_{\mu_1 \mu_2 \cdots\mu_J,\beta}(k)
Z^{(J-1)}_{\mu_1 \mu_2 \cdots\mu_J,\alpha}(k')\, k'_\alpha=
\alpha (J)\left(\sqrt{k^2}\sqrt{k'^2}\right)^{J}P_{J}(z),
\\
\nonumber
&&k'_\beta\, Z^{(J-1)}_{\mu_1 \mu_2 \cdots\mu_J,\beta}(k)
Z^{(J-1)}_{\mu_1 \mu_2 \cdots\mu_J,\alpha}(k')\, k'_\alpha=
k'^2 \alpha (J)\left(\sqrt{k^2}\sqrt{k'^2}\right)^{J-1}P_{J-1}(z),
\\
\nonumber
&&k'_\beta\, Z^{(J-1)}_{\mu_1 \mu_2 \cdots\mu_J,\beta}(k)
Z^{(J-1)}_{\mu_1 \mu_2 \cdots\mu_J,\alpha}(k')\, k_\alpha=
\\
\nonumber
&&=\alpha (J)\left(\sqrt{k^2}\sqrt{k'^2}\right)^{J}
\left[\frac{2J+1}{J}zP_{J-1}(z)-\frac{J+1}{J}P_{J}(z)\right],
\\
\nonumber
&&g^{\perp}_{\beta\alpha}\,
Z^{(J-1)}_{\mu_1 \mu_2 \cdots\mu_J,\beta}(k)
Z^{(J-1)}_{\mu_1 \mu_2 \cdots\mu_J,\alpha}(k')=
\frac{2J+1}{J}\alpha (J)
\left(\sqrt{k^2}\sqrt{k'^2}\right)^{J-1}P_{J-1}(z);
\label{C.12}
\ee

and for the factors $\left(K_\beta \epsilon_{\beta \nu_1\nu_2\nu_3} P_{\nu_1}
Z^{(J)}_{\nu_2 \mu_1\cdots\mu_J,\nu_3}(k)\,\,
\epsilon_{\alpha \lam_1\lam_2\lam_3} P'_{\lam_1}\times \right .
$\\
 $\left .
 \ \ \ \  \times
Z^{(J)}_{\lam_2 \mu_1\cdots\mu_J,\lam_3}(k') K_\alpha\right)$:
\be
&&k_\beta\, \epsilon_{\beta \nu_1\nu_2\nu_3} P_{\nu_1}
Z^{(J)}_{\nu_2 \mu_1\cdots\mu_J,\nu_3}(k)\,\,
\epsilon_{\alpha \lam_1\lam_2\lam_3} P'_{\lam_1}
Z^{(J)}_{\lam_2 \mu_1\cdots\mu_J,\lam_3}(k')\, k_\alpha=0,
\\
\nonumber
&&k_\beta\, \epsilon_{\beta \nu_1\nu_2\nu_3} P_{\nu_1}
Z^{(J)}_{\nu_2 \mu_1\cdots\mu_J,\nu_3}(k)\,\,
\epsilon_{\alpha \lam_1\lam_2\lam_3} P'_{\lam_1}
Z^{(J)}_{\lam_2 \mu_1\cdots\mu_J,\lam_3}(k')\, k'_\alpha=0,
\\
\nonumber
&&k'_\beta\, \epsilon_{\beta \nu_1\nu_2\nu_3} P_{\nu_1}
Z^{(J)}_{\nu_2 \mu_1\cdots\mu_J,\nu_3}(k)\,\,
\epsilon_{\alpha \lam_1\lam_2\lam_3} P'_{\lam_1}
Z^{(J)}_{\lam_2 \mu_1\cdots\mu_J,\lam_3}(k')\, k'_\alpha=0,
\\
\nonumber
&&k'_\beta\, \epsilon_{\beta \nu_1\nu_2\nu_3} P_{\nu_1}
Z^{(J)}_{\nu_2 \mu_1\cdots\mu_J,\nu_3}(k)\,\,
\epsilon_{\alpha \lam_1\lam_2\lam_3} P'_{\lam_1}
Z^{(J)}_{\lam_2 \mu_1\cdots\mu_J,\lam_3}(k')\, k_\alpha=
\\
\nonumber
&&=\frac{(2J+3)^2}{(J+1)^3}\alpha(J)
\left(\sqrt{k^2}\sqrt{k'^2}\right)^{J+1}(PP')
\left[zP_J(z)-P_{J+1}(z)\right],
\\
\nonumber
&&g^{\perp}_{\beta\alpha}\,\epsilon_{\beta \nu_1\nu_2\nu_3} P_{\nu_1}
Z^{(J)}_{\nu_2 \mu_1\cdots\mu_J,\nu_3}(k)\,\,
\epsilon_{\alpha \lam_1\lam_2\lam_3} P'_{\lam_1}
Z^{(J)}_{\lam_2 \mu_1\cdots\mu_J,\lam_3}(k')=
\\
\nonumber
&&=-\frac{J(2J+3)^2}{(J+1)^3}\alpha(J)
\left(\sqrt{k^2}\sqrt{k'^2}\right)^{J}(PP')
P_J(z).
\label{C.13}
\ee

\section{Appendix:
The Bethe--Salpeter equations for the
$\omega$, $\phi$, $a_2$ and $f_2$ trajectories}
\setcounter{equation}{0}
Here we present the Bethe--Salpeter equations for the
$\omega$, $\phi$, $a_2$, and $f_2$ trajectories.
Though the explicit form of these equations is rather cumbersome,
the investigation of these trajectories is informative for the
reconstruction of quark-antiquark forces.

\subsection{ Equations for the $a_2$ ($M^2,n$)-trajectories}

The following states are located on two
($n,M^2$)-trajectories
for the $a_2$ states $M\leq 2400$ MeV \cite{syst}:\\
1) $a_2 (1320)$ with $n =1$, $a_2 (1660)$ with $n =2$,
$a_2 (1950)$ with $n =3$, $a_2 (2255)$ with $n =4$; \\
2) $a_2$-trajectory: $a_2 (2030)$ with $n =1$, $a_2 (2310)$ with $n =2$.\\
Correspondingly, we have two coupled equations for two wave functions:
\be
(s-M^2) \left[ 4\psi^{(1,1,2)}_{a_2,n}(s)A_j  +
      \left(\frac53 s+4k^2\right)
        \psi^{(1,3,2)}_{a_2,n}(s)B_j\right]=
\ee
$$
=\int\limits_{4m^2}^{\infty}\frac{ds'}{\pi}\rho(s')
\, 8\, (-k'^2)
 \psi^{(1,1,2)}_{a_2,n}(s')A_j\times
$$
$$
\times
\left[
  \frac{5}{14}\left(s+\frac{12}{5} k^2 \right) k'^4 \,
     \widetilde V^{(3)}_V(s,s')
  + \frac{3}{10}\sqrt{ss'}k'^2\, \widetilde V^{(2)}_A(s,s')-
\right .
$$
$$
\left .
  -\frac65 m^2\, k'^2 \, \widetilde V^{(2)}_S(s,s')
  +\frac{1}{3}\left( s'+\frac{8}{5} k'^2\right)\,
     \widetilde V^{(1)}_V(s,s')
\right]-
$$
$$
-\int\limits_{4m^2}^{\infty}\frac{ds'}{\pi}\rho(s')
\,2 \, k'^6
 \psi^{(1,3,2)}_{a_2,n}(s')B_j\times
$$
$$
\times
\left[
  m^2\frac{100}{21}\sqrt{ss'}\, \widetilde V^{(3)}_T(s,s')
  +\frac{5}{14}\left(
    \frac53 ss'+4s'k^2+4sk'^2+\frac{48}{5} k^2k'^2\right)
\times
\right .
$$
$$
\left .
  \times \widetilde V^{(3)}_V(s,s') -m^2\,\frac{24}{5}\, \widetilde V^{(2)}_S(s,s')
   -\frac45 \,\sqrt{ss'}\widetilde V^{(2)}_A(s,s')
   +\frac{32}{15} \, \widetilde V^{(1)}_V(s,s')
    \right] ,
$$
and
\be
(s-M^2) \left[\left(\frac52 s+4k^2\right)
     \psi^{(1,1,2)}_{a_2,n}(s)A_j
     +4k^4\psi^{(1,3,2)}_{a_2,n}(s)B_j\right]=
\ee
$$
=\int\limits_{4m^2}^{\infty}\frac{ds'}{\pi}\rho(s')
2\, (-k'^2)
 \psi^{(1,1,2)}_{a_2,n}(s')A_j \times
$$
$$
\times
\left[
  m^2\,\frac{20}{3}\,
  \sqrt{ss'}\, \widetilde V^{(1)}_T(s,s')
  +\frac13\left(\frac52 ss'+4s'k^2+4sk'^2+\frac{32}{5} k^2k'^2\right)
    \widetilde  V^{(1)}_V(s,s')-
\right .
$$
$$
\left .
 -m^2\,\frac{24}{5}\,k^2k'^2 \widetilde V^{(2)}_S(s,s')
 -\frac95\,\sqrt{ss'}k^2k'^2
     \widetilde V^{(2)}_A(s,s')
   +\frac{24}{7}\, k^4k'^4
     \widetilde V^{(3)}_V(s,s') \right ]-
$$
$$
-\int\limits_{4m^2}^{\infty}\frac{ds'}{\pi}\rho(s')
\, 8\, k'^6 \psi^{(1,3,2)}_{a_2,n}(s')B_j \times
$$
$$
\nonumber
\times \left[
 \frac{5}{14}\, \left(s'+\frac{12}{5} k'^2 \right) k^4 \,
    \widetilde V^{(3)}_V(s,s')
 +\frac{3}{10}\, \sqrt{ss'}k^2 \, \widetilde V^{(2)}_A(s,s')-
\right .
$$
$$
\nonumber
\left .
 -m^2\,\frac{6}{5} k^2\, \widetilde V^{(2)}_S(s,s')
 +\frac13 \,\left( s+\frac85 k^2\right)\,
   \widetilde V^{(1)}_V(s,s')
\right].
$$
Normalization and orthogonality
conditions read:
\be
\int\limits_{4m^2}^{\infty}\frac{ds}{\pi}\rho(s)
\left [
A^2_j  \left(\psi^{(1,1,2)}_{a_2,n}(s)\right)^2
2\alpha (2)(-k^2) \left(\frac52 s+4k^2\right)+
\right .
\ee
$$
\left .
+2A_j B_j  \psi^{(1,1,2)}_{a_2,n}(s)\psi^{(1,3,2)}_{a_2,n}(s)
\, 8\, \alpha (2) (-k^6)+
\right .
$$
$$
\left .
+
B_j^2 \left( \psi^{(1,3,J)}_{a_2,n}(s) \right)^2
\, 2\, \alpha (2)(-k^6)\left(\frac53 s+4k^2 \right)
\right] =1
$$
and
\be
\int\limits_{4m^2}^{\infty}\frac{ds}{\pi}\rho(s)
\left [
A_1 A_2\left( \psi^{(1,1,2)}_{a_2,n}(s)\right)^2
\, 2\, \alpha (2)(-k^2)\left(\frac52 s+4k^2\right)+
\right .
\ee
$$
\left .
+(A_1 B_2+A_2 B_1)\psi^{(1,1,2)}_{a_2,n}(s)\psi^{(1,3,2)}_{a_2,n}(s)
\, 8\, \alpha (2) (-k^6)+
\right .
$$
$$
\left .
+
B_1B_2 \left( \psi^{(1,3,2)}_{a_2,n}(s) \right)^2
\, 2\, \alpha (2)(-k^6)\left(\frac53 s+4k^2 \right)
\right] =0.
$$

\subsection{ Equations for the $\omega$ and $\phi$
($M^2,n$)-trajectories}

We have four trajectories in this sector with the
following states located on the
($n, M^2$)-trajectories \cite{syst}:\\
1) $S$-wave dominant states: $\omega (780)$ with $n =1$,
$\omega (1420)$ with
$n=2$, $\omega (1800)$ with $n =3$, $\omega (2150)$ with $n =4$;\\
2) $D$-wave dominant states: $\omega (1640)$ with $n =1$,
$\omega (1920)$ with $n =2$, $\omega (2295)$ with $n =3$;\\
3) $S$-wave dominant states: $\phi (1020)$ with $n =1$, $\phi (1660)$
with $n=2$, $\phi (1950)$ with $n =3$;\\ 4) $D$-wave dominant states:
$\phi (1700)$ with $n =1$.\\
Correspondingly, we have four coupled equations for four wave
functions. The first one reads as follows:
\be
(s-M^2)
\left[4\psi^{(1,0,1)}_{\omega,(n\bar n),n}(s)A_j\cos\Theta_n
      +\left(\frac32s+4k^2\right)
        \psi^{(1,2,1)}_{\omega,(n\bar n),n}(s)\times \right .
\ee
$$
\left .
\times B_j\cos\Theta'_n\right]=\int\limits_{4m^2}^{\infty}\frac{ds'}{\pi}\rho(s')
\, 8 \, \psi^{(1,0,1)}_{\omega,(n\bar n),n}(s')A_j\cos\Theta_n \times
$$
$$
\times
\left[
  \frac{3}{10}\left(s+\frac83k^2 \right)k'^4 \,
     \widetilde V^{(2)}_{(n\bar n \to n\bar n), V}(s,s')
  +\frac13 \sqrt{ss'}k'^2\,
     \widetilde V^{(1)}_{(n\bar n \to n\bar n), A}(s,s')-
\right .
$$
$$
\left .
  -\frac43 m^2\, k'^2\,
     \widetilde V^{(1)}_{(n\bar n \to n\bar n), S}(s,s')
  +\left(s'+\frac43 k'^2\right)\,
     \widetilde V^{(0)}_{(n\bar n \to n\bar n), V}(s,s')
\right]+
$$
$$
+\int\limits_{4m^2}^{\infty}\frac{ds'}{\pi}\rho(s')\,
2\, k'^4 \psi^{(1,2,1)}_{\omega,(n\bar n),n}(s')B_j\cos\Theta'_n \times
$$
$$
\times
\left[
   m^2\frac{18}{5}\,\sqrt{ss'}\,
     \widetilde V^{(2)}_{(n\bar n \to n\bar n), T}(s,s')
    +\frac{3}{10}\,\left(\frac32 ss'+4s'k^2+4sk'^2
    +\frac{32}{3} k^2k'^2\right) \times
\right .
$$
$$
\left .
 \times
 \widetilde V^{(2)}_{(n\bar n \to n\bar n), V}(s,s')-
  m^2\frac{16}{3} \,
    \widetilde V^{(1)}_{(n\bar n \to n\bar n), S}(s,s')
  -\frac23 \, \sqrt{ss'}
    \widetilde V^{(1)}_{(n\bar n \to n\bar n), A}(s,s')+
\right .
$$
$$
\left .
+\frac{16}{3} \,
    \widetilde V^{(0)}_{(n\bar n \to n\bar n), V}(s,s')\right]
+\int\limits_{4m^2_s}^{\infty}\frac{ds'}{\pi}\rho_s(s')
\, 8 \, \psi^{(1,0,1)}_{\omega,(s\bar s),n}(s')A_j\sin\Theta_n\times
$$
$$
\times
\left[
  \frac{3}{10}\left(s+\frac83k^2 \right)k'^4_s \,
     \widetilde V^{(2)}_{(s\bar s \to n\bar n), V}(s,s')
  +\frac13 \sqrt{ss'}k'^2_s\,
     \widetilde V^{(1)}_{(s\bar s \to n\bar n), A}(s,s')-
\right .
$$
$$
\left .
  -\frac43 mm_s\, k'^2_s\,
     \widetilde V^{(1)}_{(s\bar s \to n\bar n), S}(s,s')
  +\left(s'+\frac43 k'^2_s\right)\,
     \widetilde V^{(0)}_{(s\bar s \to n\bar n), V}(s,s')
\right]+
$$
$$
+\int\limits_{4m^2_s}^{\infty}\frac{ds'}{\pi}\rho_s(s')\,
2\, k'^4_s \psi^{(1,2,1)}_{\omega,(s\bar s),n}(s')B_j\sin\Theta'_n\times
$$
$$
\times
\left[
   mm_s\frac{18}{5}\,\sqrt{ss'}\,
     \widetilde V^{(2)}_{(s\bar s \to n\bar n), T}(s,s')
   +\frac{3}{10}\,\left(\frac32 ss'+4s'k^2+4sk'^2_s
    +\frac{32}{3} k^2k'^2_s\right)
    \times
\right .
$$
$$
\left .
  \times \widetilde V^{(2)}_{(s\bar s \to n\bar n), V}(s,s')
  -mm_s\frac{16}{3} \,
    \widetilde V^{(1)}_{(s\bar s \to n\bar n), S}(s,s')
  -\frac23 \, \sqrt{ss'}
    \widetilde V^{(1)}_{(s\bar s \to n\bar n), A}(s,s')+
    \right .
$$
$$
\left .
 +\frac{16}{3} \,
    \widetilde V^{(0)}_{(s\bar s \to n\bar n), V}(s,s')
    \right] ,
$$
the second one,
\be
(s-M^2)
\left[4\psi^{(1,0,1)}_{\omega,(s\bar s),n}(s)A_j\sin\Theta_n
      +\left(\frac32s+4k^2_s\right)
        \psi^{(1,2,1)}_{\omega,(s\bar s),n}(s) \times \right .
\ee
$$
\left .
\times B_j\sin\Theta'_n\right]=\int\limits_{4m^2}^{\infty}\frac{ds'}{\pi}\rho(s')
\, 8 \, \psi^{(1,0,1)}_{\omega,(n\bar n),n}(s')A_j\cos\Theta_n\times
$$
$$
\times
\left[
  \frac{3}{10}\left(s+\frac83k^2_s \right)k'^4 \,
     \widetilde V^{(2)}_{(n\bar n \to s\bar s), V}(s,s')
  +\frac13 \sqrt{ss'}k'^2\,
     \widetilde V^{(1)}_{(n\bar n \to s\bar s), A}(s,s')-
\right .
$$
$$
\left .
  -\frac43 mm_s\, k'^2\,
     \widetilde V^{(1)}_{(n\bar n \to s\bar s), S}(s,s')
  +\left(s'+\frac43 k'^2\right)\,
     \widetilde V^{(0)}_{(n\bar n \to s\bar s), V}(s,s')
\right]+
$$
$$
+\int\limits_{4m^2}^{\infty}\frac{ds'}{\pi}\rho(s')\,
2\, k'^4 \psi^{(1,2,1)}_{\omega,(n\bar n),n}(s')B_j\cos\Theta'_n\times
$$
$$
\times
\left[
   mm_s\frac{18}{5}\,\sqrt{ss'}\,
     \widetilde V^{(2)}_{(n\bar n \to s\bar s), T}(s,s')
   +\frac{3}{10}\,\left(\frac32 ss'+4s'k^2_s+4sk'^2
    +\frac{32}{3} k^2_sk'^2\right) \times
\right .
$$
$$
\left .
\times \widetilde V^{(2)}_{(n\bar n \to s\bar s), V}(s,s')-
mm_s\frac{16}{3} \, \widetilde V^{(1)}_{(n\bar n \to s\bar s), S}(s,s')
  -\frac23 \, \sqrt{ss'}
    \widetilde V^{(1)}_{(n\bar n \to s\bar s), A}(s,s')+
\right .
$$
$$
\left .
  +\frac{16}{3} \,
    \widetilde V^{(0)}_{(n\bar n \to s\bar s), V}(s,s')
    \right] +
\int\limits_{4m^2_s}^{\infty}\frac{ds'}{\pi}\rho_s(s')
\, 8 \, \psi^{(1,0,1)}_{\omega,(s\bar s),n}(s')A_j\sin\Theta_n\times
$$
$$
\times
\left[
  \frac{3}{10}\left(s+\frac83 k^2_s \right)k'^4_s \,
     \widetilde V^{(2)}_{(s\bar s \to s\bar s), V}(s,s')
  +\frac13 \sqrt{ss'}k'^2_s\,
     \widetilde V^{(1)}_{(s\bar s \to s\bar s), A}(s,s')-
\right .
$$
$$
\left .
  -\frac43 m^2_s\, k'^2_s\,
     \widetilde V^{(1)}_{(s\bar s \to s\bar s), S}(s,s')
  +\left(s'+\frac43 k'^2_s\right)\,
     \widetilde V^{(0)}_{(s\bar s \to s\bar s), V}(s,s')
\right]+
$$
$$
+\int\limits_{4m^2_s}^{\infty}\frac{ds'}{\pi}\rho_s(s')\,
2\, k'^4_s \psi^{(1,2,1)}_{\omega,(s\bar s),n}(s')B_j\sin\Theta'_n\times
$$
$$
\times
\left[
   m^2_s\frac{18}{5}\,\sqrt{ss'}\,
     \widetilde V^{(2)}_{(s\bar s \to s\bar s), T}(s,s')
   +\frac{3}{10}\,\left(\frac32 ss'+4s'k^2_s+4sk'^2_s
    +\frac{32}{3} k^2_sk'^2_s\right)
    \times
\right .
$$
$$
\left .
  \times  \widetilde V^{(2)}_{(s\bar s \to s\bar s), V}(s,s')-m^2_s\frac{16}{3} \,
    \widetilde V^{(1)}_{(s\bar s \to s\bar s), S}(s,s')
  -\frac23 \, \sqrt{ss'}
    \widetilde V^{(1)}_{(s\bar s \to s\bar s), A}(s,s')+
\right .
$$
$$
\left .
+\frac{16}{3} \,
    \widetilde V^{(0)}_{(s\bar s \to s\bar s), V}(s,s')
    \right] ,
$$
the third one,
\be
(s-M^2)\left[\left(3s+4k^2\right)
     \psi^{(1,0,1)}_{\omega,(n\bar n),n}(s)A_j\cos\Theta_n
+4k^4\psi^{(1,2,1)}_{\omega,(n\bar n),n}(s) \times
\right .
\ee
$$
\left .
\times B_j\cos\Theta'_n
\right]=\int\limits_{4m^2}^{\infty}\frac{ds'}{\pi}\rho(s')\,
2\, \psi^{(1,0,1)}_{\omega,(n\bar n),n}(s')A_j\cos\Theta_n\times
$$
$$
\times
\left[
   24\, m^2\,\sqrt{ss'}\,
     \widetilde V^{(0)}_{(n\bar n \to n\bar n), T}(s,s')
  +\left(3ss'+4s'k^2+4sk'^2+\frac{16}{3} k^2k'^2\right) \times
\right .
$$
$$
\left .
  \times
  \widetilde  V^{(0)}_{(n\bar n \to n\bar n), V}(s,s')
  -m^2\,\frac{16}{3} \,k^2k'^2
    \widetilde V^{(1)}_{(n\bar n \to n\bar n), S}(s,s')
  -\frac83 \,\sqrt{ss'}k^2k'^2
    \times
\right .
$$
$$
\left .
\times
\widetilde V^{(1)}_{(n\bar n \to n\bar n), A}(s,s')
  +\frac{16}{5}\, k^4k'^4
    \widetilde V^{(2)}_{(n\bar n \to n\bar n), V}(s,s')\right ]
+\int\limits_{4m^2}^{\infty}\frac{ds'}{\pi}\rho(s')\,
8\, k'^4 \psi^{(1,2,1)}_{\omega,(n\bar n),n}(s')\times
$$
$$
\times B_j\cos\Theta'_n \left[
  \frac{3}{10}\,\left(s'+\frac83 k'^2 \right)k^4 \,
    \widetilde V^{(2)}_{(n\bar n \to n\bar n), V}(s,s')
  +\frac13 \,\sqrt{ss'}k^2 \,
    \widetilde V^{(1)}_{(n\bar n \to n\bar n), A}(s,s')-
\right .
$$
$$
\left .
  -m^2\,\frac43 k^2\,
    \widetilde V^{(1)}_{(n\bar n \to n\bar n), S}(s,s')
  +\left(s+\frac43 k^2\right)\,
    \widetilde V^{(0)}_{(n\bar n \to n\bar n), V}(s,s')
\right]+
$$
$$
+\int\limits_{4m^2_s}^{\infty}\frac{ds'}{\pi}\rho_s(s')\,
2\, \psi^{(1,0,1)}_{\omega,(s\bar s),n}(s')A_j\sin\Theta_n\times
$$
$$
\times
\left[
   24\, mm_s\,\sqrt{ss'}\,
     \widetilde V^{(0)}_{(s\bar s \to n\bar n), T}(s,s')
  +\left(3ss'+4s'k^2+4sk'^2_s+\frac{16}{3} k^2k'^2_s\right)
       \times
\right .
$$
$$
\left .
  \times
  \widetilde  V^{(0)}_{(s\bar s \to n\bar n), V}(s,s')
  -mm_s\,\frac{16}{3} \,k^2k'^2_s
    \widetilde V^{(1)}_{(s\bar s \to n\bar n), S}(s,s')
  -\frac83 \,\sqrt{ss'}k^2k'^2_s
    \times
 \right .
$$
$$
\left .
\times
\widetilde V^{(1)}_{(s\bar s \to n\bar n), A}(s,s')
  +\frac{16}{5}\, k^4k'^4_s
    \widetilde V^{(2)}_{(s\bar s \to n\bar n), V}(s,s') \right ]
+\int\limits_{4m^2_s}^{\infty}\frac{ds'}{\pi}\rho_s(s')\,
8\, k'^4_s \psi^{(1,2,1)}_{\omega,(s\bar s),n}(s')\times
$$
$$
\times  B_j\sin\Theta'_n\left[
  \frac{3}{10}\,\left(s'+\frac83 k'^2_s \right)k^4 \,
    \widetilde V^{(2)}_{(s\bar s \to n\bar n), V}(s,s')
  +\frac13 \,\sqrt{ss'}k^2 \,
    \widetilde V^{(1)}_{(s\bar s \to n\bar n), A}(s,s')-
\right .
$$
$$
\left .
  -mm_s\,\frac43 k^2\,
    \widetilde V^{(1)}_{(s\bar s \to n\bar n), S}(s,s')
  +\left(s+\frac43 k^2\right)\,
    \widetilde V^{(0)}_{(s\bar s \to n\bar n), V}(s,s')
\right],
$$
and the fourth equation:
\be
(s-M^2)\left[\left(3s+4k^2_s\right)
     \psi^{(1,0,1)}_{\omega,(s\bar s),n}(s)A_j\sin\Theta_n
+4k^4_s\psi^{(1,2,1)}_{\omega,(s\bar s),n}(s)\times
\right .
\ee
$$
\left .
\times B_j\sin\Theta'_n\right]=\int\limits_{4m^2}^{\infty}\frac{ds'}{\pi}\rho(s')\,
2\, \psi^{(1,0,1)}_{\omega,(n\bar n),n}(s')A_j\cos\Theta_n\times
$$
$$
\times
\left[
   24\, mm_s\,\sqrt{ss'}\,
     \widetilde V^{(0)}_{(n\bar n \to s\bar s), T}(s,s')
  +\left(3ss'+4s'k^2_s+4sk'^2+\frac{16}{3} k^2_sk'^2\right)
       \times
\right .
$$
$$
\left .
 \times \widetilde  V^{(0)}_{(n\bar n \to s\bar s), V}(s,s')
  -mm_s  \,\frac{16}{3} \,k^2_sk'^2
    \widetilde V^{(1)}_{(n\bar n \to s\bar s), S}(s,s')
  -\frac83 \,\sqrt{ss'}k^2_sk'^2
    \widetilde V^{(1)}_{(n\bar n \to s\bar s), A}(s,s')+
    \right .
$$
$$
\left .
+\frac{16}{5}\, k^4_sk'^4
    \widetilde V^{(2)}_{(n\bar n \to s\bar s), V}(s,s') \right ]
+\int\limits_{4m^2}^{\infty}\frac{ds'}{\pi}\rho(s')\,
8\, k'^4 \psi^{(1,2,1)}_{\omega,(n\bar n),n}(s')B_j\cos\Theta'_n\times
$$
$$
\times \left[
  \frac{3}{10}\,\left(s'+\frac83 k'^2 \right)k^4_s \,
    \widetilde V^{(2)}_{(n\bar n \to s\bar s), V}(s,s')
  +\frac13 \,\sqrt{ss'}k^2_s \,
    \widetilde V^{(1)}_{(n\bar n \to s\bar s), A}(s,s')-
\right .
$$
$$
\left .
  -mm_s\,\frac43 k^2_s\,
    \widetilde V^{(1)}_{(n\bar n \to s\bar s), S}(s,s')
  +\left(s+\frac43 k^2_s\right)\,
    \widetilde V^{(0)}_{(n\bar n \to s\bar s), V}(s,s')
\right]+
$$
$$
+\int\limits_{4m^2_s}^{\infty}\frac{ds'}{\pi}\rho_s(s')\,
2\, \psi^{(1,0,1)}_{\omega,(s\bar s),n}(s')A_j\sin\Theta_n\times
$$
$$
\times
\left[
   24\, m^2_s\,\sqrt{ss'}\,
     \widetilde V^{(0)}_{(s\bar s \to s\bar s), T}(s,s')
  +\left(3ss'+4s'k^2_s+4sk'^2_s+\frac{16}{3} k^2_sk'^2_s\right)
    \times
\right .
$$
$$
\left .
 \times \widetilde  V^{(0)}_{(s\bar s \to s\bar s), V}(s,s')
  -m^2_s\,\frac{16}{3} \,k^2_sk'^2_s
    \widetilde V^{(1)}_{(s\bar s \to s\bar s), S}(s,s')
  -\frac83 \,\sqrt{ss'}k^2_sk'^2_s
    \widetilde V^{(1)}_{(s\bar s \to s\bar s), A}(s,s')+
\right .
$$
$$
\left .
+\frac{16}{5}\, k^4_sk'^4_s
    \widetilde V^{(2)}_{(s\bar s \to s\bar s), V}(s,s') \right ]
+\int\limits_{4m^2_s}^{\infty}\frac{ds'}{\pi}\rho_s(s')\,
8\, k'^4_s \psi^{(1,2,1)}_{\omega,(s\bar s),n}(s')B_j\sin\Theta'_n\times
$$
$$
\times \left[
  \frac{3}{10}\,\left(s'+\frac83 k'^2_s \right)k^4_s \,
    \widetilde V^{(2)}_{(s\bar s \to s\bar s), V}(s,s')
  +\frac13 \,\sqrt{ss'}k^2_s \,
    \widetilde V^{(1)}_{(s\bar s \to s\bar s), A}(s,s')-
\right .
$$
$$
\left .
  -m^2_s\,\frac43 k^2_s\,
    \widetilde V^{(1)}_{(s\bar s \to s\bar s), S}(s,s')
  +\left(s+\frac43 k^2_s\right)\,
    \widetilde V^{(0)}_{(s\bar s \to s\bar s), V}(s,s')
\right].
$$

\subsection{ Equations for the $f_2$ ($n,M^2$)-trajectories}

The $f_2$ mesons lay on the following four trajectories in the ($n,M^2$) plane
\cite{syst}:\\
1) dominantly $P$-wave states: $f_2 (1285)$ with $n =1$, $f_2 (1640)$ at
$n=2$, $f_2 (1950)$ with $n =3$, $f_2 (2210)$ with $n =4$;\\
2) dominantly $P$-wave states:
$f_2 (1525)$ with $n =1$, $f_2 (1790)$ with $n =2$;\\
3) dominantly $F$-wave states:
$f_2 (2020)$ with $n =1$, $f_2 (2290)$ with $n =2$;\\
4) dominantly $F$-wave state:
$f_2 (2200)$ with $n =1$. \\
We have four equations for these four sets of states:\\
the first one,
\be
(s-M^2)\left[ 4\psi^{(1,1,2)}_{f_2,(n\bar n),n}(s)A_j\cos\Theta_n
       + \left(\frac53 s+4k^2\right)
              \psi^{(1,3,2)}_{f_2,(n\bar n),n}(s)\times
\right .
\ee
$$
\left .
\times
B_j\cos\Theta'_n\right]=\int\limits_{4m^2}^{\infty}\frac{ds'}{\pi}\rho(s')
\, 8\, (-k'^2)
 \psi^{(1,1,2)}_{f_2,(n\bar n),n}(s')A_j\cos\Theta_n\times
$$
$$
\times
\left[
  \frac{5}{14}\left(s+\frac{12}{5} k^2 \right) k'^4 \,
     \widetilde V^{(3)}_{(n\bar n \to n\bar n), V}(s,s')
  + \frac{3}{10}\sqrt{ss'}k'^2\,
     \widetilde V^{(2)}_{(n\bar n \to n\bar n), A}(s,s')-
\right .
$$
$$
\left .
  -\frac{6}{5} m^2\, k'^2\,\,
     \widetilde V^{(2)}_{(n\bar n \to n\bar n), S}(s,s')
  +\frac{1}{3}\left( s'+\frac85 k'^2\right)\,
     \widetilde V^{(1)}_{(n\bar n \to n\bar n), V}(s,s')
\right]-
$$
$$
-\int\limits_{4m^2}^{\infty}\frac{ds'}{\pi}\rho(s')
\,2 \, k'^6
 \psi^{(1,3,2)}_{f_2,(n\bar n),n}(s')B_j\cos\Theta'_n
\left[
  m^2\frac{100}{21}\sqrt{ss'}\,
    \widetilde V^{(3)}_{(n\bar n \to n\bar n), T}(s,s')+
\right .
$$
$$
\left .
  +\frac{5}{14}\left(
    \frac53 ss'+4s'k^2+4sk'^2+\frac{48}{5} k^2k'^2\right)
     \widetilde V^{(3)}_{(n\bar n \to n\bar n), V}(s,s')-
\right .
$$
$$
\left .
   -m^2\,\frac{24}{5}\,
     \widetilde V^{(2)}_{(n\bar n \to n\bar n), S}(s,s')
   -\frac45 \,\sqrt{ss'}
     \widetilde V^{(2)}_{(n\bar n \to n\bar n), A}(s,s')
   +\frac{32}{15} \,
     \widetilde V^{(1)}_{(n\bar n \to n\bar n), V}(s,s')
    \right] +
$$
$$
+\int\limits_{4m^2_s}^{\infty}\frac{ds'}{\pi}\rho_s(s')
\, 8\, (-k'^2_s)
 \psi^{(1,1,2)}_{f_2,(s\bar s),n}(s')A_j\sin\Theta_n\times
$$
$$
\times
\left[
  \frac{5}{14}\left(s+\frac{12}{5} k^2 \right) k'^4_s \,
     \widetilde V^{(3)}_{(s\bar s \to n\bar n), V}(s,s')
  + \frac{3}{10}\sqrt{ss'}k'^2_s\,
     \widetilde V^{(2)}_{(s\bar s \to n\bar n), A}(s,s')-
\right .
$$
$$
\left .
  -\frac65 mm_s\, k'^2_s\,\,
     \widetilde V^{(2)}_{(s\bar s \to n\bar n), S}(s,s')
  +\frac{1}{3}\left( s'+\frac85 k'^2_s\right)\,
     \widetilde V^{(1)}_{(s\bar s \to n\bar n), V}(s,s')
\right]-
$$
$$
-\int\limits_{4m^2_s}^{\infty}\frac{ds'}{\pi}\rho_s(s')
\,2 \, k'^6_s
 \psi^{(1,3,2)}_{f_2,(s\bar s),n}(s')B_j\sin\Theta'_n\times
$$
$$
\times
\left[
  mm_s\,\frac{100}{21}\sqrt{ss'}\,
    \widetilde V^{(3)}_{(s\bar s \to n\bar n), T}(s,s')
  +\frac{5}{14}\left(
    \frac53 ss'+4s'k^2+4sk'^2_s+\frac{48}{5} k^2k'^2_s\right)
    \times
\right .
$$
$$
\left .
   \times
    \widetilde V^{(3)}_{(s\bar s \to n\bar n), V}(s,s')
   -mm_s\,\frac{24}{5}\,
     \widetilde V^{(2)}_{(s\bar s \to n\bar n), S}(s,s')
   -\frac45 \,\sqrt{ss'}
     \widetilde V^{(2)}_{(s\bar s \to n\bar n), A}(s,s')+
\right .
$$
$$
\left .
+\frac{32}{15} \,
     \widetilde V^{(1)}_{(s\bar s \to n\bar n), V}(s,s')
    \right] ,
$$
the second one,
\be
(s-M^2) \left[ 4\psi^{(1,1,2)}_{f_2,(s\bar s),n}(s)A_j\sin\Theta_n
+ \left(\frac53 s+4k^2_s\right)
\psi^{(1,3,2)}_{f_2,(s\bar s),n}(s)\times
\right .
\ee
$$
\left .
\times
B_j\sin\Theta'_n\right]=\int\limits_{4m^2}^{\infty}\frac{ds'}{\pi}\rho(s')
\, 8\, (-k'^2)
 \psi^{(1,1,2)}_{f_2,(n\bar n),n}(s')A_j\cos\Theta_n\times
$$
$$
\times
\left[
  \frac{5}{14}\left(s+\frac{12}{5} k^2_s \right) k'^4 \,
     \widetilde V^{(3)}_{(n\bar n \to s\bar s), V}(s,s')
  + \frac{3}{10}\sqrt{ss'}k'^2\,
     \widetilde V^{(2)}_{(n\bar n \to s\bar s), A}(s,s')-
\right .
$$
$$
\left .
  -\frac65 mm_s\, k'^2\,\,
     \widetilde V^{(2)}_{(n\bar n \to s\bar s), S}(s,s')
  +\frac{1}{3}\left( s'+\frac85 k'^2\right)\,
     \widetilde V^{(1)}_{(n\bar n \to s\bar s), V}(s,s')
\right]-
$$
$$
-\int\limits_{4m^2}^{\infty}\frac{ds'}{\pi}\rho(s')
\,2 \, k'^6
 \psi^{(1,3,2)}_{f_2,(n\bar n),n}(s')B_j\cos\Theta'_n
 \left[
  mm_s\frac{100}{21}\sqrt{ss'}\,
    \widetilde V^{(3)}_{(n\bar n \to s\bar s), T}(s,s')+
\right .
$$
$$
\left .
 +\frac{5}{14}\left(
    \frac53 ss'+4s'k^2_s+4sk'^2+\frac{48}{5} k^2_sk'^2\right)
     \widetilde V^{(3)}_{(n\bar n \to s\bar s), V}(s,s')-
\right .
$$
$$
\left .
   -mm_s\,\frac{24}{5}\,
     \widetilde V^{(2)}_{(n\bar n \to s\bar s), S}(s,s')
   -\frac45 \,\sqrt{ss'}
     \widetilde V^{(2)}_{(n\bar n \to s\bar s), A}(s,s')
   +\frac{32}{15} \,
     \widetilde V^{(1)}_{(n\bar n \to s\bar s), V}(s,s')
    \right] +
$$
$$
+\int\limits_{4m^2_s}^{\infty}\frac{ds'}{\pi}\rho_s(s')
\, 8\, (-k'^2_s)
 \psi^{(1,1,2)}_{f_2,(s\bar s),n}(s')A_j\sin\Theta_n\times
$$
$$
\times
\left[
  \frac{5}{14}\left(s+\frac{12}{5} k^2_s \right) k'^4_s \,
     \widetilde V^{(3)}_{(s\bar s \to s\bar s), V}(s,s')
  + \frac{3}{10}\sqrt{ss'}k'^2_s\,
     \widetilde V^{(2)}_{(s\bar s \to s\bar s), A}(s,s')-
\right .
$$
$$
\left .
  -\frac65 m^2_s\, k'^2_s\,\,
     \widetilde V^{(2)}_{(s\bar s \to s\bar s), S}(s,s')
  +\frac{1}{3}\left( s'+\frac85 k'^2_s\right)\,
     \widetilde V^{(1)}_{(s\bar s \to s\bar s), V}(s,s')
\right]-
$$
$$
-\int\limits_{4m^2_s}^{\infty}\frac{ds'}{\pi}\rho_s(s')
\,2 \, k'^6_s
 \psi^{(1,3,2)}_{f_2,(s\bar s),n}(s')B_j\sin\Theta'_n\times
$$
$$
\times
\left[
  m^2_s\,\frac{100}{21}\sqrt{ss'}\,
    \widetilde V^{(3)}_{(s\bar s \to s\bar s), T}(s,s')
  +\frac{5}{14}\left(
    \frac53 ss'+4s'k^2_s+4sk'^2_s+\frac{48}{5} k^2_sk'^2_s\right)
\times
\right .
$$
$$
\left .
\times
\widetilde V^{(3)}_{(s\bar s \to s\bar s), V}(s,s')-
   m^2_s\,\frac{24}{5}\,
     \widetilde V^{(2)}_{(s\bar s \to s\bar s), S}(s,s')
   -\frac45 \,\sqrt{ss'}
     \widetilde V^{(2)}_{(s\bar s \to s\bar s), A}(s,s')+
\right .
$$
$$
\left .
 +\frac{32}{15} \,
     \widetilde V^{(1)}_{(s\bar s \to s\bar s), V}(s,s')
    \right],
$$
the third one,
\be
(s-M^2) \left[\left(\frac52 s+4k^2\right)
     \psi^{(1,1,2)}_{f_2,(n\bar n),n}(s)A_j\cos\Theta_n
     +4k^4\psi^{(1,3,2)}_{f_2,(n\bar n),n}(s)\times
\right .
\ee
$$
\left .
\times
B_j\cos\Theta'_n\right]=\int\limits_{4m^2}^{\infty}\frac{ds'}{\pi}\rho(s')
2\, (-k'^2)
 \psi^{(1,1,2)}_{f_2,(n\bar n),n}(s')A_j\cos\Theta_n\times
$$
$$
\times
\left[
  m^2\,\frac{20}{3}\,
  \sqrt{ss'}\,
    \widetilde V^{(1)}_{(n\bar n \to n\bar n), T}(s,s')
  +\frac13\left(\frac52 ss'+4s'k^2+4sk'^2+\frac{32}{5} k^2k'^2\right)
\times
\right .
$$
$$
\left .
\times
  \widetilde V^{(1)}_{(n\bar n \to n\bar n), V}(s,s')
  -m^2\,\frac{24}{5}\,k^2k'^2
    \widetilde V^{(2)}_{(n\bar n \to n\bar n), S}(s,s')
  -\frac95\,\sqrt{ss'}k^2k'^2
    \widetilde V^{(2)}_{(n\bar n \to n\bar n), A}(s,s')+
 \right .
$$
$$
\left .
 +\frac{24}{7}\, k^4k'^4
    \widetilde V^{(3)}_{(n\bar n \to n\bar n), V}(s,s') \right ]
-\int\limits_{4m^2}^{\infty}\frac{ds'}{\pi}\rho(s')
\, 8\, k'^6 \psi^{(1,3,2)}_{f_2,(n\bar n),n}(s')B_j\cos\Theta'_n\times
$$
$$
\times
\left[
  \frac{5}{14}\, \left(s'+\frac{12}{5} k'^2 \right) k^4 \,
    \widetilde V^{(3)}_{(n\bar n \to n\bar n), V}(s,s')
  +\frac{3}{10}\, \sqrt{ss'}k^2 \,
    \widetilde V^{(2)}_{(n\bar n \to n\bar n), A}(s,s')-
\right .
$$
$$
\left .
  -m^2\,\frac{6}{5} k^2\,
    \widetilde V^{(2)}_{(n\bar n \to n\bar n), S}(s,s')
  +\frac13 \,\left( s+\frac85 k^2\right)\,
    \widetilde V^{(1)}_{(n\bar n \to n\bar n), V}(s,s')
\right]+
$$
$$
+\int\limits_{4m^2_s}^{\infty}\frac{ds'}{\pi}\rho_s(s')
2\, (-{k'}_s^2)_s
 \psi^{(1,1,2)}_{f_2,(s\bar s),n}(s')A_j\sin\Theta_n
 \left[
  mm_s\,\frac{20}{3}\,
  \sqrt{ss'}\,
    \widetilde V^{(1)}_{(s\bar s \to n\bar n), T}(s,s')+
\right .
$$
$$
\left .
  +\frac13\left(\frac52 ss'+4s'k^2+4sk'^2_s+\frac{32}{5}
   k^2k'^2_s\right)
    \widetilde V^{(1)}_{(s\bar s \to n\bar n), V}(s,s')-
\right .
$$
$$
\left .
  -mm_s\,\frac{24}{5}\,k^2k'^2_s
    \widetilde V^{(2)}_{(s\bar s \to n\bar n), S}(s,s')
  -\frac95\,\sqrt{ss'}k^2k'^2_s
    \widetilde V^{(2)}_{(s\bar s \to n\bar n), A}(s,s')+
 \right .
$$
$$
\left .
\nonumber
 +\frac{24}{7}\, k^4k'^4_s
    \widetilde V^{(3)}_{(s\bar s \to n\bar n), V}(s,s') \right ]
-\int\limits_{4m^2_s}^{\infty}\frac{ds'}{\pi}\rho_s(s')
\, 8\, k'^6_s \psi^{(1,3,2)}_{f_2,(s\bar s),n}(s')B_j\sin\Theta'_n\times
$$
$$
\nonumber
\times \left[
  \frac{5}{14}\, \left(s'+\frac{12}{5} k'^2_s \right) k^4 \,
    \widetilde V^{(3)}_{(s\bar s \to n\bar n), V}(s,s')
  +\frac{3}{10}\, \sqrt{ss'}k^2 \,
    \widetilde V^{(2)}_{(s\bar s \to n\bar n), A}(s,s')-
\right .
$$
$$
\nonumber
\left .
  -mm_s\,\frac{6}{5} k^2\,
    \widetilde V^{(2)}_{(s\bar s \to n\bar n), S}(s,s')
  +\frac13 \,\left( s+\frac85 k^2\right)\,
    \widetilde V^{(1)}_{(s\bar s \to n\bar n), V}(s,s')
\right],
$$
and the fourth equation,
\be
(s-M^2) \left[\left(\frac52 s+4k^2_s\right)
     \psi^{(1,1,2)}_{f_2,(s\bar s),n}(s)A_j\sin\Theta_n
     +4k_s^4\psi^{(1,3,2)}_{f_2,(s\bar s),n}(s) \times
\right .
\ee
$$
\left .
\times
B_j\sin\Theta'_n\right]=\int\limits_{4m^2}^{\infty}\frac{ds'}{\pi}\rho(s')
2\, (-k'^2)
 \psi^{(1,1,2)}_{f_2,(n\bar n),n}(s')A_j\cos\Theta_n\times
$$
$$
\times
\left[
  mm_s\,\frac{20}{3}\,\sqrt{ss'}\,
    \widetilde V^{(1)}_{(n\bar n \to s\bar s), T}(s,s')
  +\frac13\left(\frac52 ss'+4s'k^2_s+4sk'^2+\frac{32}{5}
    k^2_sk'^2\right)
\times
\right .
$$
$$
\left .
\times
\widetilde V^{(1)}_{(n\bar n \to s\bar s), V}(s,s')
  -mm_s\,\frac{24}{5}\,k^2_sk'^2
    \widetilde V^{(2)}_{(n\bar n \to s\bar s), S}(s,s')
  -\frac95\,\sqrt{ss'}k^2_sk'^2
    \widetilde V^{(2)}_{(n\bar n \to s\bar s), A}(s,s')+
\right .
$$
$$
\left .
 +\frac{24}{7}\, k^4_sk'^4
    \widetilde V^{(3)}_{(n\bar n \to s\bar s), V}(s,s') \right ]
-\int\limits_{4m^2}^{\infty}\frac{ds'}{\pi}\rho(s')
\, 8\, k'^6 \psi^{(1,3,2)}_{f_2,(n\bar n),n}(s')B_j\cos\Theta'_n\times
$$
$$
\nonumber
\times \left[
  \frac{5}{14}\, \left(s'+\frac{12}{5} k'^2 \right) k^4_s \,
    \widetilde V^{(3)}_{(n\bar n \to s\bar s), V}(s,s')
  +\frac{3}{10}\, \sqrt{ss'}k^2_s \,
    \widetilde V^{(2)}_{(n\bar n \to s\bar s), A}(s,s')-
\right .
$$
$$
\left .
  -mm_s\,\frac{6}{5} k_s^2\,
    \widetilde V^{(2)}_{(n\bar n \to s\bar s), S}(s,s')
  +\frac13 \,\left( s+\frac85 k^2_s\right)\,
    \widetilde V^{(1)}_{(n\bar n \to s\bar s), V}(s,s')
\right]+
$$
$$
+\int\limits_{4m^2_s}^{\infty}\frac{ds'}{\pi}\rho_s(s')
2\, (-k'^2_s)
 \psi^{(1,1,2)}_{f_2,(s\bar s),n}(s')A_j\sin\Theta_n
 \left[
  m^2_s\,\frac{20}{3}\,\sqrt{ss'}\,
    \widetilde V^{(1)}_{(s\bar s \to s\bar s), T}(s,s')+
  \right .
$$
$$
\left .
  +\frac13\left(\frac52 ss'+4s'k^2_s+4sk'^2_s
   +\frac{32}{5}k^2_sk'^2_s\right)
 \widetilde V^{(1)}_{(s\bar s \to s\bar s), V}(s,s')-
\right .
$$
$$
\left .
  -m^2_s\,\frac{24}{5}\,k^2_sk'^2_s
    \widetilde V^{(2)}_{(s\bar s \to s\bar s), S}(s,s')
  -\frac95\,\sqrt{ss'}k^2_sk'^2_s
    \widetilde V^{(2)}_{(s\bar s \to s\bar s), A}(s,s')+
\right .
$$
$$
\left .
 +\frac{24}{7}\, k^4_sk'^4_s
    \widetilde V^{(3)}_{(s\bar s \to s\bar s), V}(s,s') \right ]
-\int\limits_{4m^2_s}^{\infty}\frac{ds'}{\pi}\rho_s(s')
\, 8\, k'^6_s \psi^{(1,3,2)}_{f_2,(s\bar s),n}(s')B_j\sin\Theta'_n\times
$$
$$
\times \left[
  \frac{5}{14}\, \left(s'+\frac{12}{5} k'^2_s \right) k^4_s \,
    \widetilde V^{(3)}_{(s\bar s \to s\bar s), V}(s,s')
  +\frac{3}{10}\, \sqrt{ss'}k_s^2 \,
    \widetilde V^{(2)}_{(s\bar s \to s\bar s), A}(s,s')-
\right .
$$
$$
\left .
  -m^2_s\,\frac{6}{5} k^2_s\,
    \widetilde V^{(2)}_{(s\bar s \to s\bar s), S}(s,s')
  +\frac13 \,\left( s+\frac85 k_s^2\right)\,
    \widetilde V^{(1)}_{(s\bar s \to s\bar s), V}(s,s')
\right].
$$

\newpage
\begin{figure}
\centerline{\epsfig{file=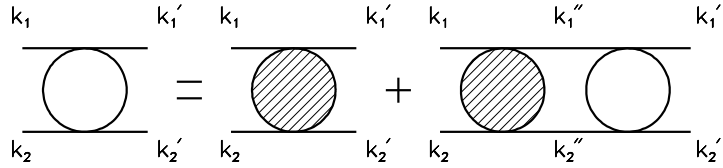,width=10cm}}
\caption{Nonhomogeneous Bethe--Salpeter equation for the scattering
amplitude;  dashed block is the interaction kernel. }
\end{figure}

\begin{figure}
\centerline{\epsfig{file=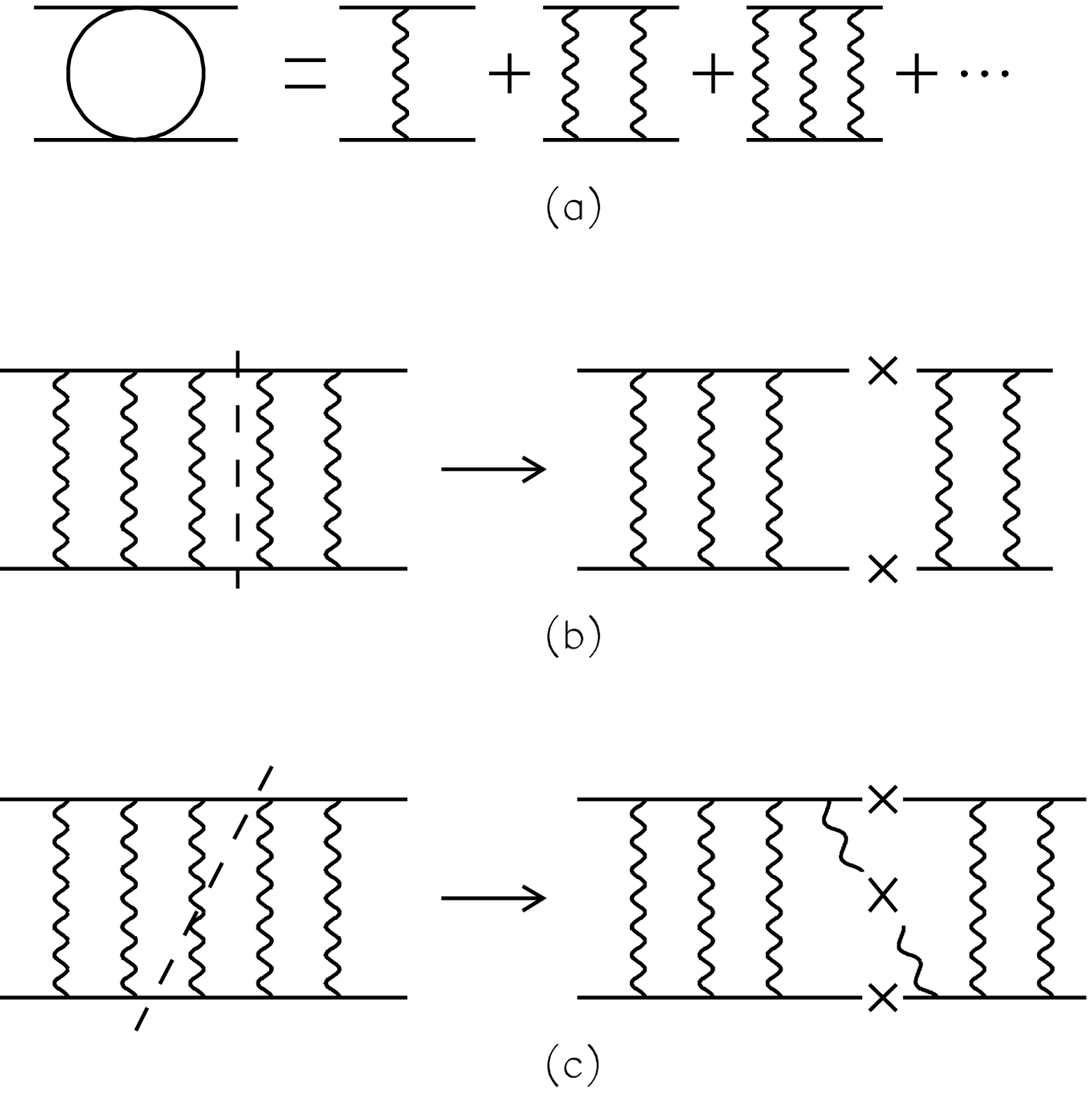,width=10cm}}
\centerline{\epsfig{file=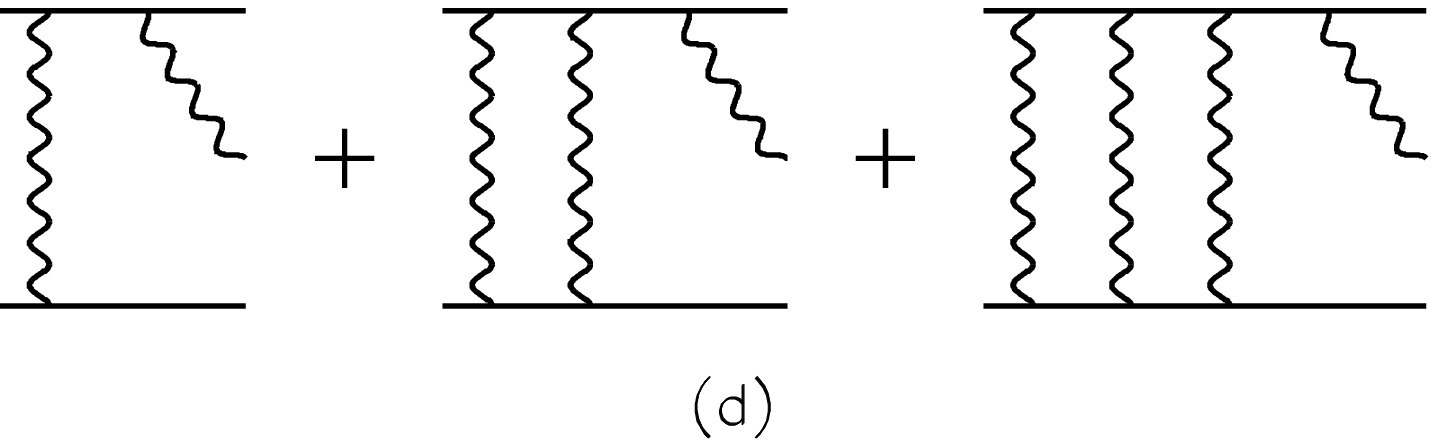,width=10cm}}
\caption{a) Presentation of the scattering amplitude as a set of the
ladder diagrams with the $t$-channel meson exchange interaction; b,c)
cuttings of the ladder diagrams; d) meson production processes which are
determined by ladder diagrams.}
\end{figure}

\begin{figure}
\centerline{\epsfig{file=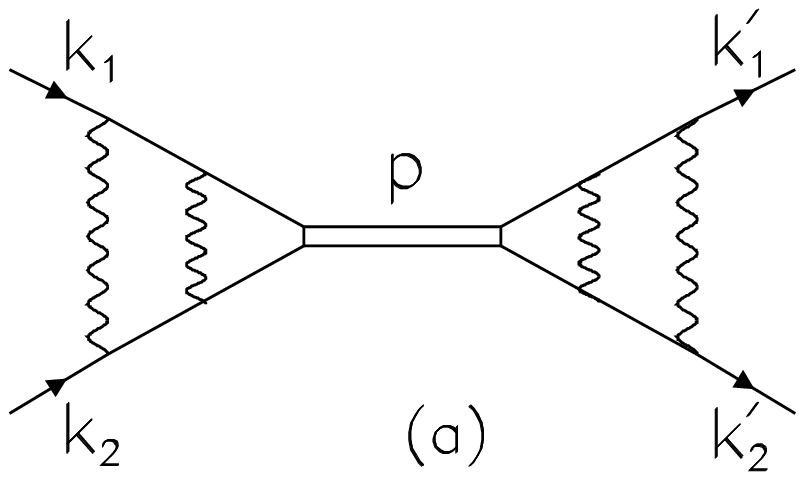,width=7cm}}
\centerline{\epsfig{file=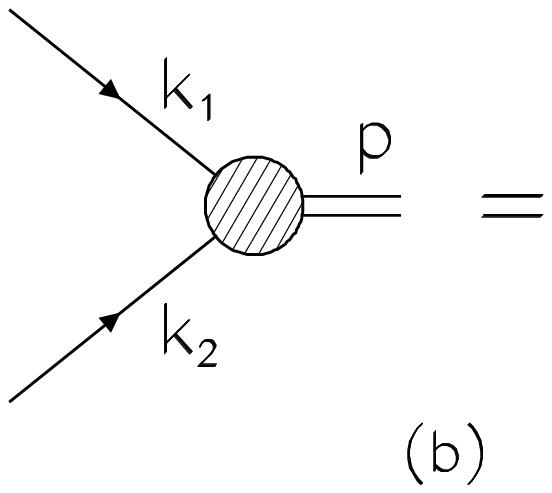,width=5cm}
            \epsfig{file=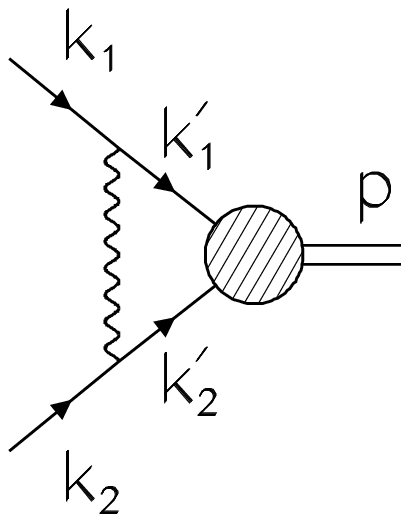,width=5cm}}
\caption{a) Scattering amplitude near the pole related to the bound
state; b) Bethe--Salpeter equation for the bound state}
\end{figure}

\begin{figure}
\centerline{\epsfig{file=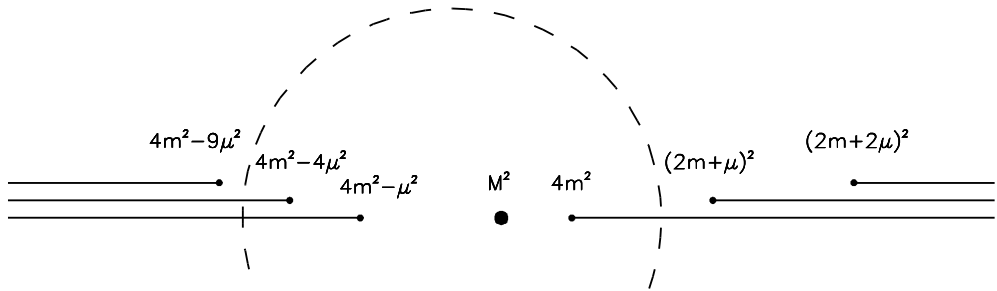,width=15cm}}
\caption{The partial-wave amplitude singularities in the complex-$s$
plane.}
\end{figure}

\begin{figure}
\centerline{\epsfig{file=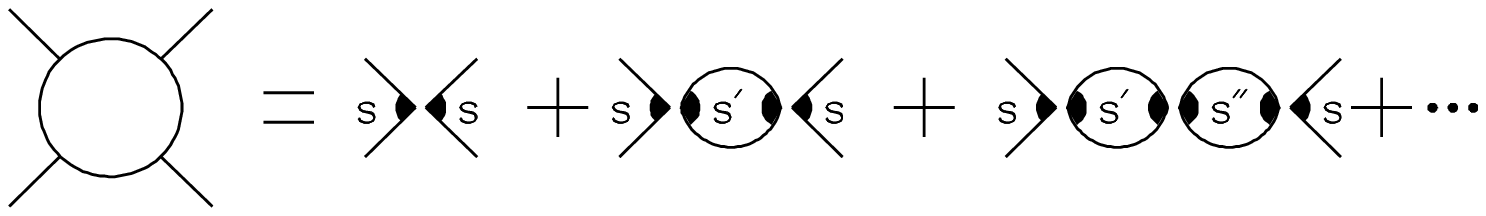,width=15cm}}
\caption{ Scattering amplitude in the dispersion-relation approach
as a set of loop diagrams with separable vertices; there is no
energy conservation in the intermediate states, $s\ne s'\ne s''$, and so
on.}
\end{figure}


\begin{thebibliography}{fussy}
\bibitem{bethe}
E. Salpeter and H.A. Bethe, Phys. Rev. {\bf 84}, 1232 (1951);\\
E. Salpeter, Phys. Rev. {\bf 91}, 994 (1953).
\bibitem{chew}
 G.F. Chew, "The Analytic S-Matrix",
(W.A. Benjamin, New York, 1966);\\
G.F. Chew and S. Mandelstam, Phys. Rev. {\bf 119}, 467 (1960).
\bibitem{deut1}
V.V. Anisovich, M.N. Kobrinsky, D.I. Melikhov, and
A.V. Sarantsev, Nucl. Phys. A{\bf 544}, 747 (1992).
\bibitem{deut2}
A.V. Anisovich and V.A. Sadovnikova, Yad. Fiz.
{\bf 55}, 2657 (1992); {\bf 57}, 75 (1994); Eur. Phys. J. {\bf A2}, 199
(1998).
\bibitem{AS} A.V. Anisovich and A.V. Sarantsev, Yad. Fiz. {\bf
55}, 2163 (1992) [Sov. J. Nucl. Phys. {\bf 55}, 1200 (1992)].
\bibitem{kapitanov}
A.V. Kapitanov, A.V. Sarantsev, Yad. Fiz. {\bf 56}, 156 (1993).
\bibitem{petry}
 V.V. Anisovich, D.I. Melikhov, B.Ch. Metsch,
H.R. Petry, Nucl. Phys. A{\bf 563}, 549 (1993).
\bibitem{nato}
V.V. Anisovich,
in {\it "Hadron Spectroscopy and the Confinement Problem"},
Ed. by D.V. Bugg,( Plenum Press, New York , 1996).
\bibitem{operators} A.V. Anisovich, V.V. Anisovich, V.N. Markov, M.A.
Matveev, V.A. Sarantsev, J. Phys. G. {\bf 28},
15 (2002).
\bibitem{kienzele}
M.N. Kinzle-Focacci,in {\it Proceedings of the VIIIth Blois Workshop, Protvino, Russia, 28
June-2 July 1999}, ed. by  V.A.~Petrov, A.V.~Prokudin, (World
Scientific, 2000).
\bibitem{schegelsky}
 V.A. Schegelsky, in Open Session of HEP Division of PNPI: {\it "On
the Eve of the XXI Century"}, 25-29 December 2000.
\bibitem{novosibirsk}
CMD-2 Collab.(R.R. Akhmetshin {\it et al.}),
Phys.  Lett. B {\bf 462}, 371 (1999); {\bf 462}, 380 (1999);\\
SND Collab.(M.N. Achasov {\it et al.}), Phys. Lett. B {\bf 485}, 349
(2000).
\bibitem{raddecay1}
V.V. Anisovich, D.I. Melikhov, V.A.~Nikonov, Phys. Rev.
D{\bf52}, 5295 (1995); D{\bf55}, 2918 (1997);\\
A.V. Anisovich, V.V. Anisovich, D.V. Bugg, and
V.A. Nikonov, Phys. Lett. B{\bf456} 80 (1999);\\
A.V. Anisovich, V.V. Anisovich, L. Montanet,
 and V.A. Nikonov, Eur. Phys. J. A{\bf 6}, 247 (1999).
\bibitem{raddecay2}
A.V. Anisovich, V.V. Anisovich, and V.A. Nikonov, Eur.
Phys. J. A{\bf 12} 103 (2001);\\
A.V. Anisovich, V.V. Anisovich, V.N. Markov, and V.A. Nikonov,
Yad. Fiz., in press; \\
A.V. Anisovich, V.V. Anisovich, M.A. Matveev, and V.A. Nikonov,
Yad. Fiz., in press. \\
\bibitem{syst} A.V. Anisovich, V.V. Anisovich, and A.V. Sarantsev,
Phys. Rev. D {\bf62}, 051502 (2000);\\
A.V. Anisovich, V.V. Anisovich, V.A. Nikonov, and A.V. Sarantsev,
in: {\it "PNPI XXX, Scientific Highlights, Theoretical Physics
Division"},Gatchina (2001),p. 58;\\
V.V. Anisovich, "Systematics of $q \bar q$ states and glueball", hep-ph/
0110326
\bibitem{RAL} A.V. Anisovich, C.A. Baker, C.J. Batty {\em et al.},
Phys. Lett. B{\bf449}, 114 (1999); B{\bf452}, 173 (1999);
B{\bf452}, 180 (1999); B{\bf452}, 187 (1999);
B{\bf472}, 168 (2000);  B{\bf476}, 15 (2000);  B{\bf477}
, 19 (2000); B{\bf491}, 40 (2000);  B{\bf491}, 47 (2000);
B{\bf496}, 145 (2000); B{\bf507}, 23 (2001);
B{\bf508}, 6 (2001); B{\bf513}, 281 (2001); B{\bf517}, 261 (2001);
{\bf B517}, 273 (2001); \\
Nucl. Phys. A{\bf651}, 253 (1999);
A{\bf662}, 319 (2000); A{\bf662}, 344 (2000).
\bibitem{PDG-00}PDG Group(D.E. Groom {\it et al.}),
Eur. Phys. J. C {\bf 15}, 1 (2000).
\bibitem{petry2} R. Ricken, M. Koll, D. Merten, B.C. Metsch, and
H.R. Petry, Eur. Phys. J. A {\bf 9}, 221 (2000).
\bibitem{inst}
R.D. Carlitz and D.B. Creamer, Ann. Phys. {\bf 118}, 429 (1979);\\
E.V. Shuryak, Nucl. Phys. {\bf B203}, 93 (1982);\\
V.V. Anisovich, S.M. Gerasyuta, and A.V. Sarantsev,
Int. J. Mod. Phys. {\bf A6}, 625 (1991).
\bibitem{K}
 V.V. Anisovich and A.V. Sarantsev,
 hep-ph/0204328.
\bibitem{CDD}
L. Castillejo, F.J. Dyson,and R.H. Dalitz, Phys. Rev. {\bf 101}, 453
(1956).


\end{thebibliography}
\end{document}